\documentclass[12pt]{article}
\pdfoutput=1
\usepackage{amsmath, amsthm,comment}
\usepackage{amsfonts}
\usepackage{amssymb,graphics,psfrag}
\usepackage{array,epsfig,stmaryrd,graphicx}
\usepackage{cite}
\usepackage{graphicx}
\usepackage{makeidx}
\usepackage{multicol}
\usepackage{geometry}
\usepackage{amsfonts}
\usepackage{mathrsfs}
\usepackage{amssymb}
\usepackage{amsmath}
\usepackage{slashed}
\usepackage{heck}%
\providecommand{\U}[1]{\protect\rule{.1in}{.1in}}

\newcommand{\beq}{\begin{equation}}
\newcommand{\eeq}{\end{equation}}
\newcommand{\be}{\begin{equation}}
\newcommand{\ee}{\end{equation}}
\newcommand{\bea}{\begin{eqnarray}}
\newcommand{\eea}{\end{eqnarray}}
\newcommand{\ben}{\begin{eqnarray*}}
\newcommand{\een}{\end{eqnarray*}}
\newcommand{\ba}{\begin{aligned}}
\newcommand{\ea}{\end{aligned}}
\newcommand{\bt}{\begin{tabular}}
\newcommand{\et}{\end{tabular}}
\newcommand{\bc}{\begin{center}}
\newcommand{\ec}{\end{center}}

\newcommand{\cref}{{\bf [check ref]}}

\newcommand{\bs}{\begin{subarray}{c}}
\newcommand{\es}{\end{subarray}}

\begin{document}

\date{January, 2010}
\title{F-theory and the LHC: Stau Search}

\preprint{arXiv:1001.4084}

\institution{IAS}{\centerline{${}^{1}$School of Natural Sciences, Institute for Advanced Study, Princeton, NJ 08540, USA}}

\institution{SyracuseU}{\centerline{${}^{2}$Department of Physics, Syracuse University, Syracuse, NY 13244, USA}}%

\institution{HarvardU}{\centerline{${}^{3}$Jefferson Physical
Laboratory, Harvard University, Cambridge, MA 02138, USA}}

\authors{Jonathan J. Heckman\worksat{\IAS}%
\footnote{e-mail: \texttt{jheckman@sns.ias.edu}%
}, Jing Shao\worksat{\SyracuseU}%
\footnote{e-mail: \texttt{jishao@syr.edu}%
} and Cumrun Vafa\worksat{\HarvardU}%
\footnote{e-mail: \texttt{vafa@physics.harvard.edu}%
}}

\abstract{F-theory GUT models favor a relatively narrow range of soft supersymmetry breaking parameters in the
MSSM Lagrangian. This leads to the specific predictions that a $10-100$ MeV mass gravitino is the LSP,
and the NLSP is quasi-stable, with a lifetime between a second to an hour.
In a wide range of parameter space, the NLSP turns out to be a stau, though a bino-like
lightest neutralino is also possible. Focusing on F-theory GUTs with a stau NLSP, we
study the discovery potential at the LHC for such scenarios. Models with a quasi-stable
stau predict a striking signature of a heavy charged particle passing
through the detector. As a function of the parameters
of minimal F-theory GUTs, we study how many of such events to expect,
and additional signatures correlated with the presence of quasi-stable staus. We
also study the prospects for staus to become stopped in or near
the detector, as well as potential ways to distinguish such models from minimal
gauge mediation models with similar spectra.}

\maketitle

\enlargethispage{\baselineskip}
\tableofcontents

\section{Introduction}

A few decades of theoretical research has led to many novel ideas for
what to expect at the TeV energy scale. Perhaps the most well motivated idea
is that supersymmetry will soon be discovered. In addition to its aesthetic
appeal, supersymmetry provides a relatively simple
resolution of the hierarchy problem. There is also circumstantial evidence
for supersymmetry both from the prescient prediction of a heavy top
mass, and the remarkable unification of the gauge coupling constants in the MSSM,
which does not work with only Standard Model degrees of freedom.
Though the lack of detection necessarily constrains the available parameter
space, leading to a mild degree of fine-tuning (the mini-hierarchy problem),
supersymmetry remains among the most robust, and motivated options
for beyond the Standard Model physics.

Another motivation for supersymmetry is that at high energies it is predicted
by string theory. Given the existence of meta-stable vacua where supersymmetry
is broken, this suggests the attractive possibility that at least for matter
charged under the Standard Model gauge group, supersymmetry persists down
to the TeV scale.

Even so, this by itself is not particularly predictive, because the details
of supersymmetry breaking strongly affect the superpartner masses and interactions
with the Standard Model. This in turn directly affects the expected collider
signatures of a particular model.

In the past few years, F-theory based models have emerged as promising
candidates for particle phenomenology \cite{DonagiWijnholt,BHVI,WatariTATARHETF,IbanezSUSYFTHEORY,BuchbinderSUSY,BHVII,HMSSNV,MarsanoGMSB,HVGMSB,DonagiWijnholtBreak,
MarsanoToolbox,HVLHC,Font:2008id,HVCKM,Blumenhagen:2008zz,Blumenhagen:2008aw,FGUTSCosmo,Bourjaily:2009vf,
Hayashi:2009ge,Andreas:2009uf,Chen:2009me,HKSV,DonagiWijnholtIII,BHSV,Randall:2009dw,HVCP,Bourjaily:2009ci,
Tatar:2009jk,Jiang:2009za,Collinucci:2009uh,Blumenhagen:2009up,EPOINT,Marsano:2009gv,Conlon:2009qa,Font:2009gq,
Blumenhagen:2009yv,FGUTSNC,Li:2009fq,Conlon:2009qq,Choi:2009tp,Hayashi:2009bt,Cordova:2009fg,Marchesano:2009rz,
Chung:2009ib,Vafa:2009se,Marsano:2009wr,Dudas:2009hu,Heckman:2010bq}. In these models, the Standard Model
gauge group and matter originate in a compact region of the
six extra dimensions of string theory. Physically, this corresponds to a regime
where particle physics can be treated in a limit decoupled from gravity. Gauge groups localize on four
real-dimensional subspaces, chiral matter localizes on two real-dimensional
Riemann surfaces, and cubic interaction terms between chiral matter
localize at points of the geometry. Flavor physics in particular appears
to play a key role in constraining candidate geometries \cite{HVCKM,BHSV,EPOINT}, requiring a
point of $E_8$ enhancement \cite{EPOINT}.

Combining this with the assumption that supersymmetry persists down to the TeV scale
severely limits consistent scenarios. A deformation of minimal gauge mediated supersymmetry breaking
seems to be the most natural option in F-theory GUTs \cite{BHVII,MarsanoGMSB,HVGMSB}, though other
approaches are also possible.\footnote{See for example \cite{Dermisek:2007qi,Verlinde:2007qk,Grimm:2008ed,BuicanINST} and in the context of moduli mediation \cite{Conlon:2005ki,Conlon:2006tj,Conlon:2007dw,Conlon:2007gk,Conlon:2006wz,IbanezSUSYFTHEORY}.} Parameterizing the effects of
supersymmetry breaking through the vev of a GUT singlet $\langle X \rangle = x + \theta^2 F$, in F-theory GUT
deformations of minimal gauge mediation models, the $\mu$-term is generated through the Giudice-Masiero type operator
\begin{equation}\label{eq:GM}
\int d^4\theta {X^\dagger H_uH_d\over \Lambda_{UV}},
\end{equation}
where $\Lambda_{UV}$ is fixed by the internal scale of the geometry to be near the
GUT scale. In order to generate a weak scale value for $\mu$, this requires a
specific scale of supersymmetry breaking
$$F\sim M_{weak} M_{GUT}\sim 10^{17} \text{ GeV}^2.$$
This fixes a narrow range of values for $\sqrt{F} \sim 10^{8} - 10^{9}$ GeV, which
implies that the gravitino is the lightest superpartner (LSP) of mass
$10-100$ MeV, and that the next lightest superpartner (NLSP) is
quasi-stable on timescales probed by collider detectors, with a lifetime on
the order of one second to an hour. The primary decay chains of all
MSSM superpartners involve the NLSP, which strongly influences the experimental signatures of an F-theory
GUT.

This begs the question: Which particle is the NLSP? In F-theory GUTs, constraints from other
sectors of the model force the lightest stau $\widetilde{\tau}_1$ to be the NLSP over much of parameter space. This
is due to two effects: First, in minimal models with a point of $E_8$ unification, the messengers of the gauge mediation model
tend to be in vector-like pairs transforming in the $10 \oplus \overline{10}$ of $SU(5)$. In comparison with messengers in the
$5 \oplus \overline{5}$, the higher dimension of the $10$ representation lowers the mass of the stau relative to the
gauginos. Requiring that gravity can be decoupled requires an asymptotically free $SU(5)$ GUT group which constrains the number of vector-like pairs in the $10 \oplus \overline{10}$ to $N_{10} \leq 2$. Moreover, there is also a stringy ``PQ deformation'' of the spectrum due to heavy gauge boson exchange of an anomalous $U(1)_{PQ}$ \cite{HVGMSB}. This also has the effect of lowering the stau mass by an amount $\Delta_{PQ}$ which tends to be on the order of the weak scale. Even so, for a smaller range of
parameter space a bino-like lightest neutralino $\widetilde{\chi}^{0}_1$ may be the NLSP. In a very particular
Dirac neutrino scenario the right-handed sneutrino can also be the NLSP, but in this case the next to next to lightest superpartner (NNLSP)
is also quasi-stable, and is typically the stau.

If the NLSP is a stau, it will have dramatic consequences for the LHC.  It is a heavy charged particle leaving the detector!
Moreover, since R-parity is preserved, the number of such staus will always come in pairs. As a massive
quasi-stable charged particle, the stau will initially register
as a muon in the detector. The primary challenge is therefore to distinguish such ``fake muons'' from actual muons. Such stau
events have extremely low Standard Model background, and can easily be isolated from actual muons through suitable selection cuts.
Stau NLSP scenarios have been extensively studied in the literature \cite{Martin:1998vb,Hinchliffe:1998ys,Ambrosanio:2000ik,
Ambrosanio:2000zu,Ellis:2006vu,Ito:2009xy}. One of the most attractive phenomenological features of
such scenarios is that because it is relatively easy to detect, it is possible to extract
its mass. Correlating this with the four-momenta of other final state products,
this provides a means to proceed up a decay chain, extracting the masses of other
superpartners \cite{Ellis:2006vu}.

In this paper we perform a parameter space scan over F-theory GUTs, showing that over a broad range of parameter space the stau is an NLSP.
Next, we study the prospects for detecting signatures of stau NLSP F-theory GUT scenarios at the LHC. See \cite{HKSV} for a
study of related F-theory GUT scenarios with a bino NLSP. The expected signatures in stau NLSP scenarios greatly depend on both
the center-of-mass energy and the integrated luminosity, and with increases in each, the prospects for detecting staus improve. We find
that with energies which will soon be achieved at the LHC, such signatures will indeed be within reach!

Though the primary signature of stau NLSP scenarios is the presence of ``fake muons'',
the parameters of an F-theory GUT also strongly determine the additional signatures expected. We find that the number of leptons in the final state also depends on the number of messengers $N_{10}$, and the size of $\Delta_{PQ}$. For a single $10 \oplus \overline{10}$ pair of messengers and moderate PQ deformation, many of the staus are unaccompanied by other final states. For larger values of the PQ deformation, and for more messengers, additional sleptons become lighter than gauginos, providing new decay channels. This tends to increase the number of leptons expected with a given stau signature.

We also study the prospects for generating staus which can become stuck in or near the detector and then decay at some later time. When these staus eventually decay to a gravitino and tau, observing such events would provide additional evidence for quasi-stable staus. Further, for staus stopped in the detector, correlating the presence of the tau produced from the decay of the stau with the initial collision event can in principle provide a means to measure the lifetime of the NLSP, and consequently the scale of supersymmetry breaking.

Another remarkable feature of such scenarios is that because a stau NLSP is detectable, it is possible to reconstruct with high precision the masses of sparticles which decay into a stau.  We review the expected level of precision in such mass fits, and then explain how this level of accuracy provides a means to break degeneracies between F-theory GUTs and minimal gauge mediation models with a stau NLSP which might otherwise be present with cruder signatures.

The organization of the rest of this paper is as follows. In section \ref{sec:ParamSpace} we review the salient features of F-theory GUTs. In section \ref{sec:NNLSP} we show that for a wide range of parameter space, the stau is the NLSP, while over a smaller range, the bino is the NLSP. We next move to collider signatures of F-theory GUTs with a quasi-stable stau NLSP. In section \ref{sec:LongLive} we discuss details of the simulation performed and compute the production cross sections for supersymmetric events, and in section \ref{sec:SEARCH} we study search channels of potential interest. Section \ref{sec:STOPPED} discusses the prospects for observing stopped staus, and in section \ref{sec:distinguish} we study whether mass reconstruction based on the stau NLSP can be used to distinguish F-theory GUTs from minimal gauge mediation models with similar spectra. Section \ref{sec:CONCLUDE} contains our conclusions. Appendix A contains additional details on the simulation, Appendix B contains some additional discussion on the selectron mass in F-theory GUTs and minimal gauge mediation, and Appendix C contains miscellaneous plots.

\section{Parameter Space of F-theory GUTs}\label{sec:ParamSpace}

Our aim in this paper will be to study the phenomenological consequences of a
particularly minimal class of F-theory GUT models based on a point of $E_{8}$
unification \cite{EPOINT}, and the corresponding supersymmetry breaking
scenarios. See \cite{IbanezSUSYFTHEORY} for other supersymmetry breaking scenarios in
F-theory GUTs, and their experimental signatures.

In an F-theory GUT, all of the particle content of the Standard Model
localizes on a spacetime filling seven-brane which wraps a complex
two-dimensional surface. The matter fields of the MSSM\ localize on complex
one-dimensional curves formed by pairwise intersections of the GUT\ stack with
other seven-branes, and the Yukawa couplings of the MSSM\ localize at points
of the geometry formed by triple intersections of seven-branes. In this type
of setup, achieving realistic flavor hierarchies requires that these Yukawa
points be close together. In minimal setups, there is a single point where the
geometry enhances up to a point of $E_{8}$ symmetry which is broken down to
$SU(6)$ and $SO(10)$ along curves supporting respectively $5$'s and $10$'s of
$SU(5)$, and is broken down to $SU(5)_{GUT}$ over the bulk of the seven-brane
worldvolume. Minimal F-theory GUT models are especially predictive, and these
are the scenarios we study in this paper.

The supersymmetry breaking sector localizes on another seven-brane, which supports
an anomalous $U(1)_{PQ}$ gauge boson. This seven-brane intersects
the GUT seven-brane, leading to the matter of the MSSM, all of which is
charged under $U(1)_{PQ}$. Supersymmetry breaking is triggered by the vev of a
GUT\ singlet $X$:%
\begin{equation}
\left\langle X\right\rangle =x+\theta^{2}F.
\end{equation}
Supersymmetry breaking is communicated to the visible sector through gauge mediation \cite{DineFischlerSusyTech,DimopoulosRabySupercolor,DineFischlerII,NappiOvrut,
DineFischlerIII,AlvarezGaumeClaudsonWiseLow,DimopoulosRabyGeometric,
DineNelsonGMSB,DineNelsonShirman,DineNelsonNirShirman} as well as through the
exchange of a heavy anomalous $U(1)$ gauge boson. See \cite{GiudiceSUSYReview}
for a review of gauge mediation. In the deformation of minimal gauge
mediation considered in \cite{HVGMSB} (see also \cite{MarsanoGMSB}), the effects of supersymmetry
breaking are transmitted to the messengers through the cubic interaction term:%
\begin{equation}
\int d^{2}\theta\text{ }XYY^{\prime},
\end{equation}
for messenger fields $Y$ and $Y^{\prime}$ in vector-like pairs of complete
$SU(5)$ GUT\ multiplets. Once $X$ develops a supersymmetry breaking vev, loops with gauge fields will
then communicate this to the visible sector. Assuming that the interaction $X Y Y^{\prime}$ also
localizes at the same point of $E_{8}$ unification, it can be shown that in all but one Dirac
neutrino scenario, the messenger fields are \textit{forced} to transform in the
$10\oplus\overline{10}$ of $SU(5)_{GUT}$ \cite{EPOINT}. We denote the number of such vector-like pairs by
$N_{10}$. Heuristically, the requirement that we can decouple gravity from the gauge theory requires that $SU(5)_{GUT}$
is an asymptotically free gauge theory, which implies $N_{10} = 1$ or $2$.

When the $\mu$-term is generated
through the higher-dimension operator of equation \eqref{eq:GM}, the scale of supersymmetry breaking is fixed to be:
\begin{equation}
F \sim M_{weak} \cdot M_{GUT}.
\end{equation}
Introducing the characteristic scale:
\begin{equation}
\Lambda \equiv \frac{F}{x}
\end{equation}
of minimal gauge mediation, we can express the messenger scale $M_{mess}$ in terms of $x$ and the remaining parameters as:
\begin{equation}
x \equiv M_{mess} \sim \frac{M_{weak} \cdot M_{GUT}}{\Lambda}.
\end{equation}
Phenomenologically, we require $\Lambda \sim 100$ TeV to achieve a realistic mass spectrum. Including all
order one coefficients, we have \cite{HVGMSB}:
\begin{align}
F  &  \sim 10^{17}\text{ GeV}^{2}\\
x  &  \sim10^{12}\text{ GeV.}%
\end{align}
The specific range of supersymmetry breaking and messenger scales, as well as
the constrained messenger content constitutes one important prediction of this
class of models. The vev $x$ also doubles as the effective axion decay
constant, which is an accord with the present window of viable values.

At the messenger scale, localization on seven-branes allows an F-theory
GUT to generate a weak scale value for the $\mu$-parameter. Localization on seven-branes also implies
that both $B \mu$ and the A-terms vanish at the messenger scale:
\begin{align}
B\mu(M_{mess})  &  =0\\
A_{ijk}(M_{mess})  &  =0.
\end{align}
At lower energy scales the parameter $B\mu$ is radiatively generated, and the
ratio of the Higgs up and down vevs, $v_{u}/v_{d} \equiv \tan\beta\sim20-30$
is large \cite{RattazziSaridLargeTanBeta}. Another consequence of generating $B\mu$ in this way
is that proper electro-weak symmetry breaking then requires $\mu M_{1/2} \tan \beta > 0$,
where $M_{1/2}$ denotes the gaugino masses (see for example
\cite{BaggerMatchevPierceGMSB,RattazziSaridLargeTanBeta,SignMartin}). Our sign
convention is the same as \cite{SOFTSUSYAllanach} where $M_{1/2}, \tan \beta >0$,
which in turn implies $\mu > 0$. To a large extent, the value of $\Lambda$
determines the amount of fine-tuning in the Higgs sector. In particular, the
current bound on the Higgs mass of $\sim114$ GeV translates to a lower bound
on $\Lambda$ which we denote as $\Lambda_{\min}$. The value of $\Lambda_{\min
}$ also depends on the messenger content, so that for $N_{10}=1$ and $2$, we
have:%
\begin{align}
\Lambda_{\min}(N_{10}  &  =1)=50\text{ TeV}\\
\Lambda_{\min}(N_{10}  &  =2)=28\text{ TeV.}%
\end{align}
To illustrate the broader features of this parameter space scan, we shall
sometimes allow slightly lower values of $\Lambda$. The
maximal value of $\Lambda$ we consider is based on the requirement that we not
overly fine-tune the Higgs sector, which to some extent is a matter of
aesthetics. The upper bound on $\Lambda$ we take is:%
\begin{align}
\Lambda_{\max}(N_{10}  &  =1)=95\text{ TeV}\\
\Lambda_{\max}(N_{10}  &  =2)=53\text{ TeV.}%
\end{align}
Since the messenger scale also plays a role in setting the scale of the axion
decay constant, we shall effectively treat $M_{mess}$ as a fixed parameter. A
change in $\Lambda$ therefore corresponds to altering the scale of
supersymmetry breaking.

The soft masses at the messenger scale receive additional corrections from
integrating out the heavy PQ gauge boson:%
\begin{equation}
L\supset-4\pi\alpha_{PQ}e_{X}e_{\Psi}\int d^{4}\theta\frac{X^{\dag}X\Psi
^{\dag}\Psi}{M_{U(1)_{PQ}}^{2}}%
\end{equation}
where $e_{X}$ and $e_{\Psi}$ respectively denote the PQ charges of $X$ and the
MSSM\ chiral superfield $\Psi$, and $M_{U(1)_{PQ}}$ is the mass of the
PQ\ gauge boson. Once $X$ develops a supersymmetry breaking vev, this induces
a shift in the soft masses at the messenger scale:%
\begin{equation}
m_{soft}^{2}=m_{mGMSB}^{2}+q_{\Psi}\Delta_{PQ}^{2}. \label{pdef}%
\end{equation}

Compatibility with minimal Majorana and Dirac neutrino scenarios \cite{BHSV} yields two
different choices for the PQ charges. In a normalization where the operator%
\begin{equation}
-4\pi\alpha_{PQ}e_{X}e_{X}\int d^{4}\theta\frac{X^{\dag}X\Psi^{\dag}\Psi
}{M_{U(1)_{PQ}}^{2}}%
\end{equation}
has the same coefficient in both Majorana and Dirac scenarios, the constant of
proportionality $q$ appearing in equation (\ref{pdef}) is:%
\begin{equation}%
\begin{tabular}
[c]{|c|c|c|c|c|}\hline
& $10_{M}$ & $\overline{5}_{M}$ & $5_{H}$ & $\overline{5}_{H}$\\\hline
$q_{\text{Majorana}}$ & $-4/5$ & $-8/5$ & $+8/5$ & $+12/5$\\\hline
$q_{\text{Dirac}}$ & $-1$ & $-1$ & $+2$ & $+2$\\\hline
\end{tabular}
\ \ \ \ \ \ \ \ . \label{pqcharges}%
\end{equation}
Technically speaking, there is slightly more freedom in the Dirac scenario,
because there is also a $U(1)_{B-L}$ symmetry available in Dirac neutrino
scenarios. Since the effects of the PQ deformation are similar in Majorana and
Dirac scenarios, we shall focus on Majorana scenarios.

The PQ deformation lowers the soft masses of the chiral matter fields, while
raising the soft masses in the Higgs sector. For sufficiently large values of
$\Delta_{PQ}$, this induces a tachyonic mode in the spectrum, which imposes an
upper bound on the value of $\Delta_{PQ}$ on the order of $200-400$ GeV, the
precise value of which depends on $\Lambda$. Let us also note that there is also an experimental
upper bound on the size of $\Delta_{PQ}$ because of the present $100$ GeV bound on the mass of
the stau \cite{Abbiendi:2003yd}.
\begin{figure}
[ptb]
\begin{center}
\includegraphics[
height=3.87in,
width=6.0001in
]%
{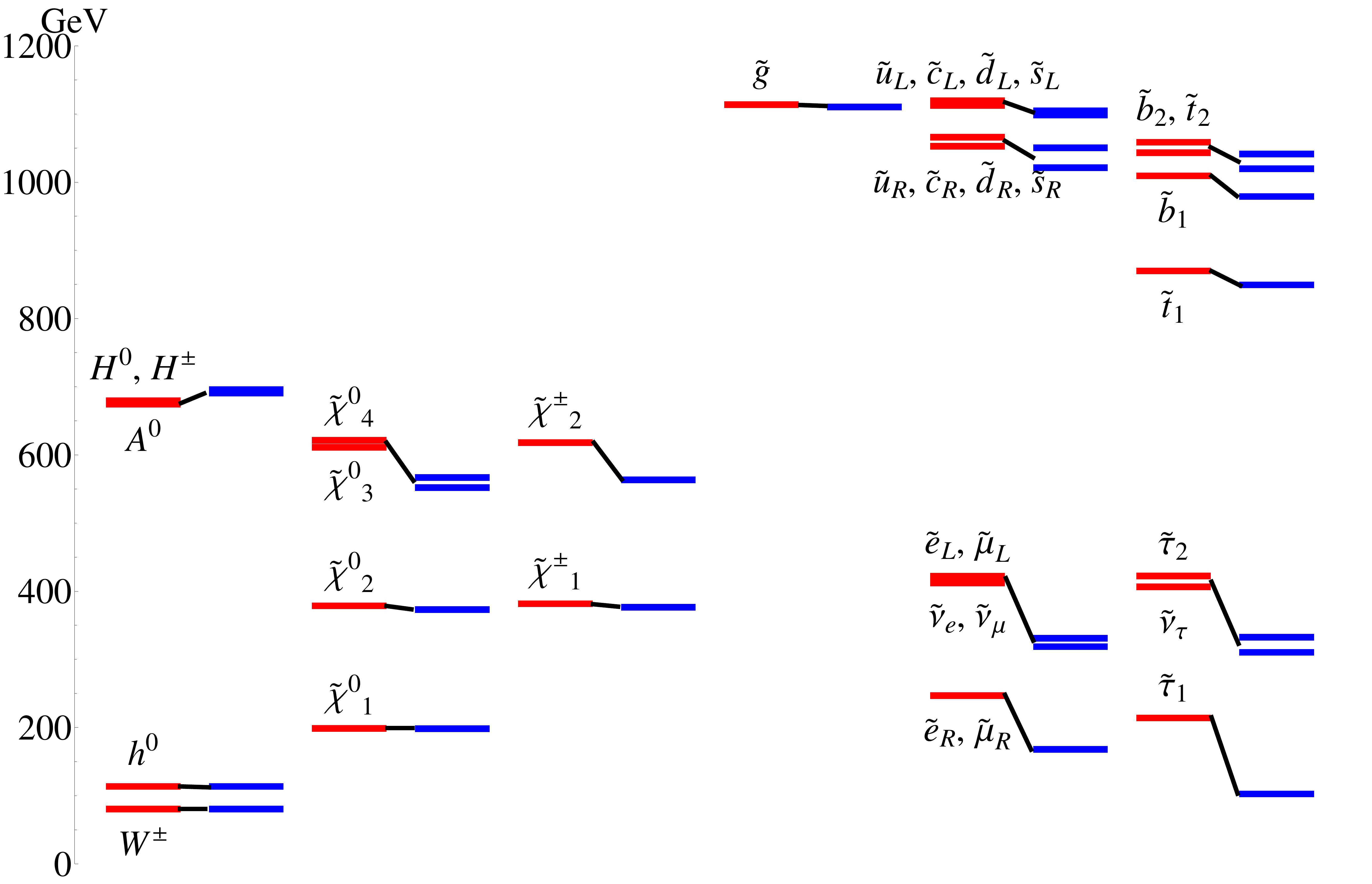}%
\caption{Spectrum of an F-theory GUT Majorana neutrino scenario with
$N_{10}=1$, $\Lambda=50$ TeV for minimal (red left columns) and
maximal (blue right columns) PQ deformation of order $200$ GeV. At small
values of the PQ deformation, the bino is the NLSP, whereas for larger values
this transitions to the stau. Further note that at larger PQ deformation, the
right-handed selectron and smuon are also lighter than the bino, and all of
the sleptons are lighter than the second neutralino and lightest chargino.}%
\label{n10eq1together}%
\end{center}
\end{figure}

Summarizing, the parameter space of an F-theory GUT which we scan over is:
\begin{itemize}
\item Fixed messenger scale $M_{mess}\sim10^{12}$ GeV

\item For $N_{10}=1:50$ TeV $\lesssim\Lambda(N_{10}=1)\lesssim
95$ TeV

\item For $N_{10}=2:28$ TeV $\lesssim\Lambda(N_{10}%
=2)\lesssim53.2$ TeV

\item $0\leq\Delta_{PQ}\lesssim200-400$ GeV
\end{itemize}

See figure \ref{n10eq1together} for a plot of the sparticle
spectrum of a representative F-theory GUT with $N_{10}=1$, $\Lambda = 50$ TeV
at minimal and maximal PQ\ deformation, and figure \ref{n10eq2together}\ of Appendix C
for the case of $N_{10}=2$ and $\Lambda = 28$ TeV.

\section{The LSP, NLSP and NNLSP}\label{sec:NNLSP}

The collider phenomenology of a supersymmetric model strongly depends on which
sparticle is the lightest superpartner (LSP), the next lightest superpartner
(NLSP), and sometimes the next to next lightest superpartner (NNLSP). In this section,
we study the spectrum of F-theory GUTs, and show that the LSP\ is always the
gravitino, and in Majorana neutrino scenarios the NLSP\ is either the bino or
the stau. In Dirac neutrino scenarios, we find that for small values of the PQ
deformation, a right-handed sneutrino is also a possible NLSP, though in such
cases the NNLSP which is either a bino or stau is still quasi-stable on the same
timescale as the NLSP. To perform our numerical analysis of the spectrum we use
the program \texttt{SOFTSUSY} \cite{SOFTSUSYAllanach}.

The narrow window for supersymmetry breaking implies that the gravitino is the LSP, with mass:%
\begin{equation}
m_{3/2}=\sqrt{\frac{4\pi}{3}}\frac{F}{M_{pl}}\sim10-100\text{ MeV.}%
\end{equation}
The gravitino is stable, and all of the other MSSM\ superpartners eventually
decay to it, plus their Standard Model counterparts. In this class of models,
such gravitinos provide a viable dark matter candidate \cite{FGUTSCosmo},
much as in the sweet spot model of supersymmetry breaking \cite{KitanoIbeSweetSpot}.

In many minimal gauge mediation models, the NLSP\ is a stau or a bino-like
neutralino, and the situation in F-theory GUTs is no different. What is
different in F-theory GUTs is that there is a specific range of values for the parameters of the model.
Even without knowing the precise identity of the NLSP, the scale of
supersymmetry breaking dictates that it will be quasi-stable, with a lifetime
on the order of one second to an hour \cite{HVLHC}, so that it will appear stable to
a collider detector.

Increasing the number of messengers or equivalently the dimension of the
representation to which they belong tends to lower the masses of
the scalars relative to the gauginos. For example, in a model with $N_{5}$
messengers in the $5 \oplus \overline{5}$, the scalar and gaugino masses at the
messenger scale are:%
\begin{align}
m_{\text{scalar}}  &  =\sqrt{N_{5}}\cdot C_{\text{scalar}}\frac{\alpha}{4\pi
}\Lambda\\
m_{\text{gaugino}}  &  =N_{5}\cdot C_{\text{gaugino}}\frac{\alpha}{4\pi
}\Lambda,
\end{align}
where schematically, the $C$'s are constants fixed by the quadratic Casimirs
of various representations, and the $\alpha$'s are the fine structure
constants of the gauge groups under which the sparticles are charged. This is
especially important for the present class of models, where the messengers are
in vector-like pairs of $10\oplus\overline{10}$'s, the number of which we
denote by $N_{10}$. The soft masses are proportional to the Dynkin index of
the messengers. This means that each messenger pair in the $10\oplus
\overline{10}$ has the same effect as three in the $5\oplus\overline{5}$, so
that $3N_{10}=N_{5}$.%

\begin{figure}
[ptb]
\begin{center}
\includegraphics[
height=5.5633in,
width=6.1947in
]%
{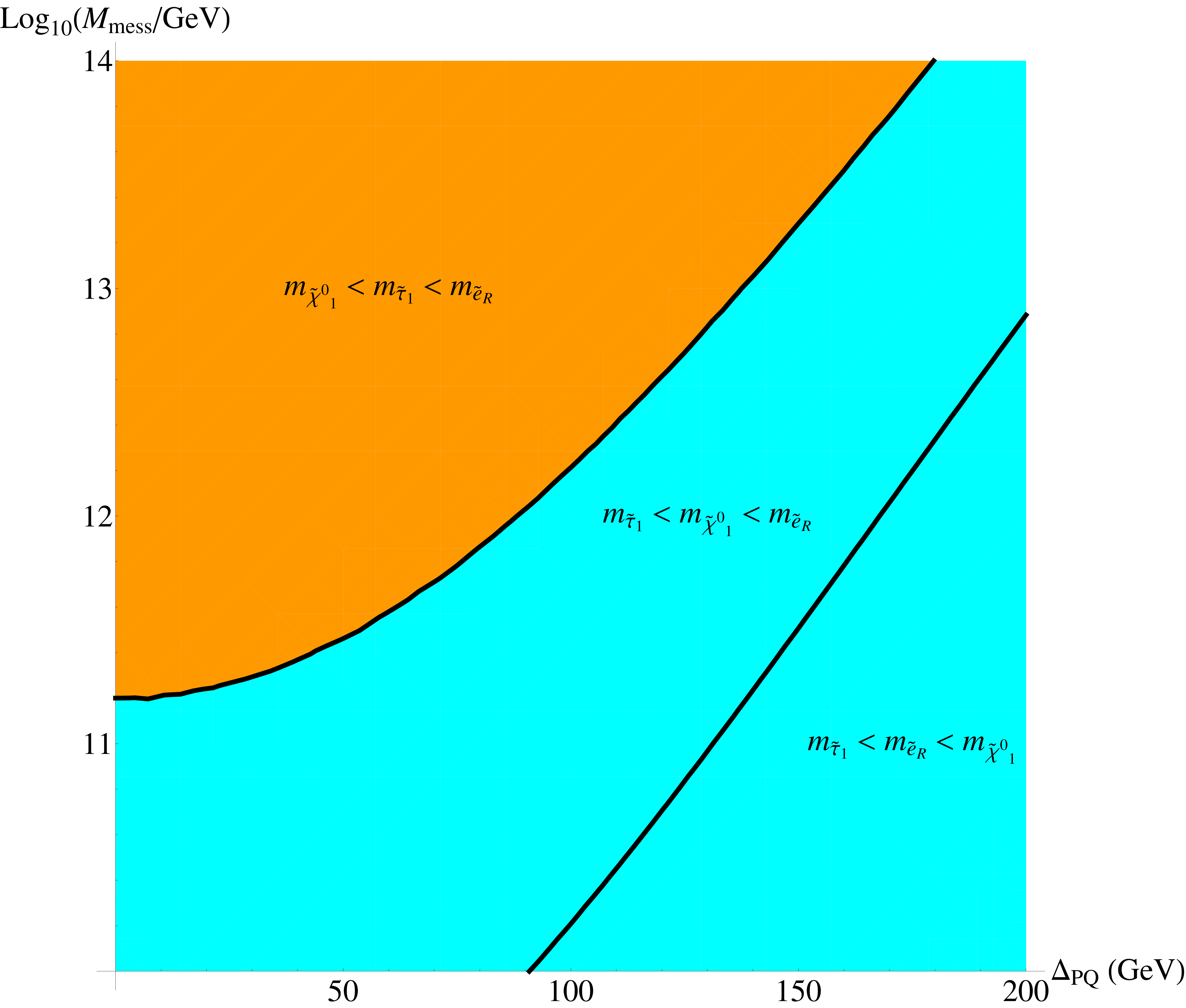}%
\caption{Contour plot indicating the relative masses of the bino, lightest stau and right-handed
selectron as function of $\Delta_{PQ}$ and $M_{mess}$ in the representative F-theory GUT
scenario with $N_{10} = 1$ and $\Lambda = 50$ TeV. There are
three qualitatively different regions separated by
the two lines where the bino is equal in mass to either the stau (upper line)
or slectron (lower line). Here we also indicate the bino NLSP (upper left orange region)
and stau NLSP (lower right cyan region) regimes of parameter space.}%
\label{imprvn10eq1nlspscan}%
\end{center}
\end{figure}

The rest of this section is organized as follows. Besides the discrete
parameter $N_{10}$, the low energy content of a minimal F-theory GUT is controlled by the
messenger scale, the characteristic scale of minimal gauge mediation $\Lambda$ and the size
of the PQ deformation $\Delta_{PQ}$. We first show that near the value of the messenger scale
favored by F-theory GUTs, small changes in $M_{mess}$ can switch the relative orderings of the bino and stau
mass. After this, we study the effects of the PQ
deformation. This latter deformation leads to qualitatively distinct mass
spectra in the case of $N_{10}=1$ models, but leads to more subtle changes in
the relative masses when $N_{10}=2$. Next, we introduce some benchmark models, which
crudely characterize these qualitatively different mass spectra. This is followed
by a discussion of how shifts in the relative
sparticle masses due to the PQ deformation affect decay topologies, which will be important later when we discuss
candidate collider signals.

\subsection{Masses and the Messenger Scale}

We now discuss F-theory GUT scenarios at zero PQ deformation, but where we vary the messenger scale.
For \textit{low scale} gauge mediation models models with $N_{10}=1$, the stau
is the NLSP. Increasing the messenger scale raises the mass of the scalars,
because of the additional renormalization group evolution from the messenger to the weak scale.
For example, in minimal gauge mediation models with $N_{10}=1$, and
$\Lambda\sim50$ TeV, the bino becomes the NLSP for $M_{mess}%
\sim10^{11}$ GeV. In minimal gauge mediation models with $N_{10}=2$, however,
the stau remains the NLSP\ until $M_{mess}\sim10^{14}$ GeV.

Even when $N_{10}=1$, the stau is the NLSP for a large region of F-theory
GUT\ parameter space. Indeed, although the messenger scale tends to increase the
mass of the stau relative to the bino, the PQ\ deformation lowers it back
down. For example, in a Majorana neutrino scenario with $N_{10}=1$,
$M_{mess}\sim10^{12}$ GeV, $\Lambda\sim50$ TeV, the bino and
stau become equal in mass around $\Delta_{PQ}\sim90$ GeV. Figure
\ref{imprvn10eq1nlspscan} shows a plot detailing how $M_{mess}$ and
$\Delta_{PQ}$ determine the relative masses of the stau and bino. This
also shows that for low enough messenger scales, and high enough PQ
deformations, the right-handed selectron (and smuon, since they are degenerate
in mass) can also become lighter than the bino.

\begin{figure}
[ptb]
\begin{center}
\includegraphics[
height=3.8692in,
width=6.9289in
]%
{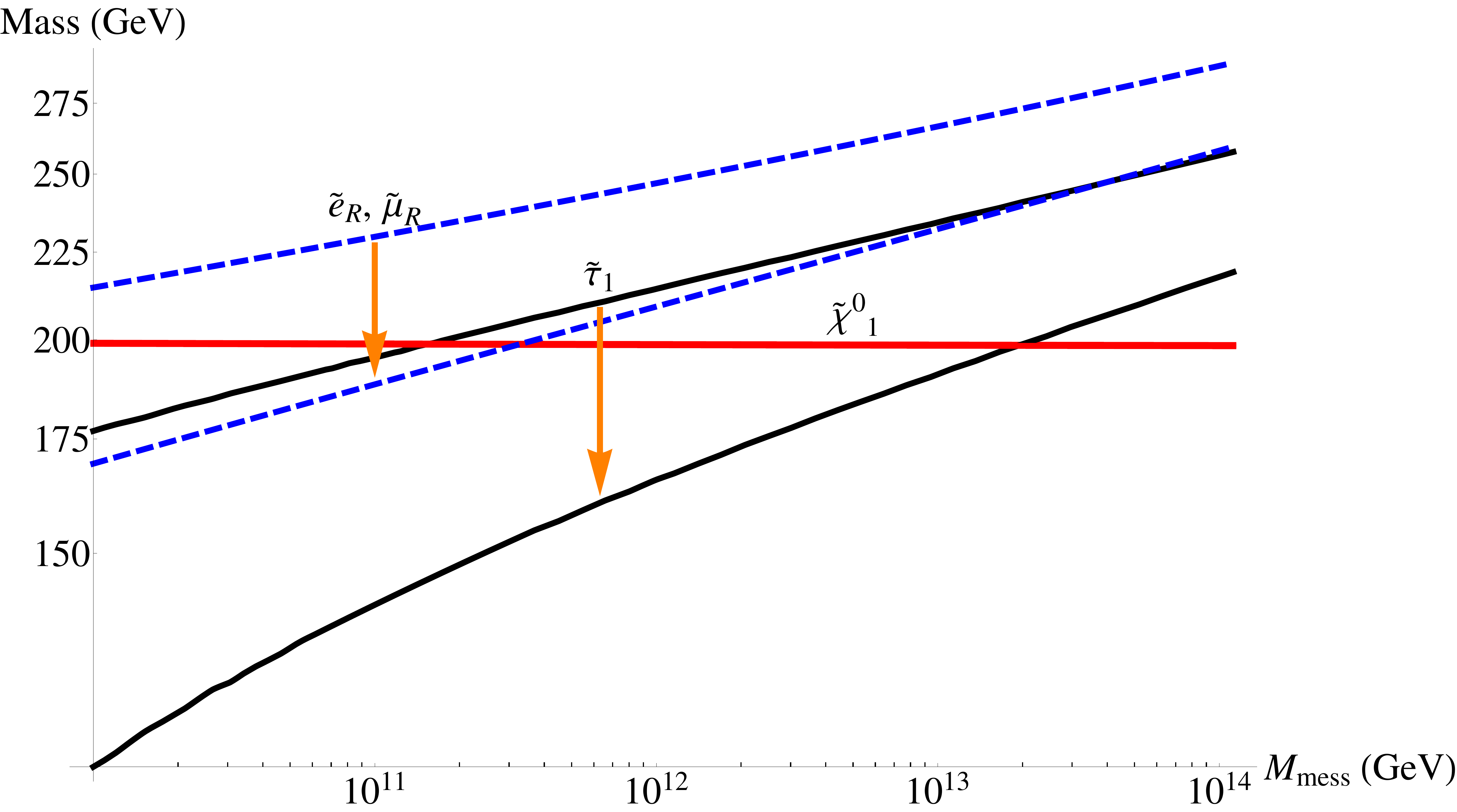}%
\caption{Plot of the four lightest sparticle masses as a function of messenger
scale $M_{mess}$. Here $N_{10}=1$, $\Lambda=50$ TeV, and we have
displayed the values for zero PQ deformation\ (upper lines) and $\Delta
_{PQ}=150$ GeV (lower lines).\ Note that the bino mass remains constant, both
as a function of messenger scale and of PQ deformation.}%
\label{messonetenscan}%
\end{center}
\end{figure}

Figure \ref{messonetenscan} shows the masses of the four lightest Standard
Model superpartners as a function of $M_{mess}$ and $\Delta_{PQ}$ for a
Majorana scenario F-theory GUT with $N_{10}=1$ and $\Lambda=50$ TeV. See
figure \ref{messtwotenscan} of Appendix C for the analogous plot
with $N_{10}=2$, $\Lambda=28$ TeV.%
Because the effects of the messenger scale induce only
logarithmic changes in the masses of the scalar partners, in the remainder of
this paper we shall fix $M_{mess}$ to be:%
\begin{equation}
M_{mess}=10^{12}\text{ GeV.}%
\end{equation}

\subsection{PQ Deformation of the Masses}

In this subsection we study in greater detail the effects of the
PQ\ deformation on the identity of the NLSP, and the other sparticle masses.
As follows from the table of line (\ref{pqcharges}), the PQ deformation tends
to lower the masses of all of the chiral matter superpartners relative to the
gauginos. This effect is more pronounced for lower mass sparticles because the
shift in the mass squared is:%
\begin{equation}
m^{2}=m_{(0)}^{2}+q\Delta_{PQ}^{2}%
\end{equation}
so that%
\begin{equation}
m=m_{(0)}\left(  1+\frac{q}{2}\frac{\Delta_{PQ}^{2}}{m_{(0)}^{2}}+O\left(
\frac{\Delta_{PQ}^{4}}{m_{(0)}^{4}}\right)  \right)  .
\end{equation}
Indeed, we find that the PQ deformation has a far greater effect on the masses
of non-colored superpartners, so we focus on their masses in this subsection.

\begin{figure}
[ptb]
\begin{center}
\includegraphics[
height=5.1733in,
width=5.8262in
]%
{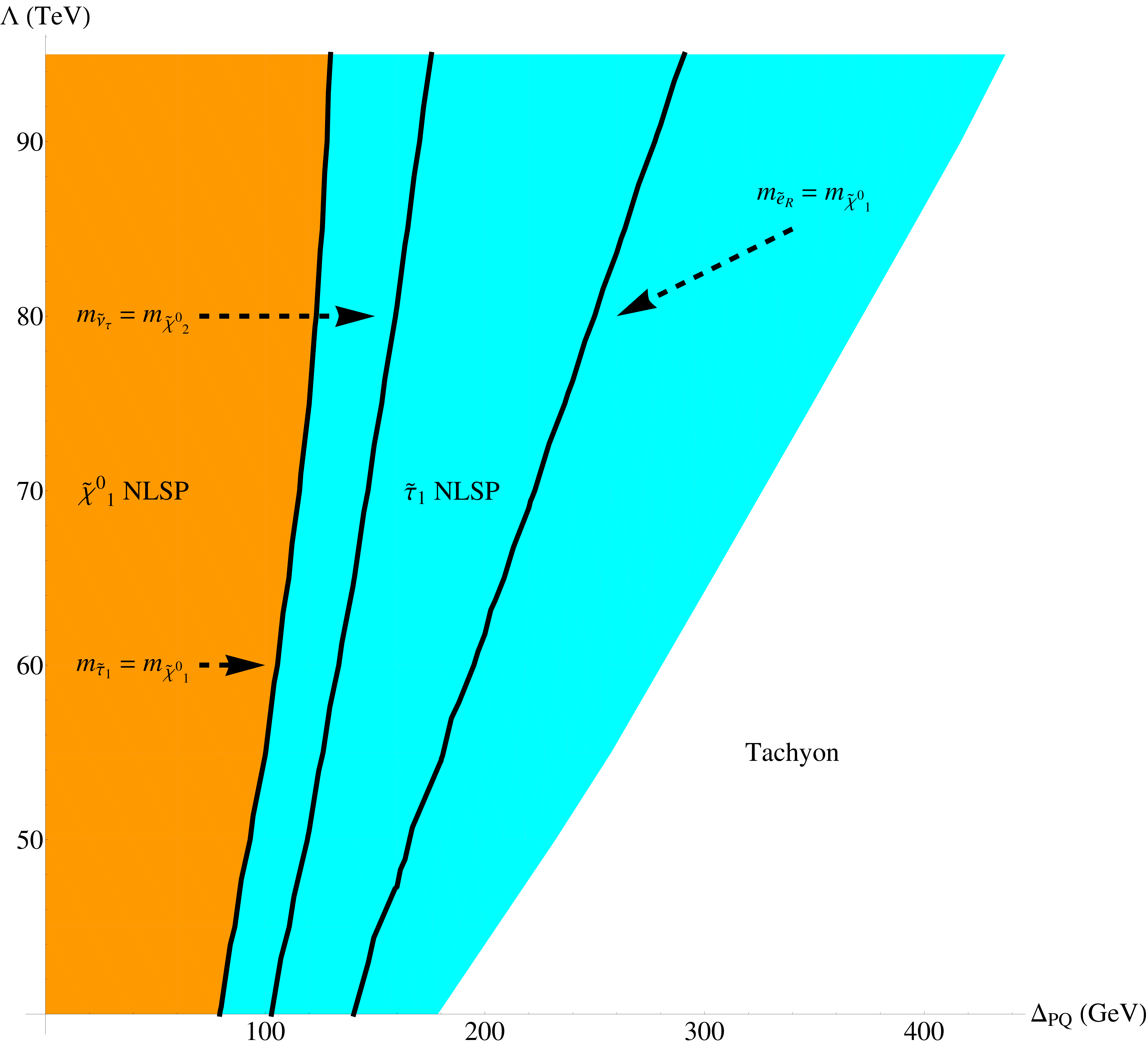}%
\caption{As a function of $\Delta_{PQ}$ and $\Lambda$, the masses of the
sleptons will change relative to the gauginos. For F-theory GUTs with
$N_{10}=1$ messengers, as $\Delta_{PQ}$ increases, there are three qualitative
crossing regions where the stau becomes lighter than the bino, all sleptons become lighter than
the second neutralino and lightest chargino, and the right-handed selectron
and smuon become lighter than the bino. In the plot, the orange region to the left denotes the range of
parameter space where the bino is the NLSP, and the cyan region to the
right indicates the region of the stau NLSP.}%
\label{improcontplotonetenscan}%
\end{center}
\end{figure}

\begin{figure}
[ptb]
\begin{center}
\includegraphics[
height=3.8692in,
width=6.6686in
]%
{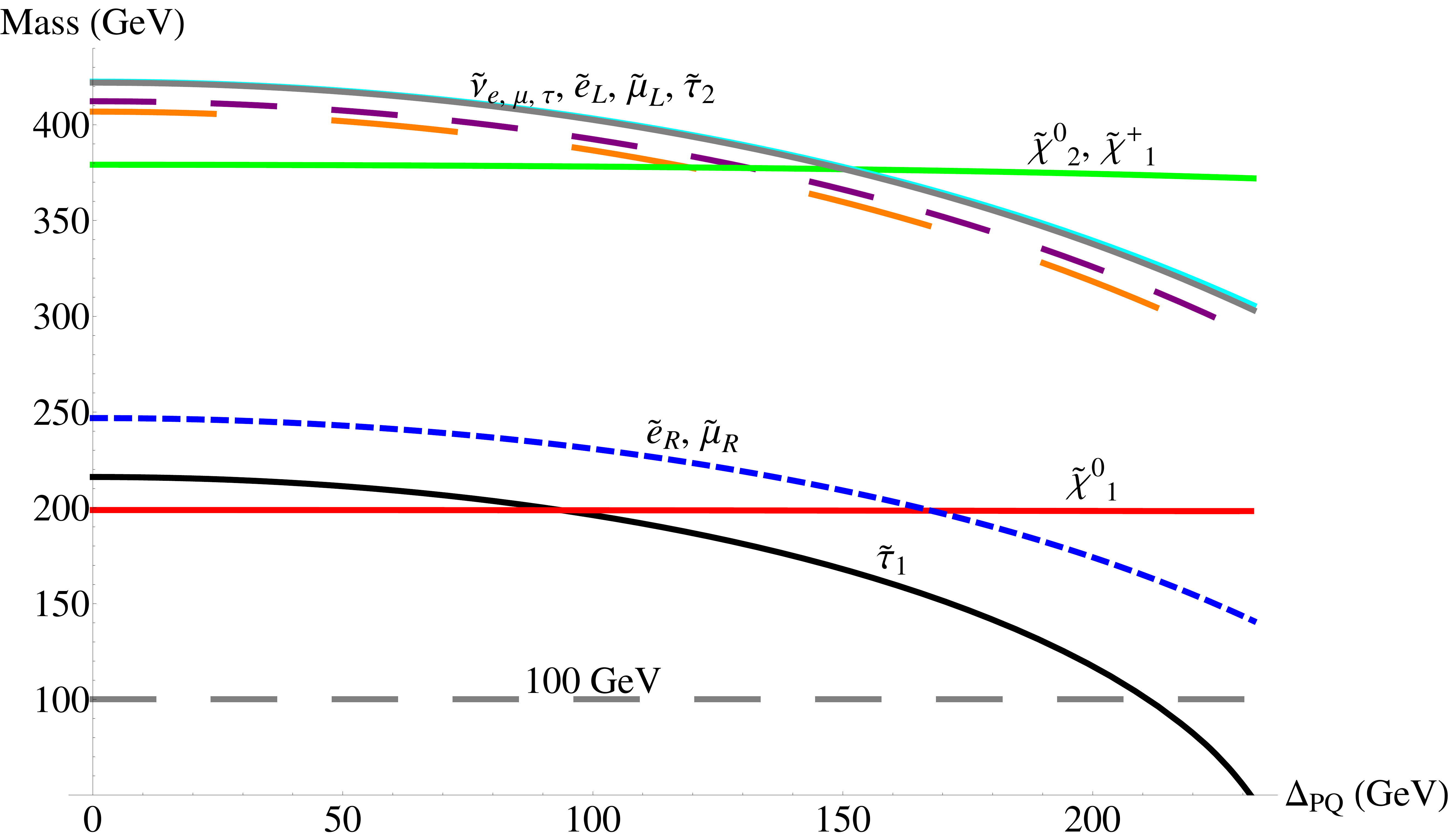}%
\caption{Mass spectrum of the sleptons and lightest chargino and two lightest neutralinos with
$N_{10}=1$, $\Lambda=50$ TeV as a function of $\Delta_{PQ}$. At
values of $\Delta_{PQ}\sim90$ GeV, the stau becomes the NLSP. At somewhat
larger values of the PQ deformation, the sneutrinos become lighter than
$\widetilde{\chi}_{2}^{0}$ and at larger values, the right-handed selectron
and smuon become lighter than the bino. The dashed grey line at $100$ GeV
denotes the present experimental bound on the mass of a quasi-stable stau.}%
\label{imprvmasslowonetenscan}%
\end{center}
\end{figure}

Fixing the messenger scale, we now examine how the relative masses of the sleptons shift as a
function of $\Delta_{PQ}$ and $\Lambda$. Figure \ref{improcontplotonetenscan}
shows the three qualitative crossing regions expected as one increases the
value of the PQ\ deformation in F-theory GUTs with $N_{10}=1$. The first
crossover from a bino NLSP scenario to a stau NLSP\ scenario occurs at
relatively low values of the PQ\ deformation. Moreover, by inspection of this
plot, we observe that the value of $\Delta_{PQ}$ for which this transition
occurs is roughly constant as a function of $\Lambda$, going from a value of
$\Delta_{PQ} \sim 90$ GeV for $\Lambda\sim\Lambda_{\min}$ to $\Delta_{PQ} \sim 100$ GeV for $\Lambda
\sim\Lambda_{\max}$. By inspection of the plot, there is a large range of
parameter space where the stau is the NLSP, and a smaller sliver where the
bino is the NLSP. Increasing $\Delta_{PQ}$ further causes the left-handed
sleptons to become lighter than the second neutralino $\widetilde{\chi}^{0}_{2}$
and lightest chargino $\widetilde{\chi}^{\pm}_{1}$. This
opens up new decay channels for these gauginos to left-handed sleptons, and
significantly alters the decay topology. Further increasing $\Delta_{PQ}$ causes the right-handed selectron and smuon
(which are mass degenerate) to become lighter than the bino. This alters the
decay topology of these particles to three body decays mediated through an off-shell $\widetilde{\chi}^{0}_{1}$.
Figure \ref{imprvmasslowonetenscan} shows a plot of the non-colored
sparticles as a function of $\Delta_{PQ}$ for models with $N_{10}=1$ and
$\Lambda = 50$ TeV.%

In $N_{10}=2$ scenarios, the stau is always the NLSP. Moreover, the right-handed selectron and smuon are also
lighter than the bino, and all of the sleptons are lighter than the lightest
chargino and second neutralino. As a consequence, there are no qualitatively
different crossing regions for the relative masses as a function of $\Delta_{PQ}$. On the other hand, the
relative masses of the sparticles still depend on $\Delta_{PQ}$, so that
there will be some changes in the branching fractions. Figure \ref{imprvmasslowtwotenscan} of Appendix C shows a plot of
the non-colored sparticles as a function of $\Delta_{PQ}$ for models with $N_{10} = 2$ and $\Lambda = 28$ TeV.

\subsection{Benchmark Models}\label{ssec:BENCHMARK}

In the previous sections we have studied how the mass spectrum depends on the
parameters of an F-theory GUT. In the stau NLSP regime of $N_{10} = 1$ models, we have seen that there
are three qualitatively different orderings of the slepton masses relative to the electro-weak gauginos.
For $N_{10} = 2$ models, there is less variation in the mass spectrum,
so that the primary qualitative differences are well represented by models with minimal and maximal PQ deformation.

We now introduce representative ``benchmark models'' which reflect this behavior. Since the Dirac and Majorana neutrino
scenarios are similar, we shall focus on the latter possibility. The benchmark models we consider are:
\begin{align}
\text{Maj}_{\text{LO}}^{(1)}  &  =(N_{10}=1\text{, }\Lambda=50\text{ TeV, }\Delta_{PQ}=0\text{ GeV})\\
\text{Maj}_{\text{MID}}^{(1)}  &  =(N_{10}=1\text{, }\Lambda=50\text{ TeV, }\Delta_{PQ}=140\text{ GeV})\\
\text{Maj}_{\text{HI}}^{(1)}  &  =(N_{10}=1\text{, }\Lambda=50\text{ TeV, }\Delta_{PQ}=209\text{ GeV})\\
\text{Maj}_{\text{LO}}^{(2)}  &  =(N_{10}=2\text{, }\Lambda=28\text{ TeV, }\Delta_{PQ}=0\text{ GeV})\\
\text{Maj}_{\text{HI}}^{(2)}  &  =(N_{10}=2\text{, }\Lambda=28\text{ TeV, }\Delta_{PQ}=175\text{ GeV}).
\end{align}

\subsection{PQ\ Deformation of the Branching Fractions}

Shifting the relative masses of the sparticles also changes their branching fractions. In this section we study how the PQ deformation
alters the branching fractions for particles which we expect to be produced by parton collisions.
This includes gluinos and squarks, as well as second neutralinos and lightest charginos.

Because the masses of the squarks are so high, the branching fractions for the
squarks and gluinos are effectively constant as a function of $\Delta_{PQ}$.
By contrast, the PQ deformation has a significant effect on the non-colored
sparticles, primarily because they are much lighter. To illustrate the effects
of the PQ deformation, we first show the relative branching fractions of a few
channels for Majorana neutrino scenarios for $N_{10}=1$ and $2$, at minimal
and nearly maximal PQ deformation for the benchmark models introduced in the
previous subsection. With respect to these models, we compare the dominant
decay channels and branching fractions of the gluino, lightest stop, lightest
chargino and second neutralino. We use \texttt{SDECAY} \cite{SDECAY} to compute the
branching fractions numerically.

First consider the decays of the gluino. Since the PQ deformation has only a
small effect on the colored sparticles, there is only a few percent shift in
the branching fractions:%
\begin{equation}%
\begin{tabular}
[c]{|l|c|c|c|c|}\hline
Channel & $\text{Maj}_{\text{LO}}^{(1)}$ & $\text{Maj}_{\text{HI}}^{(1)}$ &
$\text{Maj}_{\text{LO}}^{(2)}$ & $\text{Maj}_{\text{HI}}^{(2)}$\\\hline
$\widetilde{g}\rightarrow\widetilde{t}_{1}^{\mp}t^{\pm}$ & $40\%$ & $33\%$ &
$22\%$ & $21\%$\\\hline
$\widetilde{g}\rightarrow\widetilde{q}_{R}^{\mp}q^{\pm}$ & $27\%$ & $33\%$ &
$36\%$ & $37\%$\\\hline
$\widetilde{g}\rightarrow\widetilde{b}_{1}^{\mp}b^{\pm}$ & $21\%$ & $21\%$ &
$16\%$ & $16\%$\\\hline
$\widetilde{g}\rightarrow\widetilde{b}_{2}^{\mp}b^{\pm}$ & $10\%$ & $12\%$ &
$11\%$ & $11\%$\\\hline
\end{tabular}
\ \ ,
\end{equation}
where here $q$ refers to a first or second generation squark. Here we have
omitted the decays to left-handed squarks, which only become relevant for the
$N_{10}=2$ case. Note that in going from $N_{10}=1$ to $N_{10}=2$ messengers,
there is also a shift away from decays to stops and sbottoms in favor of first
and second generation squarks. Nevertheless, the large branching fraction to
stops in all cases suggest the presence of a large number of four top events,
which should be relatively easy to see with low integrated luminosity \cite{Acharya:2009gb}.

Next consider the branching fractions of the stop. We similarly find that there is little
shift for the two values of $N_{10}$ and as a function of $\Delta_{PQ}$:%
\begin{equation}%
\begin{tabular}
[c]{|l|c|c|c|c|}\hline
Channel & $\text{Maj}_{\text{LO}}^{(1)}$ & $\text{Maj}_{\text{HI}}^{(1)}$ &
$\text{Maj}_{\text{LO}}^{(2)}$ & $\text{Maj}_{\text{HI}}^{(2)}$\\\hline
$\widetilde{t}_{1}^{+}\rightarrow\widetilde{\chi}_{1,2,3,4}^{0}t^{\pm}$ &
$46\%$ & $47\%$ & $47\%$ & $47\%$\\\hline
$\widetilde{t}_{1}^{+}\rightarrow\widetilde{\chi}_{2}^{+}b^{-}$ & $31\%$ &
$31\%$ & $29\%$ & $28\%$\\\hline
$\widetilde{t}_{1}^{+}\rightarrow\widetilde{\chi}_{1}^{+}b^{-}$ & $22\%$ &
$23\%$ & $25\%$ & $25\%$\\\hline
\end{tabular}
\ .
\end{equation}

The branching fractions of the lightest chargino and second lightest
neutralino exhibit stronger dependence on the PQ deformation. For example, the
branching fractions of the lightest chargino are:%
\begin{equation}%
\begin{tabular}
[c]{|l|c|c|c|c|}\hline
Channel & $\text{Maj}_{\text{LO}}^{(1)}$ & $\text{Maj}_{\text{HI}}^{(1)}$ &
$\text{Maj}_{\text{LO}}^{(2)}$ & $\text{Maj}_{\text{HI}}^{(2)}$\\\hline
$\widetilde{\chi}_{1}^{+}\rightarrow\widetilde{\tau}_{1}^{+}\nu_{\tau}$ &
$82\%$ & $23\%$ & $25\%$ & $11\%$\\\hline
$\widetilde{\chi}_{1}^{+}\rightarrow\widetilde{\chi}_{1}^{0}W^{+}$ & $18\%$ &
$3\%$ & $5\%$ & $2\%$\\\hline
$\widetilde{\chi}_{1}^{+}\rightarrow\widetilde{\nu}_{l}l^{+}$ & $0\%$ & $30\%$
& $29\%$ & $33\%$\\\hline
$\widetilde{\chi}_{1}^{+}\rightarrow\widetilde{\nu}_{\tau}\tau^{+}$ & $0\%$ &
$19\%$ & $18\%$ & $19\%$\\\hline
$\widetilde{\chi}_{1}^{+}\rightarrow\widetilde{l}_{L}\nu_{l}$ & $0\%$ & $18\%$
& $17\%$ & $25\%$\\\hline
$\widetilde{\chi}_{1}^{+}\rightarrow\widetilde{\tau}_{2}^{+}\nu_{\tau}$ &
$0\%$ & $7\%$ & $7\%$ & $10\%$\\\hline
\end{tabular}
\ ,
\end{equation}
where here $l$ denotes a first or second generation lepton. Similar
$\Delta_{PQ}$ and $N_{10}$ dependence is also present in the branching fractions
of the second neutralino:%
\begin{equation}%
\begin{tabular}
[c]{|l|c|c|c|c|}\hline
Channel & $\text{Maj}_{\text{LO}}^{(1)}$ & $\text{Maj}_{\text{HI}}^{(1)}$ &
$\text{Maj}_{\text{LO}}^{(2)}$ & $\text{Maj}_{\text{HI}}^{(2)}$\\\hline
$\widetilde{\chi}_{2}^{0}\rightarrow\widetilde{\tau}_{1}^{\pm}\tau^{\mp}$ &
$80\%$ & $26\%$ & $29\%$ & $12\%$\\\hline
$\widetilde{\chi}_{2}^{0}\rightarrow\widetilde{\chi}_{1}^{0}h$ & $15\%$ &
$3\%$ & $5\%$ & $1\%$\\\hline
$\widetilde{\chi}_{2}^{0}\rightarrow\widetilde{\chi}_{1}^{0}Z$ & $2\%$ & $0\%$
& $0\%$ & $0\%$\\\hline
$\widetilde{\chi}_{2}^{0}\rightarrow\widetilde{\nu}_{l}^{\ast}\nu_{l}$ or
$\widetilde{\nu}_{l}\overline{\nu}_{l}$ & $0\%$ & $27\%$ & $26\%$ &
$30\%$\\\hline
$\widetilde{\chi}_{2}^{0}\rightarrow\widetilde{l}_{L}^{\pm}l^{\mp}$ & $2\%$ &
$19\%$ & $16\%$ & $28\%$\\\hline
$\widetilde{\chi}_{2}^{0}\rightarrow\widetilde{\nu}_{\tau}^{\ast}\nu_{\tau}$
or $\widetilde{\nu}_{\tau}\overline{\nu}_{\tau}$ & $0\%$ & $17\%$ & $16\%$ &
$17\%$\\\hline
$\widetilde{\chi}_{2}^{0}\rightarrow\widetilde{\tau}_{2}^{\pm}\tau^{\mp}$ &
$0\%$ & $8\%$ & $7\%$ & $11\%$\\\hline
\end{tabular}
.
\end{equation}

\begin{figure}
[ptb]
\begin{center}
\includegraphics[
height=3.87in,
width=6.7153in
]%
{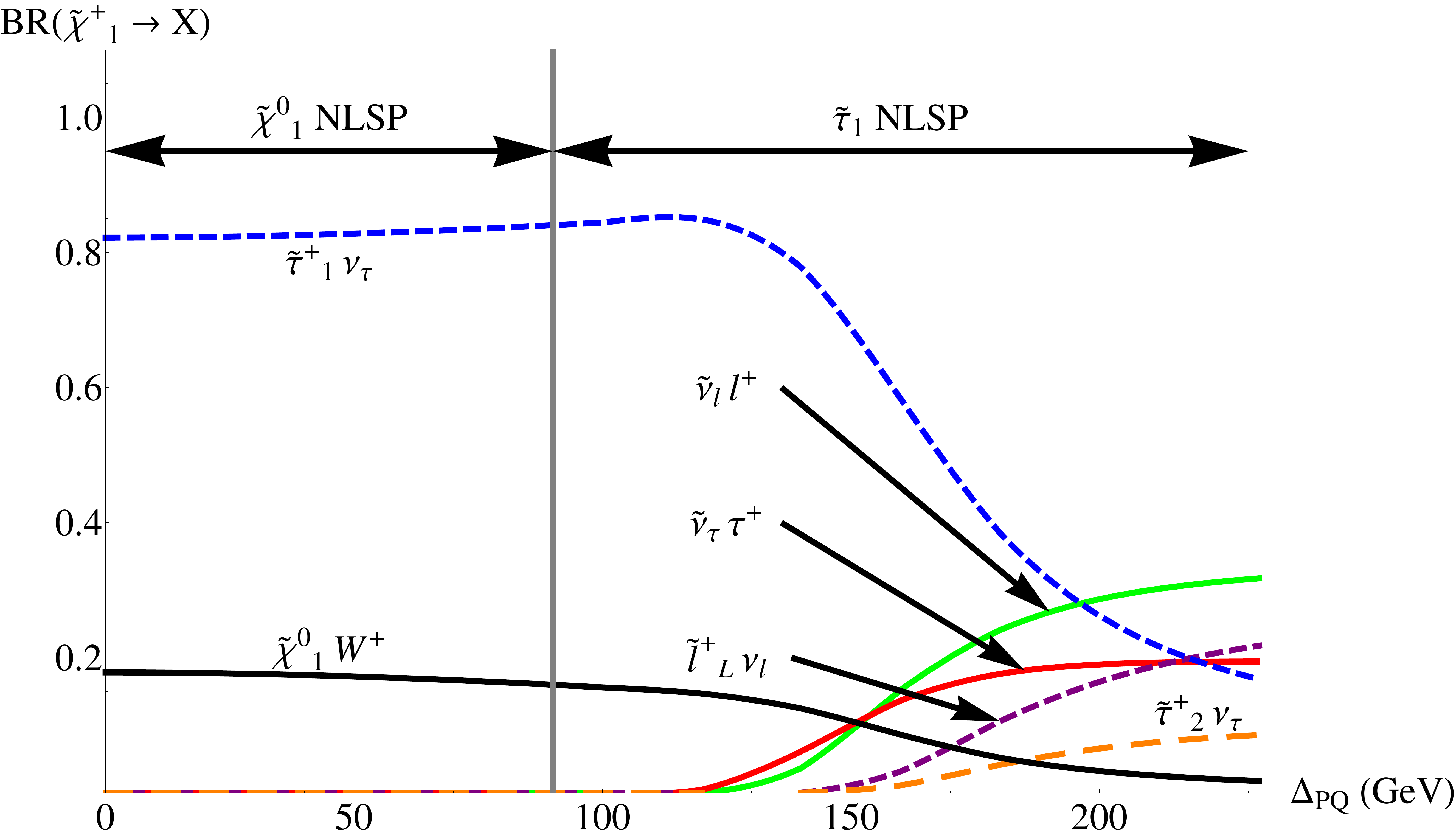}%
\caption{Dominant branching fractions of the lightest chargino as a function
of $\Delta_{PQ}$ with $N_{10}=1$ and $\Lambda=50$ TeV. As the mass
of the sneutrinos becomes lower than the chargino, additional decay channels
open up.}%
\label{imprvbranchlowonetenchp1}%
\end{center}
\end{figure}

\begin{figure}
[ptb]
\begin{center}
\includegraphics[
height=3.87in,
width=6.6928in
]%
{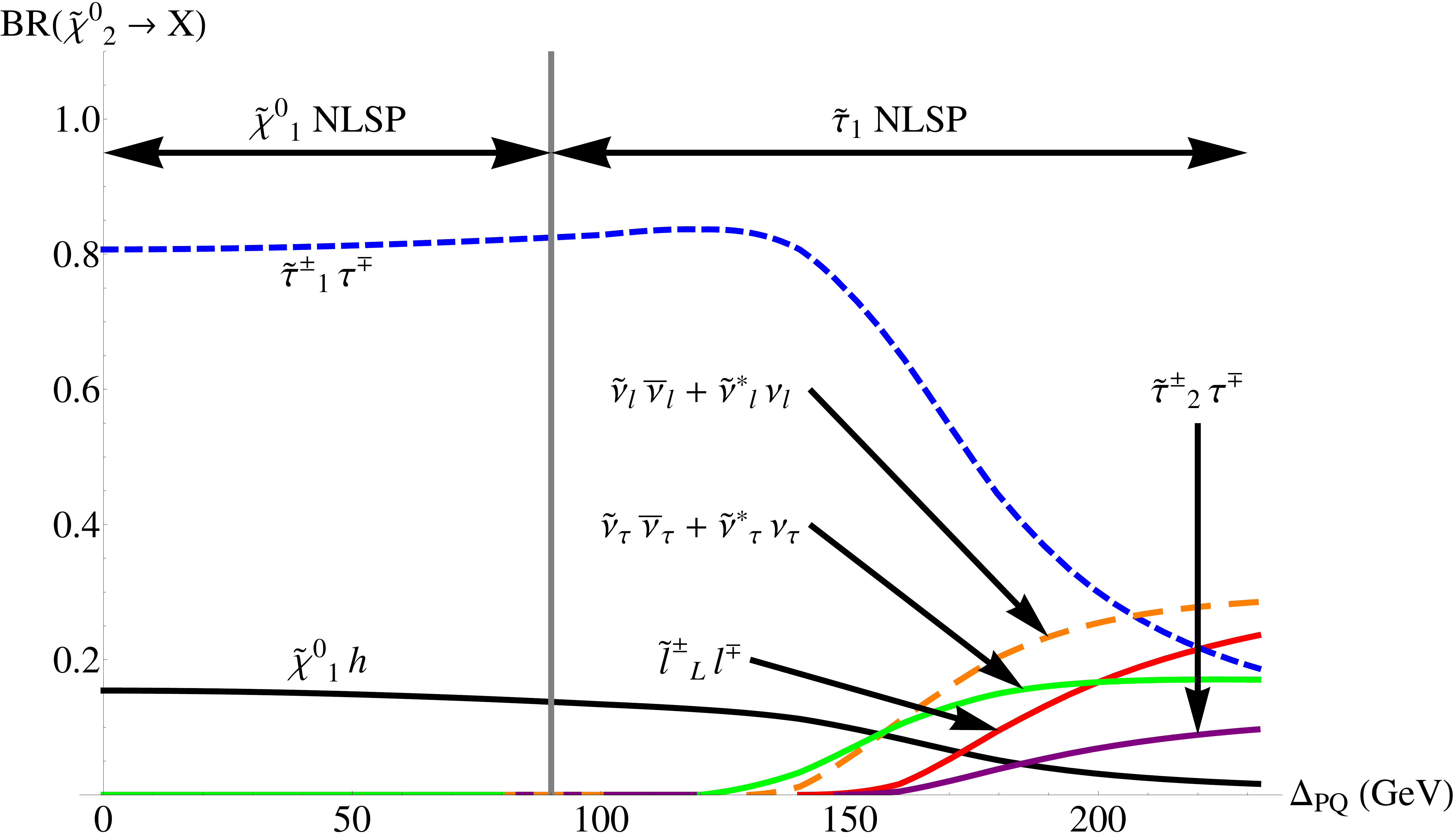}%
\caption{Dominant branching fractions of the second neutralino as a
function of $\Delta_{PQ}$ with $N_{10}=1$ and $\Lambda=50$ TeV. As
the mass of the sneutrinos becomes lower than the second neutralino, additional decay
channels open up.}%
\label{imprvbranchlowonetenzchi2}%
\end{center}
\end{figure}

Scanning over the $\Delta_{PQ}$ dependence of the branching fractions for the
second neutralino and lightest chargino reveals that significant shifts occur
for $N_{10}=1$ F-theory GUTs when the sneutrinos become lighter than the
chargino and second neutralino. Figures \ref{imprvbranchlowonetenchp1}\ and
\ref{imprvbranchlowonetenzchi2} show the respective changes in the branching
fractions for the lightest chargino and second neutralino. Figures
\ref{imprvbranchlowtwotenchp1}\ and \ref{imprvbranchlowtwotenzchi2}\ in
Appendix C show that for $N_{10}=2$, there is less of a dramatic
shift in the branching fractions, in accord with the fact that all sleptons
are already lighter than $\widetilde{\chi}_{2}^{0}$ and $\widetilde{\chi}%
_{1}^{\pm}$.

Increasing $\Delta_{PQ}$ also increases the number of leptons present in the
most likely decay chains. By inspection of all of the branching fractions, we
see that as $\Delta_{PQ}$ increases, the first and second generation sleptons
become light enough to become decay products for the lightest chargino and
second neutralino. This decreases the branching fractions $\widetilde{\chi
}_{2}^{0}\rightarrow\widetilde{\tau}_{1}^{\pm}\tau^{\mp}$ and $\widetilde
{\chi}_{1}^{+}\rightarrow\widetilde{\tau}_{1}^{+}\nu_{\tau}$, meaning the
expected decays will tend to have additional leptons.

\subsection{Right-Handed Sneutrino NLSP and Dirac Neutrinos}

In the above discussion we have assumed that the NLSP corresponds to a
superpartner of a Standard Model particle. This is appropriate in Majorana neutrino
scenarios, where there are no additional light fermions beyond those already
accounted for in the Standard Model. In F-theory GUT Dirac neutrino scenarios
\cite{BHSV}, however, the Dirac mass is induced through the higher-dimension
operator:%
\begin{equation}
\int d^{4}\theta\frac{H_{d}^{\dag}LN_{R}}{\Lambda_{UV}}\rightarrow
\frac{F_{H_{d}}^{\dag}}{\Lambda_{UV}}\int d^{2}\theta N_{L}N_{R} \label{HDLNR}%
\end{equation}
once $H_{d}$ develops a supersymmetry breaking vev on the order of the weak
scale. This type of F-term also induces an extremely small contribution to the
mass of the right-handed sneutrino states, on the order of $0.05$ eV.
In certain ranges of parameter space, the $\widetilde{\nu}_{R}$'s
can correspond to the NLSP. In this case, the NNLSP is the stau or bino. Even
so, this is rather inconsequential for collider
phenomenology because of the highly suppressed interactions of the sneutrino
with the MSSM.

First consider the contributions to the mass of the sneutrino, all of which
are correlated with supersymmetry breaking. Contributions from Planck
suppressed operators will generate a mass term on the order of the gravitino.
Assuming this is the only contribution, this suggests the possibility that the
$\widetilde{\nu}_{R}$ could be the LSP, or at least very close in mass to the gravitino.

Typically, however, the PQ\ deformation will generate larger mass terms.
Indeed, since $N_{R}$ has PQ charge $q(N_{R})=3$, there is a positive
contribution of $+3 \Delta_{PQ}^{2}$ to the sneutrino mass squared. In the
typical range of F-theory GUT\ parameter space where $\Delta_{PQ}$ is on the
order of $100$ GeV, $\widetilde{\nu}_{R}$ is simply a heavy superpartner which
is effectively decoupled from the MSSM.

In principle, contributions from a $U(1)_{B-L}$ gauge boson can also
contribute to the mass of the sneutrino, though this depends on assumptions
about this sector of the model. For simplicity, we shall assume in this paper
that the $B-L$ gauge boson is heavier than the PQ gauge boson, but for the
sake of argument, we now discuss what happens if we assume it interacts
non-trivially with the supersymmetry breaking sector. Even when the $B-L$
gauge boson has similar mass to the PQ\ gauge boson, it will have almost no
effect on the sparticle spectrum. This is because the $X$ field is not charged
under $U(1)_{B-L}$ \cite{EPOINT}, and so the analogue of the PQ\ deformation is
necessarily suppressed by additional loops. When this gauge boson is lighter than
the supersymmetry breaking scale, it can in principle induce additional
contributions, much as in ordinary gauge mediation. In such scenarios, the
mass of the sneutrino will then depend on the gauge coupling of $U(1)_{B-L}$.
Since all of the sleptons have the same magnitude $U(1)_{B-L}$ charge, the
right-handed sneutrino will have mass lower than all of the other sleptons,
and in particular the stau. Note, however, that the gauginos are not charged
under $U(1)_{B-L}$, so in principle either the bino or the right-handed
sneutrino would in this case be the NLSP.

Nevertheless, the effects of a right-handed sneutrino NLSP are fairly
inconsequential for the purposes of a collider study since the decay rate
of other sparticles to the NLSP is at most comparable to the decay to the gravitino $\widetilde{G}$.
Indeed, the decay rate for a sparticle $\widetilde{\psi}$
to decay to a gravitino plus its Standard Model counterpart $\psi$ is
\begin{equation}
\Gamma(\widetilde{\psi} \rightarrow \widetilde{G} + \psi)\sim\frac{m^{5}}{F^{2}},
\end{equation}
with $m$ the mass of $\widetilde{\psi}$. On the other hand, the decay channel provided
by the right-handed sneutrino is suppressed by the scale $\Lambda_{UV}$, and
so assuming no additional loop suppression, will lead to a decay rate on
the order of:%
\begin{equation}
\Gamma(\widetilde{\psi} \rightarrow \widetilde{\nu}_{R}+...) \sim \frac{m^{3}}{\Lambda
_{UV}^{2}}.
\end{equation}
In the context of an F-theory GUT\ where $F/\Lambda_{UV}$ is also on the order
of the weak scale, it follows that decays to the right-handed sneutrino are at most comparable
to decays to a gravitino. This implies that the NNLSP will still be
quasi-stable. By abuse of terminology, we shall therefore refer to the
NNLSP\ in such situations as the NLSP, since there is little
distinction between these possibilities for collider studies.

\section{Event Generation of Long-Lived Staus}\label{sec:LongLive}

In the previous sections we found that over much of the parameter space of minimal
F-theory GUTs with $E_8$ point unification, the stau is a quasi-stable particle on
timescales probed by a collider detector. In the remainder of this paper we focus exclusively on Majorana neutrino F-theory GUT scenarios
with a stau NLSP, and explore how associated stau signatures depend on the parameters of minimal F-theory GUTs. In addition to scanning over the parameters of F-theory GUTs, we also briefly comment on how the center-of-mass energy $\sqrt{s}$ affects the expected cross sections and distributions. Since the energy scales of operation $7$ TeV and $14$ TeV are expected to be of particular relevance, we shall often focus on these cases.

We have performed a Monte Carlo simulation of supersymmetric event production
for F-theory GUTs at the LHC using \texttt{PYTHIA} \cite{PYTHIA}, with basic detector effects included for various objects (see Appendix A for more details).
The mass spectra and decay information are generated using \texttt{SOFTSUSY} \cite{SOFTSUSYAllanach}, \texttt{SDECAY} \cite{SDECAY}
and \texttt{BRIDGE} \cite{BRIDGE}, and this output is passed to \texttt{PYTHIA} in Susy Les Houches Accord format \cite{LESHOUCHESSkands}.
In subsection \ref{ssec:CROSS}, we present the expected cross sections of F-theory GUTs as a function of
the parameters $\Lambda$, $\Delta_{PQ}$, as well as the center-of-mass energy $\sqrt{s}$. After the
initial production of a supersymmetric event, the resulting sparticles will undergo a sequence of
decays so that the final state contains an even number of staus (due to R-parity) which can be
accompanied by Standard Model particles. In subsection \ref{ssec:STAUsig} we explain how we shall
identify staus, taking into account some crude features of detectors.

\subsection{Production Cross Sections}\label{ssec:CROSS}

\begin{figure}[htbp] 
   \centering
   \includegraphics[width=5in]{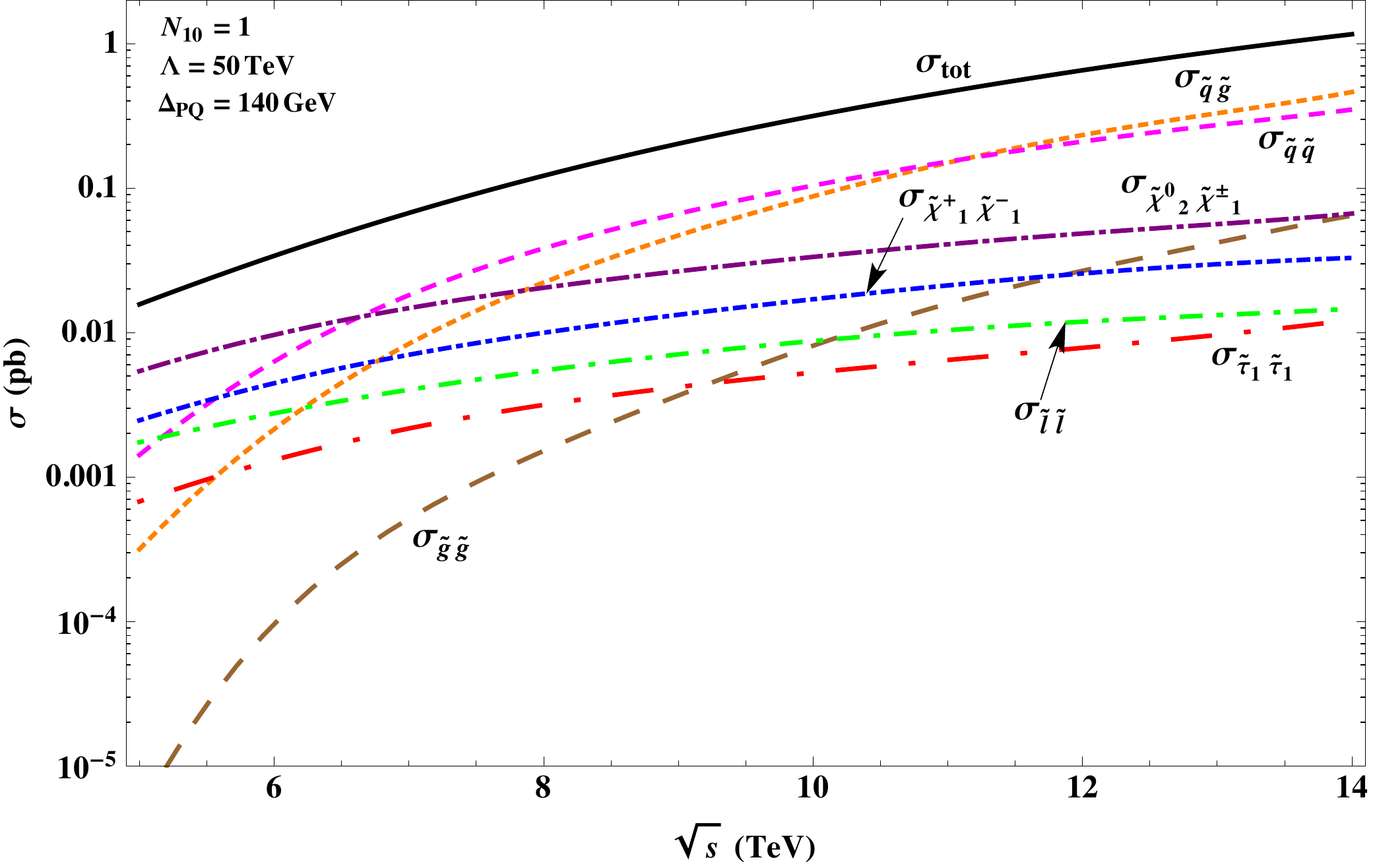}
   \caption{Plot of the leading order cross sections of various production channels as functions of center-of-mass energy $\sqrt{s}$ for the benchmark $\rm{{Maj}_{MID}^{(1)}}$ scenario.}
   \label{fig:xsec-cme}
\end{figure}

\begin{figure}[htbp] 
   \centering
   \includegraphics[width=5in]{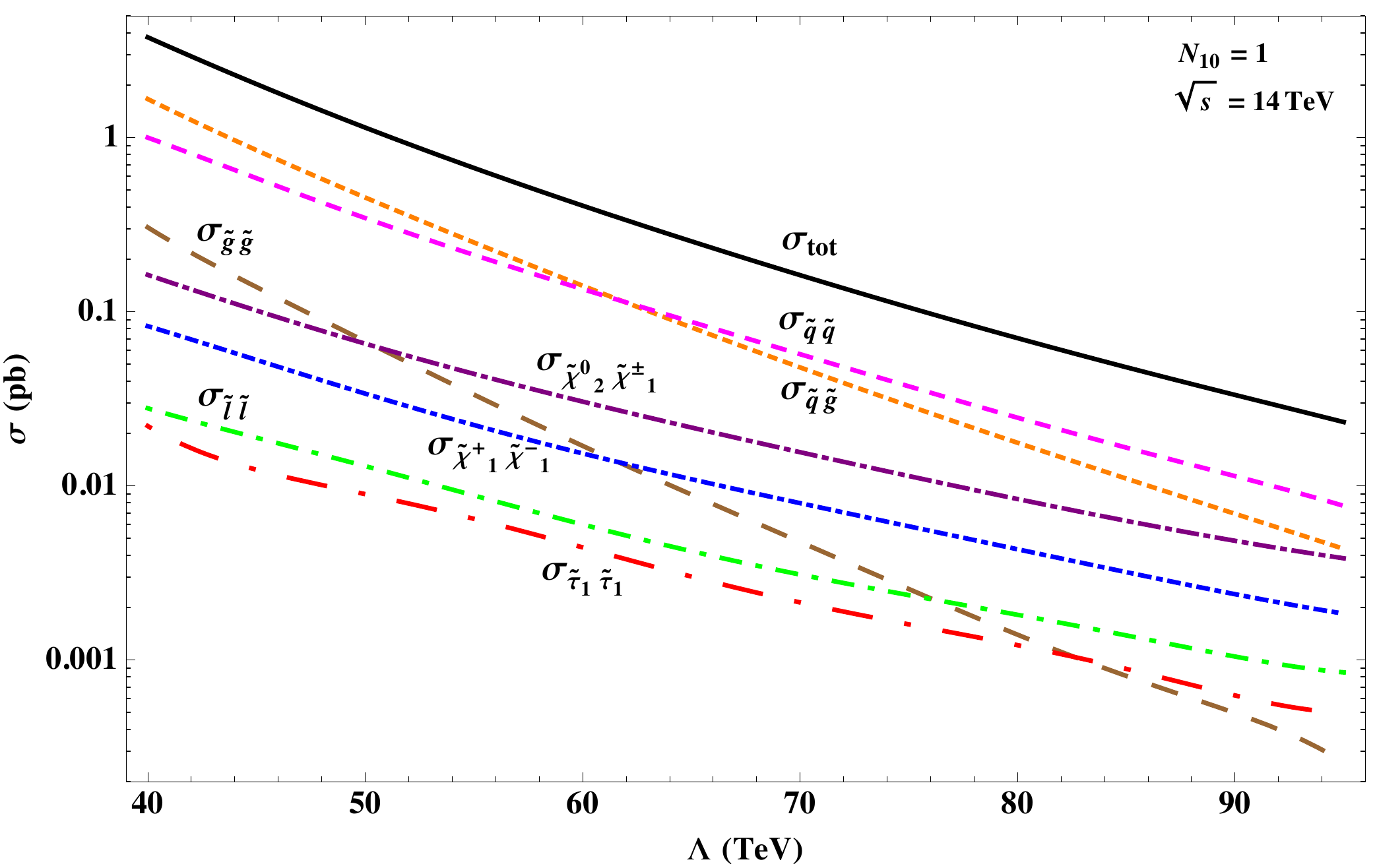}
   \caption{Plot of the leading order cross sections of various production channels for F-theory GUTs with $N_{10} = 1$ as a function of $\Lambda$ with a minimal value of $\Delta_{PQ}$ such that the stau is the NLSP. Aside from Drell-Yan production of staus and sleptons, these cross sections are independent of $\Delta_{PQ}$ (see figure \ref{fig:xsec-dpq}).}
   \label{fig:xsec-lbd-N10eq1}
\end{figure}

The number of expected supersymmetric events is determined by the production cross section of supersymmetric particles generated by parton collisions. Figure \ref{fig:xsec-cme} shows the dominant leading order cross sections of an F-theory GUT with $N_{10} = 1$ and $\Lambda \sim 50$ TeV as a function of the center-of-mass energy $\sqrt{s}$ of the LHC. See figure \ref{fig:xsec-cme2} of Appendix C for a similar plot for $N_{10} = 2$ and $\Lambda \sim 28$ TeV F-theory GUTs. Figure \ref{fig:xsec-lbd-N10eq1} shows that the production of colored objects strongly increases by roughly two to three orders of magnitude in passing from $\sqrt{s} \sim 5$ TeV to $\sqrt{s} \sim 14$ TeV. This is primarily because the mass of the gluinos and squarks are all above a TeV, so that the required energy transfer from partons must be at least of the same order of magnitude. Note, however, that even for smaller values of $\sqrt{s}$, there is also significant production of non-colored sparticles such as lightest charginos, second neutralinos, and Drell-Yan produced lightest staus.

\begin{figure}[htbp] 
   \centering
   \includegraphics[width=6.in]{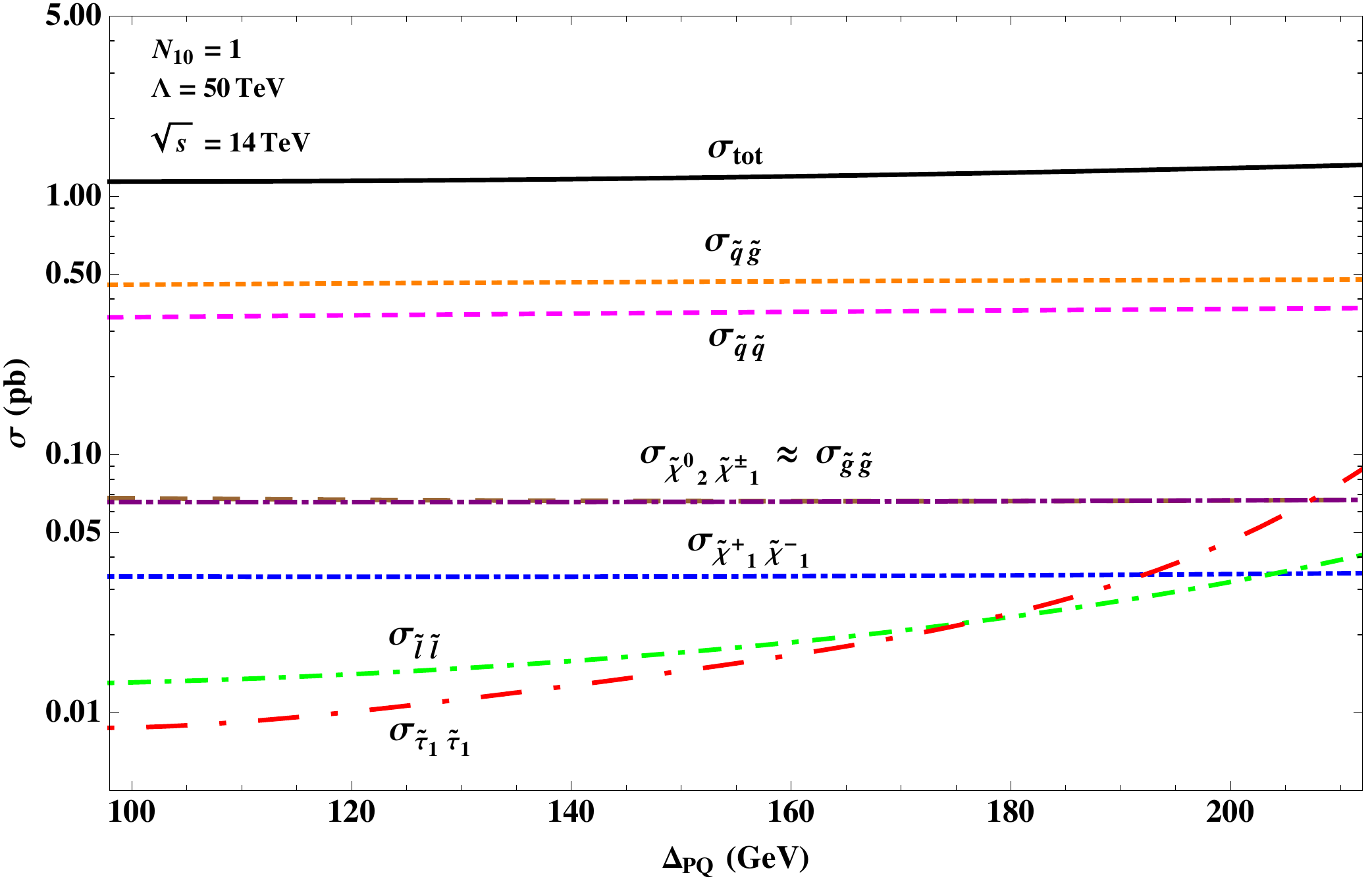}
   \caption{Plot of the leading order cross sections of the dominant production channels as functions of $\Delta_{PQ}$ for $N_{10}=1$ and $\Lambda=50$ TeV.}
   \label{fig:xsec-dpq}
\end{figure}

As a function of $\Delta_{PQ}$, there is little change in the cross section of most channels. This is to be expected, because the percentage change in the mass of the squarks as a function of $\Delta_{PQ}$ is small, and for the gauginos such as the gluino and lightest charginos and second neutralinos, the PQ deformation vanishes. The two notable exceptions to this are the Drell-Yan production of lightest staus and sleptons. This is because the mass of the lightest sleptons more strongly depends on $\Delta_{PQ}$, which in turn alters the expected production cross section. Figure \ref{fig:xsec-dpq} illustrates that for an F-theory GUT with $N_{10} = 1$ and $\Lambda = 50$ TeV and as a function of $\Delta_{PQ}$, Drell-Yan production of staus increases from $0.01$ pb at moderate values of $\Delta_{PQ} \sim 100$ GeV to roughly an order of magnitude more at $\Delta_{PQ} \sim 200$ GeV, allowing it to become comparable to the production cross section of second neutralinos and lightest charginos. Figure \ref{fig:xsec-lbd-N10eq2} of Appendix C shows a similar plot for the case $N_{10} = 2$ and $\Lambda \sim 28$ TeV.

Next consider the dependence of the cross sections on the parameter $\Lambda$. In figure \ref{fig:xsec-lbd-N10eq1} we plot the leading order production cross sections in $N_{10} = 1$ models where a minimal value of $\Delta_{PQ}$ is chosen such that the stau is the NLSP. The production cross section for gluinos is initially larger than that of the non-colored sparticles, but then becomes smaller at larger values of $\Lambda$. See figure \ref{fig:xsec-lbd-N10eq2} of Appendix C for a similar plot of the case with $N_{10} = 2$ and $\Lambda = 28$ TeV for vanishing $\Delta_{PQ}$. A related issue is that as a function of $\Lambda$, the production cross sections have different slopes. This is because the parton distribution function (pdf) for the gluons and sea quarks are different as a function of the momentum transfer fraction $x$. In particular, it is known that the gluon pdf significantly decreases at large $x$. Thus, as the sparticles increase in mass, the resulting momentum transfer from partons must be that much larger, leading to a pronounced decrease in sparticles produced from gluon collisions rather than quark collisions. At a theoretical level, increasing $\Lambda$ also tends to lead to more fine-tuning in the Higgs sector. In other words, the more motivated possibility of lower $\Lambda$ is also the one most accessible to the LHC, and this is the case we shall typically focus on.

\subsection{Candidate Staus}\label{ssec:STAUsig}

After the production of a supersymmetric particle, it will undergo a sequence of cascade decays down to the quasi-stable stau NLSP. As opposed to many supersymmetric scenarios with either a quasi-stable bino or wino, which register as missing transverse energy, a stau will leave a more prominent signal, because it is electromagnetically charged.

Though a complete detector simulation is beyond the scope of the present paper, the detector resolutions
$\sigma_{\beta}$ and $\sigma_{p}$ for the velocity $\beta \equiv v / c$ and momentum $p$ of candidate staus are
\cite{Ambrosanio:2000ik} (see also \cite{DeRoeck:2005bw}):
  \begin{equation}
  \frac{\sigma_{\beta}}{\beta} = 0.028 \beta, \quad \frac{\sigma_p}{p} =\frac{k_1 p} {\rm GeV} \oplus k_2 \sqrt{1+ \frac{m_{\tilde \tau}^2}{p^2}} \oplus \frac{k_3\; {\rm GeV}}{p} \label{eq:resolution}
  \end{equation}
where $k_1=0.0118\%$, $k_2=2\%$ and $k_3 =89\%$. Since staus are quasi-stable, and have the same charge as muons, many will pass the
level one (L1) and level 2 (L2) muon triggers, and will initially register as muons, provided they have high enough velocity.
Indeed, standard data collection and reconstruction of charged tracks in the inner tracking
chamber and muon detection system will record many muon events, of which some will actually correspond to staus.

The staus generated by cascade decays from heavier particles will typically have large velocities. On the other hand, nearly all muons will
have large $\beta$. The Standard Model background for stau candidates mainly comes from muons whose velocity $\beta$ is inaccurately measured. The typical Standard Model processes involving muons are dominantly the single boson production $W/Z+jets$, as well as top and bottom pair production $t\bar t$ and $b\bar b$. With the cut $\beta <  0.91$ on stau candidates, a rejection rate of $\sim 1000$ can be achieved on background muons.\footnote{Using 
fast-moving staus is also possible, see \cite{Chen:2009gu} for a recent study.} In 
order to differentiate muons from staus, we shall therefore impose the selection cuts for candidate staus:
\begin{itemize}
\item $0.67 < \beta < 0.91$

\item $p_{T} > 20$ GeV, $|\eta| < 2.5$.

\end{itemize}
Let us comment on the selection cut for $\beta$. As already mentioned, the upper bound on $\beta$ is necessary to reduce contamination from actual muons. The lower bound reflects the requirement that candidate staus which move too slowly will not be properly identified with the correct bunch crossing. At the ATLAS detector, for example, the muon trigger efficiency drops rapidly for slow moving particles with
velocity $\beta < 0.8$ in the barrel, and $\beta < 0.7$ in
the endcap of the detector \cite{Aad:2009wy}. In principle, the bound $\beta > 0.67$ can be relaxed to $0.6$ \cite{Aad:2009wy}, though the detector efficiency is very low. Lowering the cut in this way requires that slow-moving staus can be reconstructed through an offline analysis using the monitored drift tube (MDT) data. In section \ref{sec:SEARCH} we shall discuss how inclusive stau signals depend on this lower bound. For $\beta > 0.8$, the efficiency is very close to $100\%$. The cuts on $p_T$ and $\eta$ are included because this is the acceptance range for the detector.

\begin{figure}[htbp] 
   \centering
   \includegraphics[width=5in]{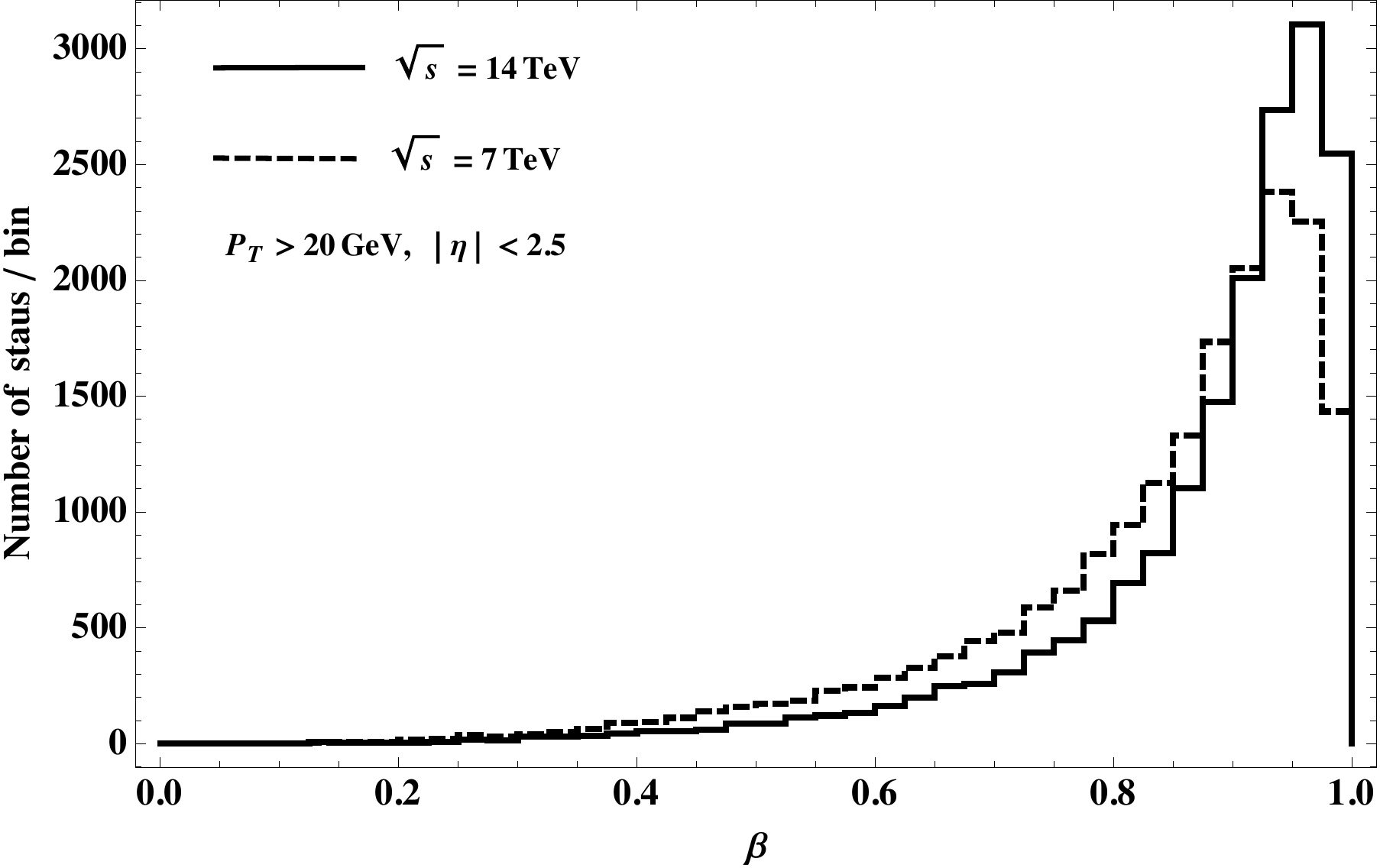}
   \caption{$\beta$ distribution of candidate staus generated with $10,000$ supersymmetric events for
   the benchmark $\rm{{Maj}_{MID}^{(1)}}$ scenario at center-of-mass energies $\sqrt{s}= 7$ TeV and $14$ TeV.}
   \label{fig:beta}
\end{figure}

Figure \ref{fig:beta} shows a plot of the $\beta$ distribution of staus for the benchmark $\rm{{Maj}_{MID}^{(1)}}$ scenario for $10,000$ supersymmetric events at $\sqrt{s} = 7$ TeV and $14$ TeV with the selection cuts on $p_T$ and $|\eta|$ imposed. This plot illustrates that the $\beta$ distribution is strongly peaked near $\beta \sim 1$, and is much lower at smaller values of $\beta$ near the lower bound of $0.67$. Another interesting feature of this plot is that increasing the center-of-mass energy from $\sqrt{s} = 7$ TeV to $14$ TeV tends to push up the number of high $\beta$ staus. This means that once the selection cut on $\beta$ is imposed, the acceptance of staus will drop slightly for the $14$ TeV case. Of course, this is counterbalanced by the fact that the overall number of supersymmetric events is expected to be higher at $\sqrt{s} = 14$ TeV. Said differently, because the distributions have roughly the same shape for $10,000$ events each, the much higher number of supersymmetric events expected at $\sqrt{s} = 14$ TeV will increase the signal, and the discovery potential.

Not all staus, however, will pass the muon trigger system. For slow enough staus, the timing of a ``muon event'' may
not correlate with the bunch crossing of the collider beam, and will therefore not pass the
initial trigger system, and must instead be detected through a dedicated offline analysis. Though the direct
detection of slow-moving staus is therefore more challenging, their low velocity will allow
some of them to become stopped in or near the detector, providing a means
to measure additional properties of the stau. We shall return to the prospects for such stopped stau scenarios later
in section \ref{sec:STOPPED}. For this reason, in the next section we focus on scenarios where candidate staus pass the muon trigger. Indeed, because
such staus are produced through the decay of heavier sparticles, a large fraction of them will have sufficiently high velocity
to pass the L1 and L2 muon triggers.

\section{Search Channels}\label{sec:SEARCH}

We now discuss potential search channels for stau NLSP scenarios in F-theory GUTs. In many supersymmetric extensions of the Standard Model, a common signature consists of searching for cascade decays which terminate with a large missing transverse energy carried by either the LSP or quasi-stable NLSP, accompanied by some number of Standard Model states. Here, this extra energy carried by the analogous quasi-stable supersymmetric state is more visible, since all cascade decays terminate with the stau NLSP.

The characteristics of a long-lived charged particle passing through the detector would already provide a striking deviation from the Standard Model.
Combined with other evidence, such as a determination of the spin of the particle, this would provide evidence in favor of a stau NLSP scenario. It is
therefore important to determine how many such stau events to expect at the LHC, and how the number of events depends on the parameters of F-theory GUTs. The background from Standard Model processes is quite low in such scenarios, and is mainly due to improperly measuring the velocity of muons. We shall return to specific selection cuts which can be used in various search channels, and how these can be used to further reduce Standard Model (SM) background.

One of the most promising discovery channels for staus is based on summing over all possible candidate events in inclusive channels which contain one or two staus.\footnote{The reason we also consider one stau events is that the other stau expected based on R-parity may not pass the selection cut imposed to differentiate staus from muons.} We study how the parameters $\Lambda$ and $\Delta_{PQ}$ of an F-theory GUT, as well as the parameters of the LHC such as the integrated luminosity and center-of-mass energy affect the discovery potential of finding staus at the LHC. We find that though the total cross section strongly depends on $\Lambda$, there is little dependence on $\Delta_{PQ}$. This means in particular that such search channels can be used as a means to constrain the value of $\Lambda$. After this has been done, other search channels can then be used as a means to constrain $\Delta_{PQ}$.

More refined information about the parameter dependence of F-theory GUTs can be extracted from exclusive channels which require a specific combination of particles beyond the ubiquitous staus. Indeed, one particularly clean signature consists of just one or two staus, and nothing else. Such signatures can arise from Drell-Yan pair production of lightest staus, or alternatively lightest charginos. In the latter case, some missing energy will also be carried away by neutrinos. Since such signatures are most directly sensitive to the mass of the stau, they provide an especially clean means to track the dependence on $\Delta_{PQ}$. Other exclusive channels with low background are based on stau events with some number of additional leptons.

Even when we cannot distinguish the stau from a muon, the corresponding events with additional leptons are difficult to mimic with Standard Model processes. Such events correspond to leptons with the same charge which we refer to as same-sign leptons, as well as events with three or more leptons.

\subsection{Inclusive Stau Channels}

In this subsection we discuss inclusive stau channels obtained by summing over all signals containing one or two staus. The inclusive $\ge 1$ stau signal has the largest cross section and therefore provides the best channel for a discovery. Even though all supersymmetric events will contain two staus because of R-parity, imposing selection cuts used to isolate a signal will lead to some loss of signal. In particular, some two stau events may instead register as one stau events, or may not register at all. Letting $\sigma^{SUSY}_{tot}$ denote the total production cross section of supersymmetric events, we define the acceptance $\epsilon_{1,2 \tilde\tau}$ of one and two stau events as the ratio of the observed cross section of one and two stau events $\sigma_{1\tilde\tau}^{obs}$ and $\sigma_{1\tilde\tau}^{obs}$ to $\sigma^{SUSY}_{tot}$:
\begin{equation}\label{eq:obscross}
\epsilon_{i\tilde\tau} \equiv \frac{\sigma_{1,2\tilde\tau}^{obs}}{\sigma^{SUSY}_{tot}}
\end{equation}
for $i = 1,2$. In the remainder of this subsection, we analyze how sensitive the observed cross sections are to the choice of $\beta$ cut imposed, and then discuss the sensitivity of this cross section to the parameters $\Lambda$ and $\Delta_{PQ}$. After this, we discuss the prospects for discovery in this channel after further selection cuts are imposed.

\subsubsection{$\beta_{min}$ and Cross Sections}

\begin{figure}[htbp] 
   \centering
   \includegraphics[width=5.in]{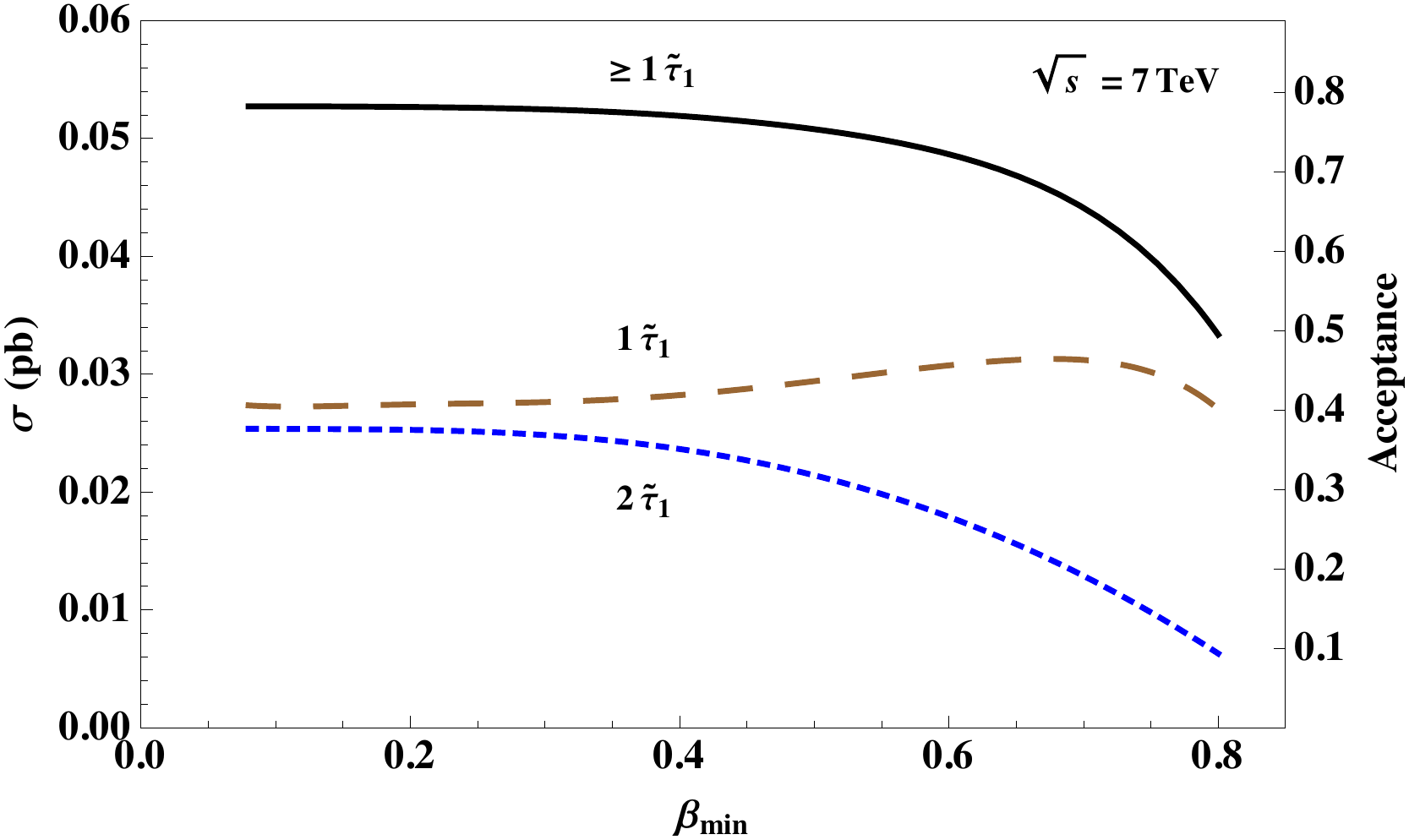}
   \includegraphics[width=5.in]{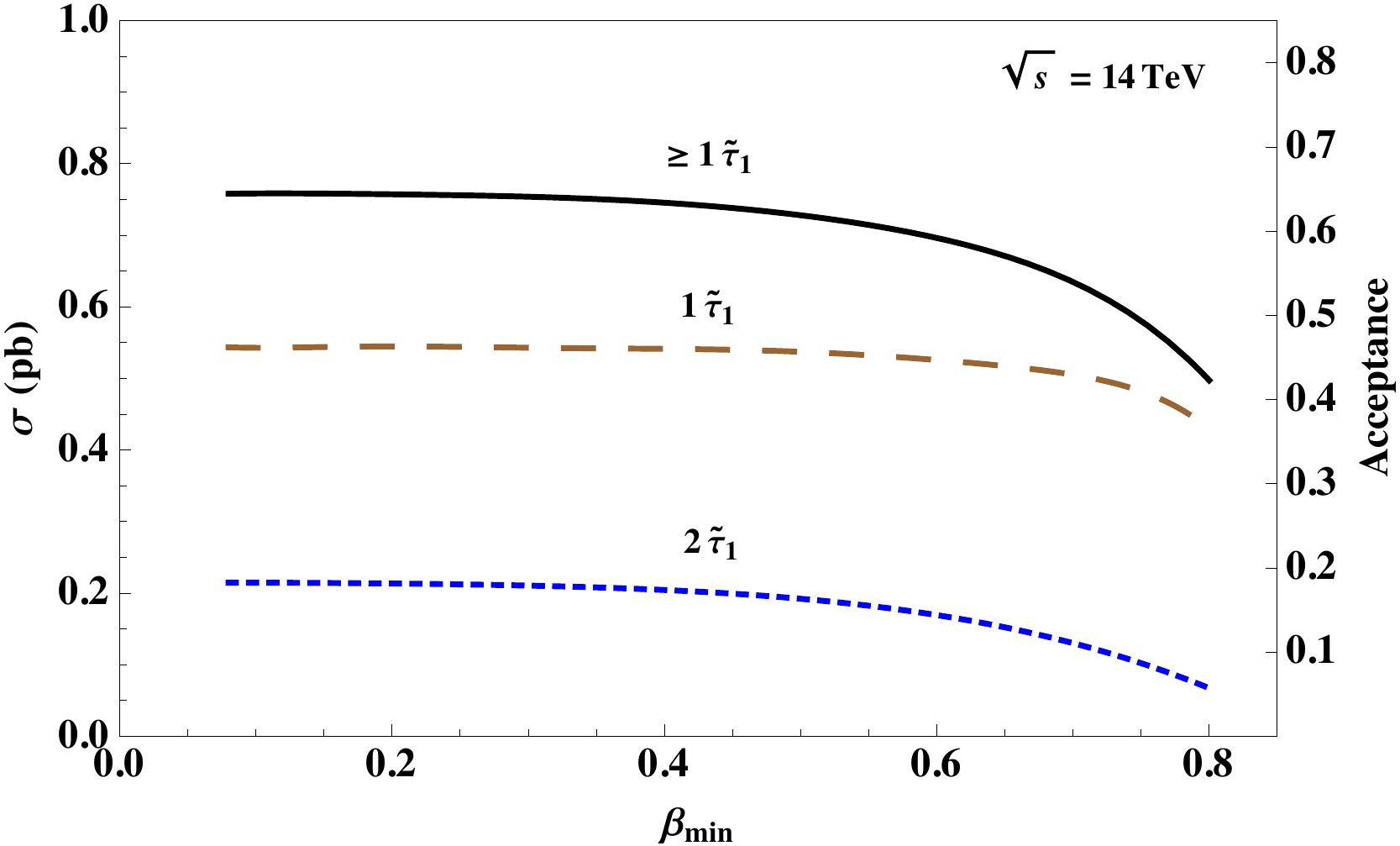}
   \caption{Observed cross sections and acceptances for events with one and two stau candidate events as functions of the lower velocity cut $\beta_{min}$ for the benchmark model ${\rm Maj}_{\rm MID}^{(1)}$ defined in subsection \ref{ssec:BENCHMARK}. The
   acceptance is by definition proportional to the observed cross section (see equation \eqref{eq:obscross}). Also displayed is the total number of one and two stau candidate events. The upper and lower panels are for $\sqrt{s}=7$~TeV and $14$~TeV respectively.}
   \label{fig:instaubt}
\end{figure}

As mentioned in section \ref{sec:LongLive}, the number of expected staus depends on the choice of selection cuts, and in particular, the lower bound $\beta_{min}$ on the velocity $\beta$. To a certain extent, this lower bound depends on how well slow-moving staus can be identified by an offline analysis. To gauge how much loss of signal to expect, we compute how $\sigma_{1\tilde\tau}^{obs}$ and $\sigma_{2\tilde\tau}^{obs}$ depend on $\beta_{min}$. In figure \ref{fig:instaubt} we plot this dependence for the benchmark ${\rm Maj}_{\rm MID}^{(1)}$ model, where we also show the acceptances $\epsilon_{1\tilde\tau}$ and $\epsilon_{2\tilde\tau}$ . This plot illustrates that for $\sqrt{s}=14$~TeV, the cross section of one stau events $\sigma_{1\tilde\tau}^{obs}$ is always a few times larger than that for two stau events $\sigma_{2\tilde\tau}^{obs}$. This is primarily because a large portion of staus have velocity $\beta > 0.91$, and are therefore eliminated by the selection cut.

Figure \ref{fig:instaubt} also illustrates that there is little dropoff in the value of $\sigma_{1\tilde\tau}^{obs}$ or $\sigma_{2\tilde\tau}^{obs}$ as a function of $\beta_{min}$ until $\beta_{min}$ becomes comparable to $0.7$, which is around the value of the selection cut taken anyway. One curious feature of this plot is that in the case where $\sqrt{s} = 7$ TeV, the one stau cross section $\sigma_{1\tilde\tau}^{obs}$ is almost the same as $\sigma_{2\tilde\tau}^{obs}$ when $\beta_{min}$ is small, but then increases as $\beta_{min}$ increases up to $\sim 0.7$. This occurs because as the cut becomes more stringent, some of the previously accepted two stau events will now be rejected, and will instead be recognized as single stau events. Comparing the total acceptance at $\sqrt{s} = 7$ TeV and $14$ TeV, note that although the total cross section at $7$ TeV is lower, the acceptance is higher. This is again because the stau velocity tends to be smaller at the lower center-of-mass energy, meaning that fewer events are rejected based on the selection cut $\beta < 0.91$.

\subsubsection{$\Lambda$ and $\Delta_{PQ}$ Dependence}

Having seen that in the case of representative benchmark models that there is little change in the observed cross section as we change the lower bound on the selection cut $\beta_{min} < \beta$, we now fix the value of this cut at $\beta_{min} = 0.67$, as in section \ref{sec:LongLive}. In this subsection we focus on how the production cross sections depend on the F-theory GUT parameters $N_{10}$, $\Lambda$ and $\Delta_{PQ}$. Figure \ref{fig:xsec-lbd-N10eq1} shows that as we increase the parameter $\Lambda$, we find that the total production cross section $\sigma_{tot}$ decreases. This is to be expected because increasing $\Lambda$ also increases the mass of all of the sparticles. Indeed, even though increasing $\Delta_{PQ}$ will tend to lower the mass of the sleptons and in particular the stau, note that in such a situation, the mass splitting between the heavier sparticles and stau will become more pronounced. This in turn means that staus generated by a cascade decay will tend to be more energetic, having higher velocity, and will therefore not pass the selection cut on $\beta$ as easily.

\begin{figure}[htbp] 
   \centering
   \includegraphics[width=5in]{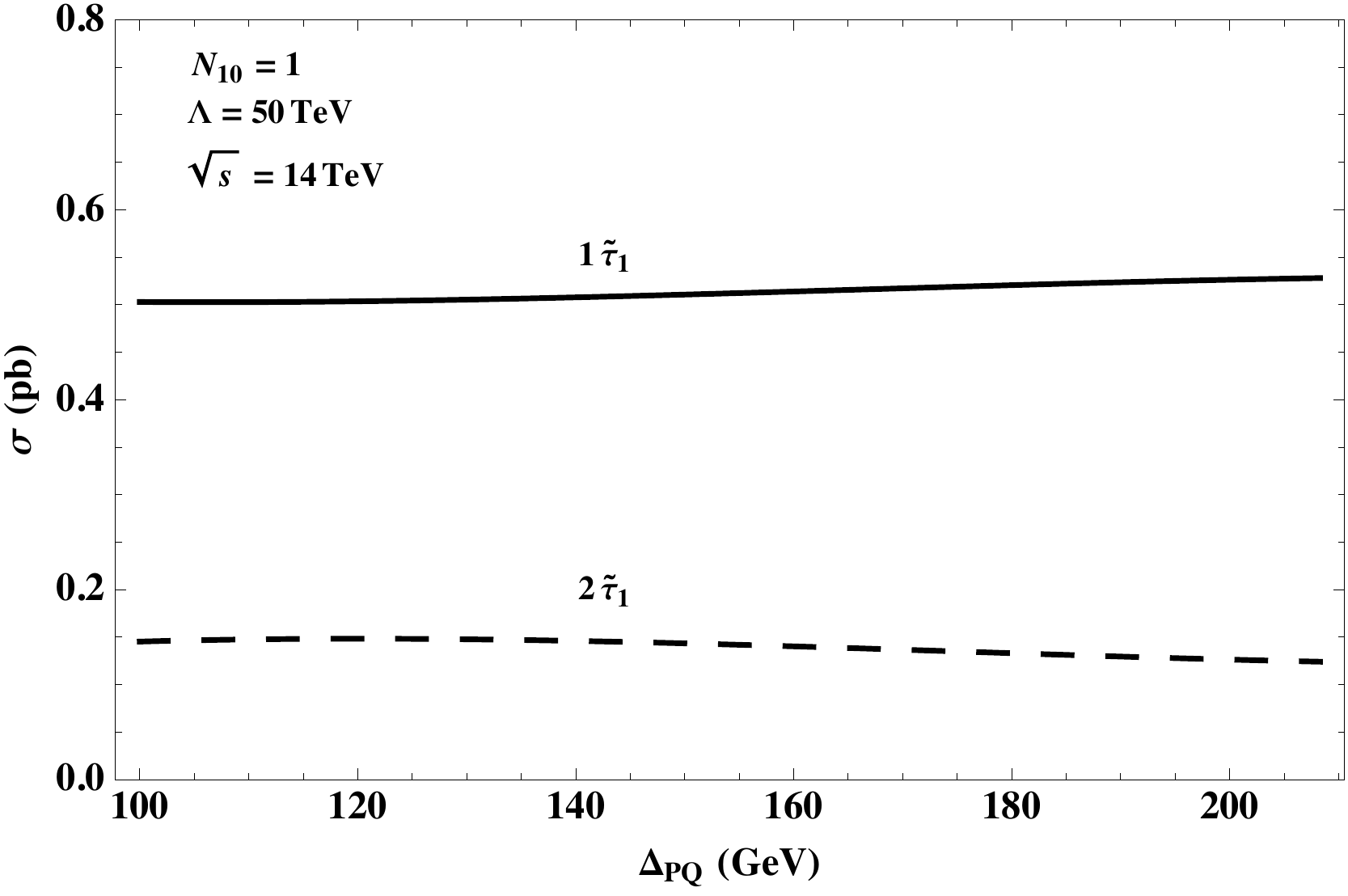}
   \caption{Plots of the observed cross sections of inclusive one and two stau signatures at $\sqrt{s}=14$~TeV as a function of $\Delta_{PQ}$ for stau NLSP F-theory GUT scenarioss with $N_{10} = 1$ and $\Lambda = 50$ TeV.}
   \label{fig:instaudpq}
\end{figure}

Figure \ref{fig:instaudpq} shows that at least for these inclusive stau signals, the dependence on the PQ deformation $\Delta_{PQ}$ is quite mild. This is mainly because the bulk of produced staus are generated from decays of colored sparticles which have masses largely independent of this deformation. If stau signals are observed, this inclusive channel can then be used as an effective means to constrain the parameter $\Lambda$, once a fit to the F-theory GUT parameters has been performed.

\subsubsection{Discovering Staus}\label{ssec:DISCOVER}

Imposing additional selection cuts on inclusive stau channels can further eliminate contamination from background muons produced through Standard Model processes, leading to a cleaner signal and a better chance of discovery. In addition to the basic cut $\beta < 0.91$ mentioned in subsection \ref{ssec:STAUsig} which we use to avoid contamination from muons, we shall also include the event selection cuts in \cite{Ambrosanio:2000ik,Ambrosanio:2000zu}:\footnote{Though we do not include it here, additional discrimination from background muons can be achieved as in \cite{Ambrosanio:2000ik,Ambrosanio:2000zu} by requiring $p > m_{{\tilde \tau}_{1}} \frac{\beta-0.05}{\sqrt{1-(\beta-0.05)^2}}$, where $m_{{\tilde \tau}_{1}}$ is obtained from the stau mass measurement (see section \ref{sec:distinguish}). This cut significantly increases the level of muon rejection, but only affects candidate stau events at the level of a few percent.}
\begin{itemize}
\item At least one hadronic jet with $p_T > 50$~GeV and a calorimetric
$E^{miss}_T > 50$ GeV (trigger requirement);
\item $m_{eff} > 800$~GeV,
\end{itemize}
where $m_{eff}$ is the total invariant mass of the event constructed starting from the transverse momentum of the high $p_T$ jets and muons (or muon-like particles)
\begin{equation}
m_{eff}= \sum_{i=1}^{min(4,N_{jet})}p_T^{jet,i} + \sum_{i=1}^{min(2,N_{\mu})} p_T^{\mu,i}.
\end{equation}
The Standard Model background after all these cuts is estimated to be $\sigma_{SM}^{bkgnd} \lesssim 1$~fb and has only a mild effect on the signal acceptance \cite{Ambrosanio:2000ik,Ambrosanio:2000zu}. Note that since this cut requires at least one hadronic jet, it will reject events generated by Drell-Yan production of staus and charginos. In practice, this means that such signals can be treated as independent constraints on the parameter space of F-theory GUTs.

\begin{figure}[htbp] 
   \centering
   \includegraphics[width=5.5in]{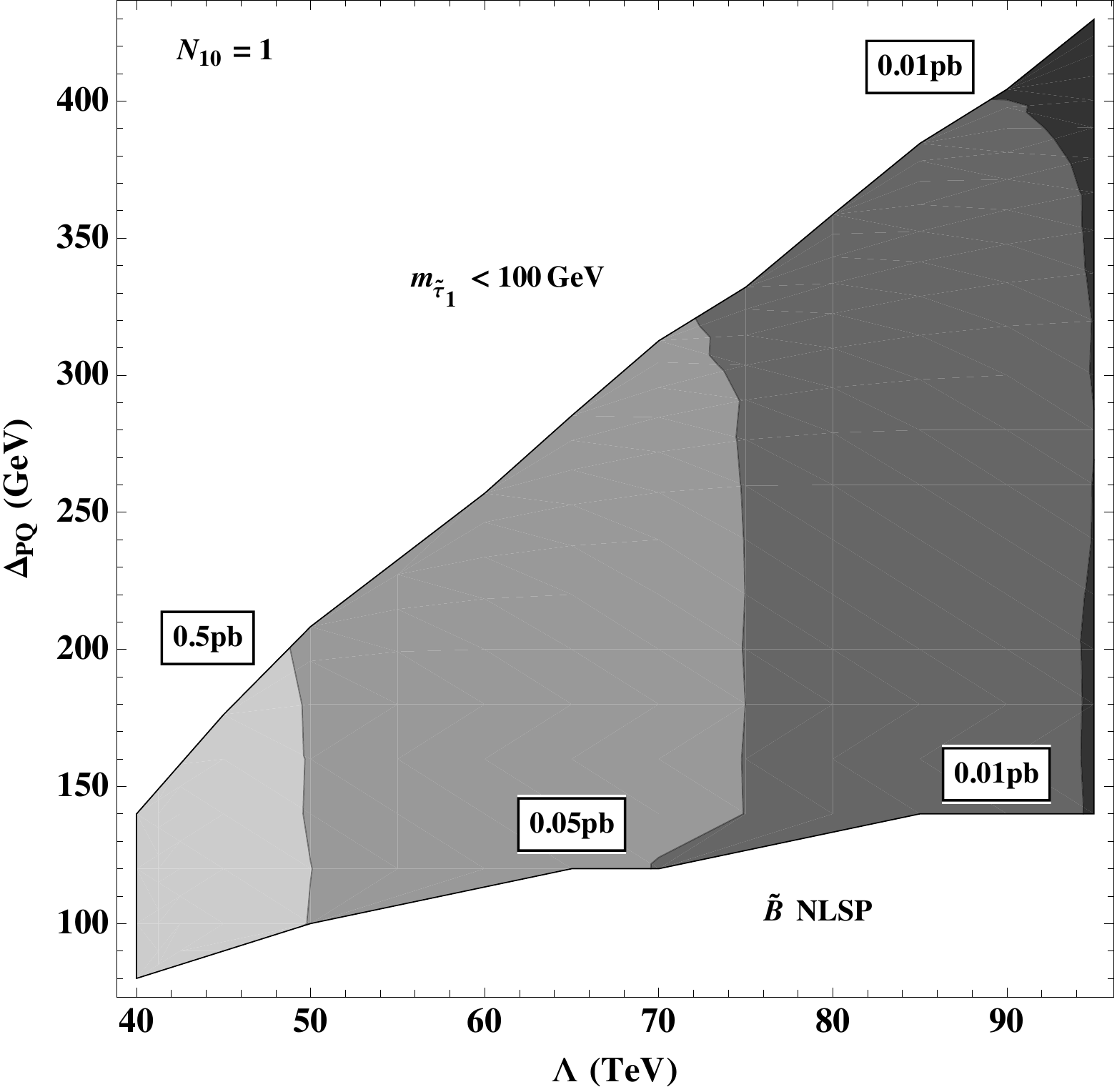}
   \caption{Contour plot of the inclusive $\geq 1 {\tilde \tau}_{1}$ signal in parameter space $\Lambda$ and $\Delta_{PQ}$ for $\sqrt{s}=14$~TeV and $N_{10} = 1$ messengers. The three contours correspond to the value of the cross section at $0.5$ pb, $0.05$ pb, and $0.01$ pb. To detect five inclusive one or two stau events, this amounts to an integrated luminosity of respectively $10$ pb$^{-1}$, $100$ pb$^{-1}$ and $0.5$ fb$^{-1}$.}
   \label{fig:dispotlONE}
\end{figure}

The estimated upper bound of $\sigma_{SM}^{bkgnd} \lesssim 1$~fb means that even for $1$ fb$^{-1}$ of integrated luminosity, we can expect around one Standard Model event. Assuming one Standard Model event as background, a $5 \sigma$ level discovery would then correspond to roughly seeing five times as many candidate stau events. To gauge the necessary integrated luminosity required to claim discovery, we have computed the inclusive cross section at $\sqrt{s} = 14$ TeV of one and two stau candidate events as a function of $N_{10}$, $\Lambda$, and $\Delta_{PQ}$ by imposing the further selection cut mentioned above. In figure \ref{fig:dispotlONE} we plot for $N_{10} = 1$ models the resulting contours for cross sections of $0.5$ pb, $0.05$ pb, and $0.01$ pb. See figure \ref{fig:dispotlTWO} of Appendix C for the analogous plot for $N_{10} = 2$ models. To detect five inclusive one or two stau events, this amounts to an integrated luminosity of respectively $10$ pb$^{-1}$, $100$ pb$^{-1}$ and $0.5$ fb$^{-1}$. Note that this in some sense overestimates the number of Standard Model background events, which upon allowing  for fractions of an event for each integrated luminosity is roughly $0.01$, $0.1$ and $0.5$.

At lower center-of-mass energies $\sqrt{s} \sim 7$ TeV, the total production cross section is smaller, so the prospects for discovery are lower. Even so, for large enough integrated luminosities, it may still be possible to discover such stau events. Indeed, returning to figures \ref{fig:xsec-cme} and \ref{fig:instaubt}, we see that the total cross section of inclusive stau events at $7$ TeV is lower by a factor of $10$ compared with the cross section at $14$ TeV, so that the required integrated luminosity for discovery is roughly a factor of $10$ higher compared with what is required in the $14$ TeV plots of figure \ref{fig:dispotlONE} and figure \ref{fig:dispotlTWO} of Appendix C. For example with a few hundreds of ${\rm pb}^{-1}$ data delivered at $7$ TeV,  we can expect to discover staus with $\Lambda=50$~TeV for $N_{10}=1$ or $\Lambda=28$~TeV for $N_{10}=2$. Depending on how long the LHC operates at $7$ TeV, higher values of $\Lambda$ may also be within reach.

\subsection{Exclusive Two Stau Channel}

Additional information about the parameter dependence of F-theory GUTs can also be extracted by examining more specialized channels. In particular, in subsection \ref{ssec:DISCOVER} we imposed a selection cut on inclusive stau signatures
requiring at least one hadronic jet. This excludes, however, signatures
such as Drell-Yan production of staus and charginos, which can lead to
two opposite sign staus. In the case of Drell-Yan produced charginos, the decay
of the charginos will also generate neutrinos which will leave the detector as missing energy. Because the stau mass depends strongly on $\Delta_{PQ}$,
this correlates the expected cross section with $\Delta_{PQ}$. Moreover, with the additional
information on the measured stau mass (see section \ref{sec:distinguish}), the stau production cross section can also be used to determine
the spin of the stau \cite{Kane:2008kw}. Though the specifics of the event selection are different, this same channel has been used in Tevatron searches for long-lived charged particles \cite{Abazov:2008qu,Aaltonen:2009kea}.

In this subsection we consider events with only two staus, and no additional
(detectable) Standard Model states. More explicitly, we require there is at least one stau candidate which passes the cut on $\beta$ and $\eta$ introduced in subsection \ref{ssec:STAUsig}, and which passes the slightly more stringent cut that $p_T > 100$ GeV for any stau candidate. For candidate events in which only one stau passes this cut, we shall still require that an additional ``lepton'' has $p_T > 100$ GeV. To exclude the presence of final states with additional particles, we also veto events with jets satisfying $p_T> 20$ GeV and $|\eta|< 3$ and extra leptons satisfying $p_T>5$ GeV and $|\eta|<2.5$. To further distinguish between staus generated by Drell-Yan production of staus and charginos, it is also helpful to impose the cut $E_T^{miss} < 10$ GeV which eliminates staus created from charginos which decay to a highly energetic neutrino plus a stau. In the following we consider exclusive two stau events with and without this cut.

\begin{figure}[htbp] 
   \centering
   \includegraphics[width=5.in]{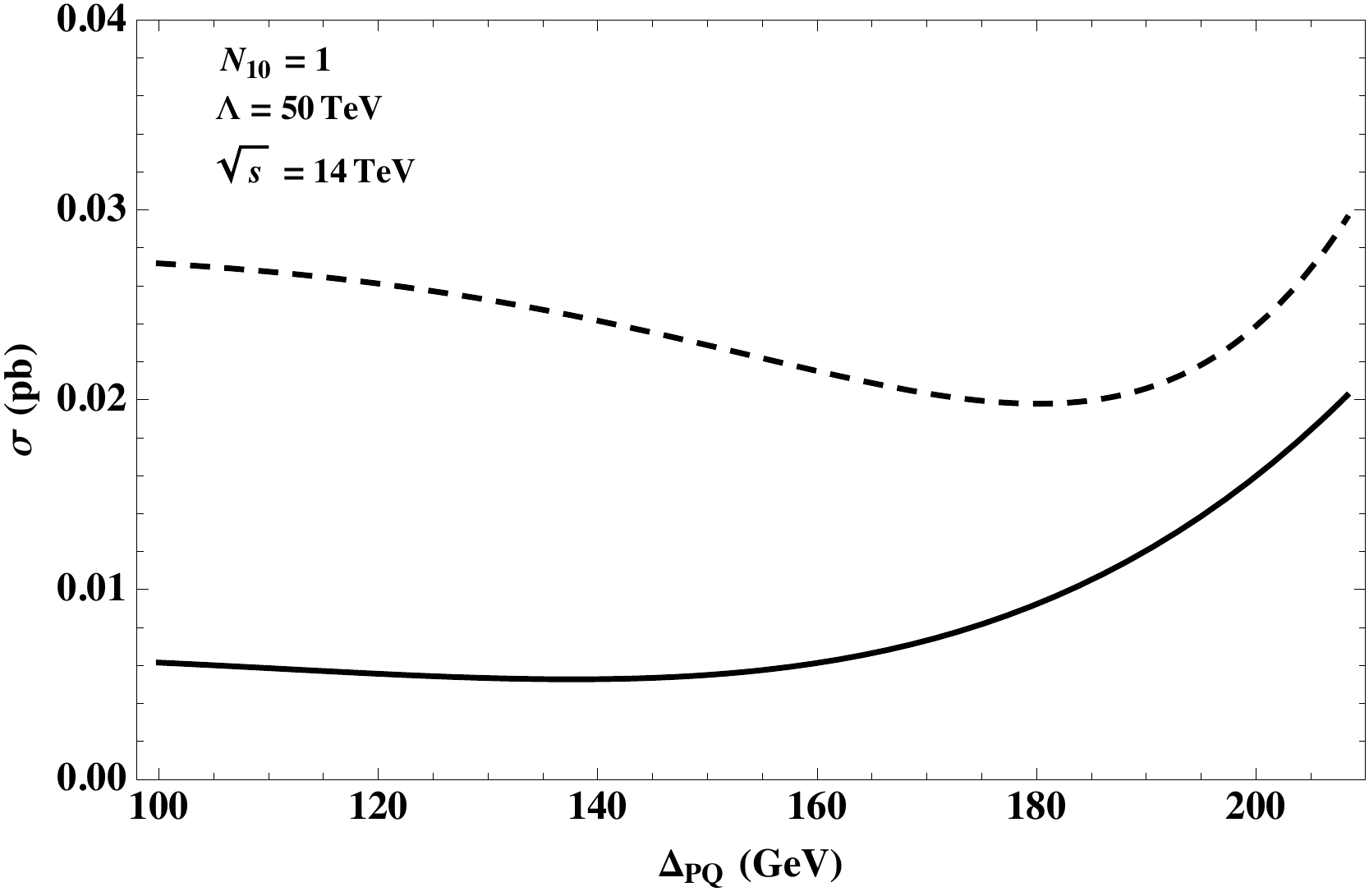}
   \includegraphics[width=5.in]{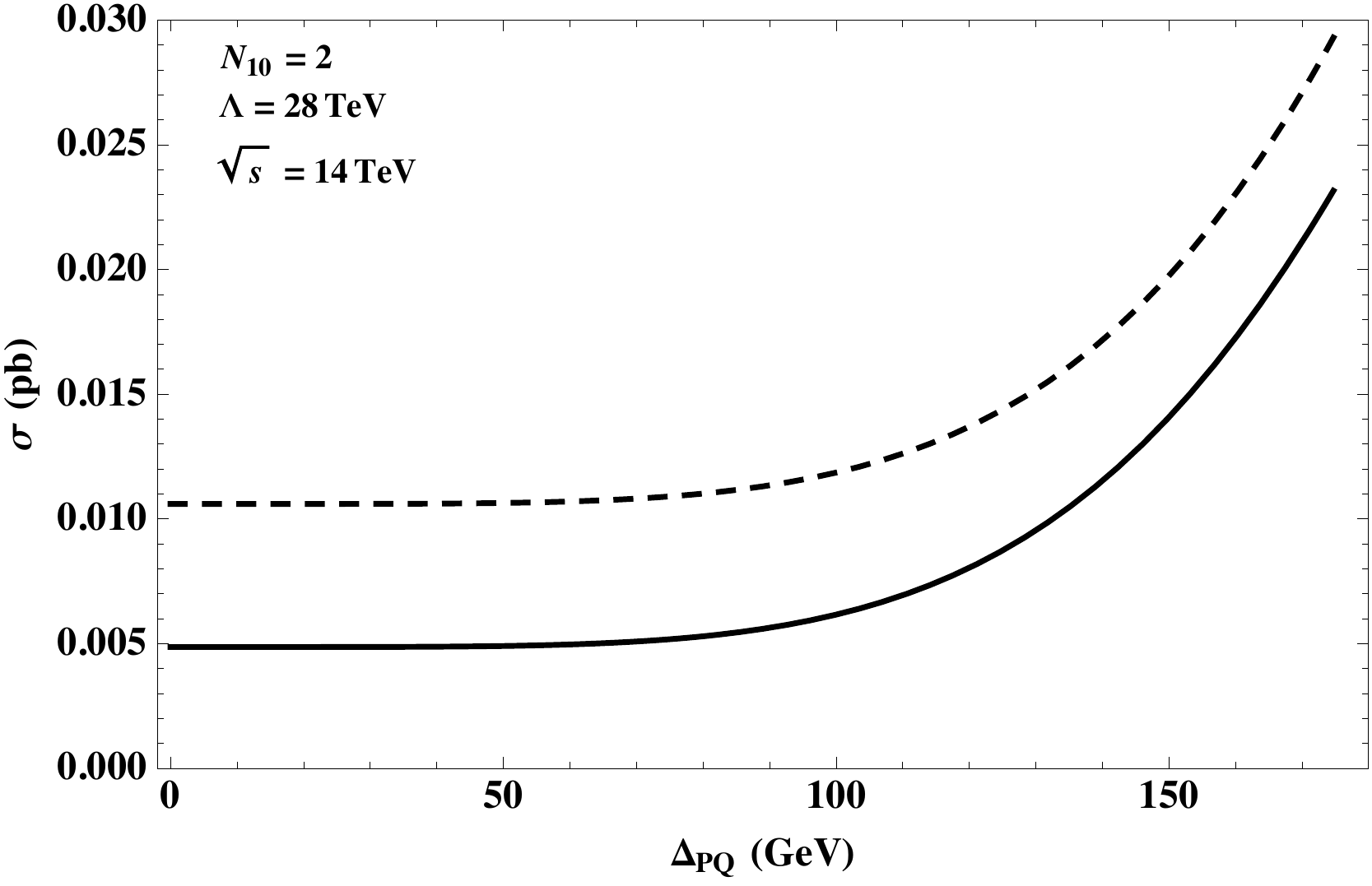}
   \caption{Cross sections of exclusive events with only two staus as functions of $\Delta_{PQ}$ in stau NLSP F-theory GUTs.
   The upper panel is for $N_{10}=1$ and $\Lambda=50$~TeV, while the lower one is for $N_{10}=2$ and $\Lambda=28$~TeV.
   The upper dashed line corresponds to the signal without a cut on  $E^{miss}_T$, while the lower solid line corresponds to the signal
   with $E_T^{miss} < 10$ GeV.}
   \label{fig:onlystaudpq}
\end{figure}

We now study the dependence of the resulting exclusive two stau events as a function of $\Delta_{PQ}$, fixing a particular representative value of $\Lambda = 50$ TeV for $N_{10} = 1$ and $\Lambda = 28$ TeV for $N_{10} = 2$ F-theory GUTs. In figure \ref{fig:onlystaudpq} we plot this dependence. First consider the case of $N_{10} = 1$ F-theory GUTs. When the further cut $E_T^{miss} < 10$ GeV is not imposed, there is a two-fold degeneracy in the expected cross section. This is because the two contributions from Drell-Yan produced staus and charginos behave differently under shifts of the PQ deformation. Whereas the PQ deformation decreases the mass of the stau, and thus leads to a monotonically increasing contribution from Drell-Yan produced staus, for Drell-Yan produced charginos, the overall production cross section of charginos is independent of $\Delta_{PQ}$, but as seen in section \ref{sec:NNLSP}, the subsequent branching fraction of charginos to a stau and neutrino decreases at larger $\Delta_{PQ}$, as new decay channels become available. This has the effect of lowering the contribution to this channel. In tandem with the contribution from Drell-Yan produced staus, the net effect is to produce a small dip in the cross section as a function of $\Delta_{PQ}$. Note, however, that once the further cut $E_T^{miss} < 10$ GeV is included, the cross section is again a monotonically increasing function of $\Delta_{PQ}$, as expected.

Next consider the case of $N_{10} = 2$ F-theory GUT models. In section \ref{sec:NNLSP} we found that the change in the chargino branching fraction is much milder in such scenarios, essentially because no new decay channels open up as $\Delta_{PQ}$ is changed. As a consequence, even without imposing a further $E_T^{miss} < 10$ GeV cut, we see that the exclusive two stau channel is a monotonically increasing function of $\Delta_{PQ}$. Imposing this cut leads to a shift of about $0.005$ pb in the overall cross section, which is largely independent of $\Delta_{PQ}$.

Since the stau is the sparticle with mass most sensitive to the PQ deformation, it is especially interesting to use the expected number of events
to extract more detailed information on the PQ dependence. Imposing the further selection cut $E_T^{miss} < 10$ GeV, we note that in figure \ref{fig:onlystaudpq}, the overall cross section ranges from roughly $0.005$ pb at low $\Delta_{PQ}$ in the $N_{10} = 2$ scenario, to about $0.02$ pb in both $N_{10} = 1$ and $N_{10} = 2$ scenarios. For $10$ fb$^{-1}$ of integrated luminosity, this translates to $50 - 200$ expected events.



\newpage
\subsection{Two Stau Plus Lepton(s) Channels}

In this subsection we study two stau events which include one or two leptons. Such final states typically originate from the decay of the second neutralino and lightest chargino as in the decay processes:
\begin{eqnarray}
\widetilde{\chi}_2^0 &\rightarrow &{\tilde l}^{\pm} + l^{\mp} \rightarrow  {\tilde \tau}^{\pm}_{1} + \tau^{\mp} + l^{+} + l^{-}, \nonumber\\
                               && \tilde \nu + \nu \rightarrow {\tilde \tau}^{\pm} + (\tau^{\mp} \text{ or } l^{\mp}) + 2 \nu, \nonumber\\
                               && {\tilde \tau}_1^{\pm} + \tau^{\mp} \nonumber\\
\widetilde{\chi}_1^{+} &\rightarrow &{\tilde \tau}_1^{+} + \nu_{\tau}, \nonumber\\
&& {\tilde \nu}_{\tau} + \tau^{+} \rightarrow {\tilde \tau}_1^{\pm} + \tau^{\mp} + \tau^{+} + \nu \nonumber
\end{eqnarray}
where as usual in collider studies, we have distinguished the somewhat different $\tau$ and other leptonic final states ($l$'s). $\tau$'s which decay leptonically to $e$ and $\mu$ will contribute isolated leptons to the
final state. Given the fact that the lepton efficiency is high, the relative size of different signatures with different number of leptons encodes the information about the branching factions of chargino and neutralino decays, and therefore can be used
to probe the size of $\Delta_{PQ}$, as well as to serve in reconstructing the mass of the second neutralino and lightest chargino, a point we shall return to in section \ref{sec:distinguish}.

\begin{figure}[htbp] 
   \centering
   \includegraphics[width=5.in]{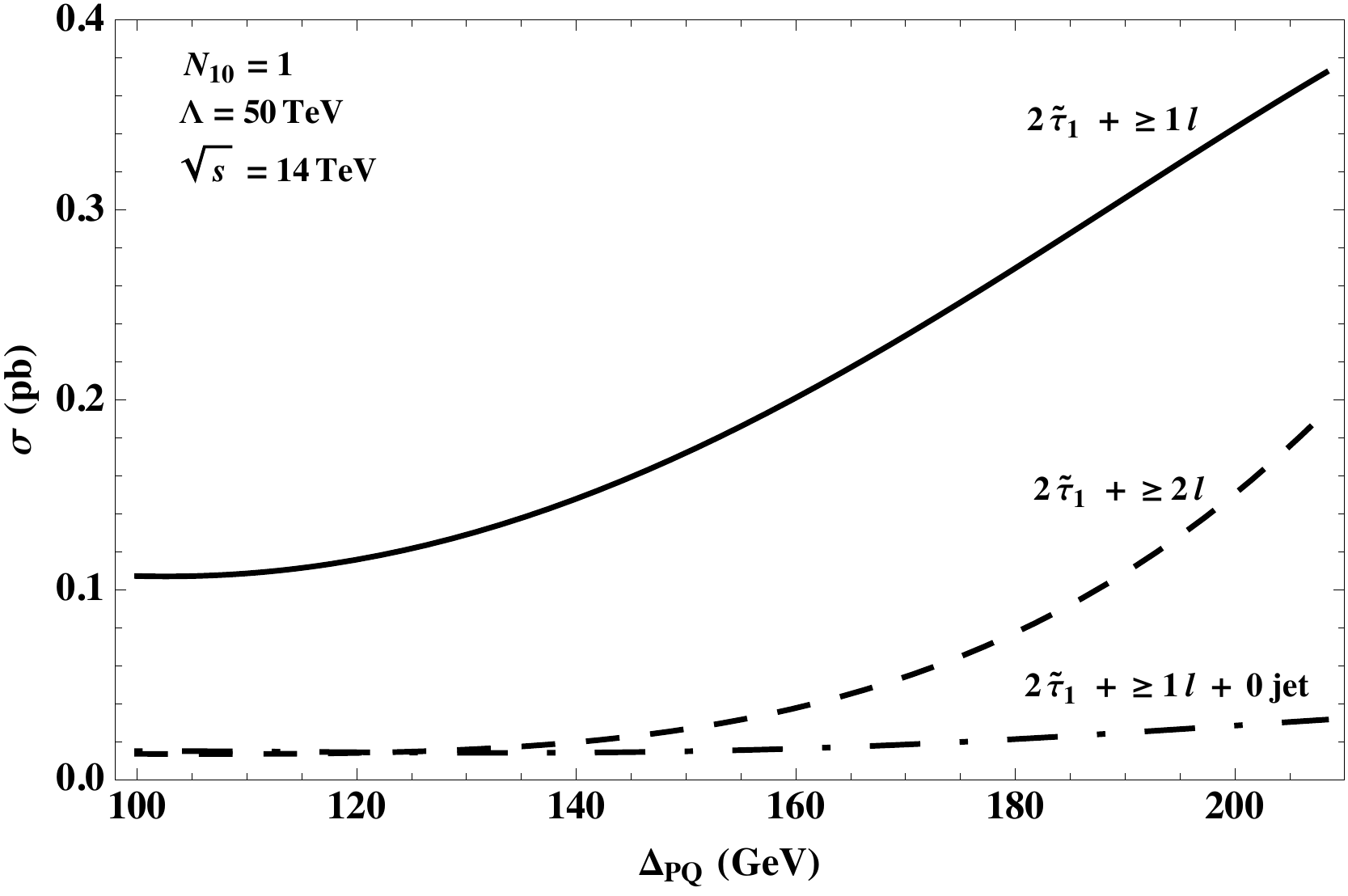}
   \includegraphics[width=5.in]{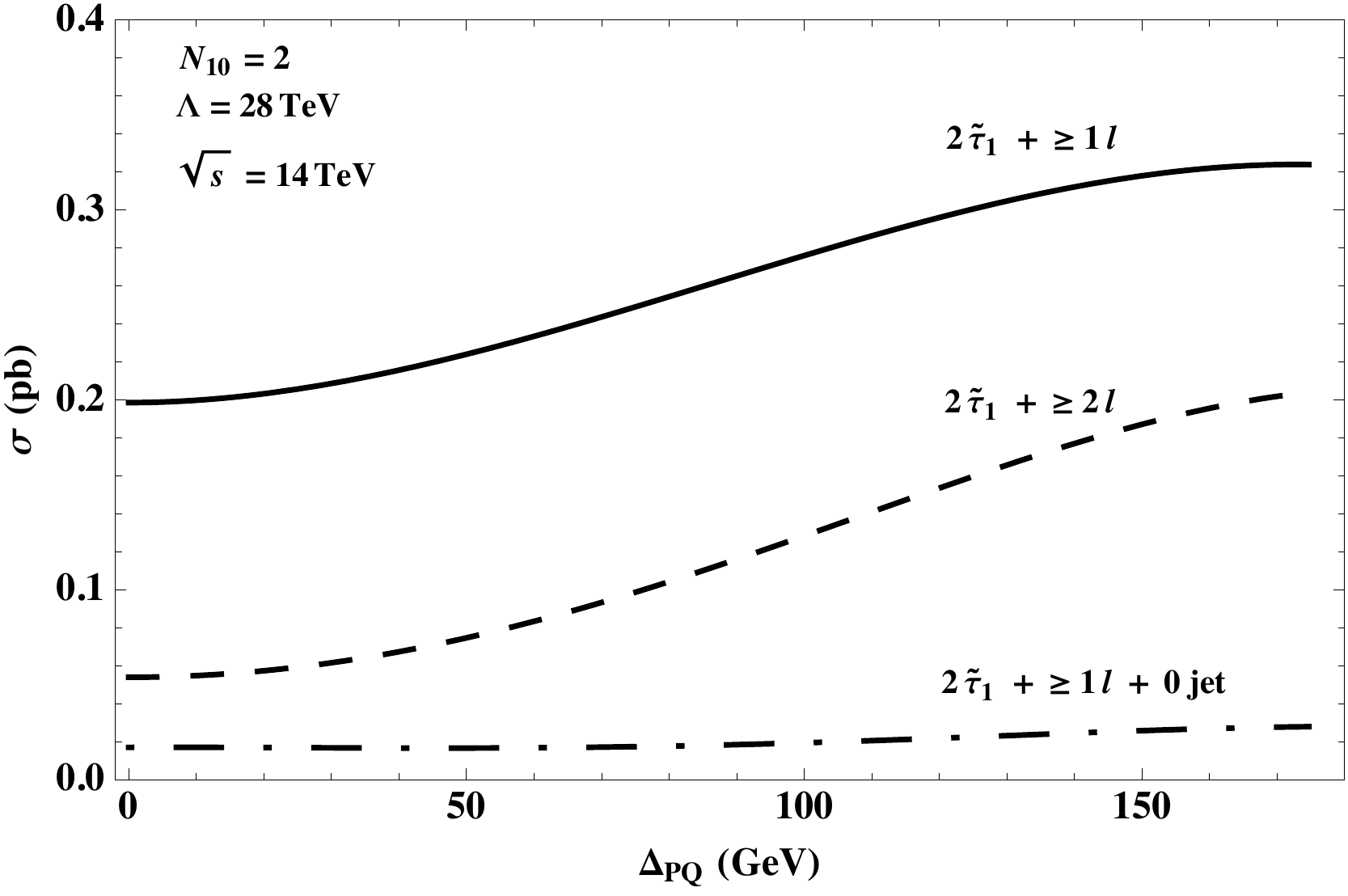}
   \caption{Plots of the cross sections of candidate events with two staus and $\ge 1$ lepton + $0$ jets, $\ge 1$ lepton and $\ge 2$ leptons as functions of $\Delta_{PQ}$ in the stau NLSP regime of F-theory GUTs with $N_{10}=1$ and $\Lambda=50$~TeV (upper panel) and $N_{10}=2$ and $\Lambda=28$~TeV (lower panel).}
   \label{fig:staulepdpq}
\end{figure}

Here we consider the following signatures:
\begin{itemize}
\item  $2\;\tilde \tau_1 + (\ge 1 l) + 0$ jet
\item  $2\;\tilde \tau_1 + (\ge 1 l)$
\item  $2\;\tilde \tau_1 + (\ge 2 l)$
\end{itemize}
where the $2\tilde \tau$ events are selected as in the previous subsection, and ``$0$ jet'' refers
to events which are vetoed if they contain a jet with $p_T> 20$~GeV and $|\eta|< 3$. For these channels, the background is again suppressed by
the strong stau selection cuts, as well as by requiring additional leptons. The dependence of these
signals on $\Delta_{PQ}$ is shown in figure \ref{fig:staulepdpq} for F-theory GUTs with $N_{10}=1$ and $\Lambda=50$~TeV, as well as $N_{10}=2$ and $\Lambda=28$~TeV. By inspection, we see that the inclusive stau signals with leptons and jets
have large cross section as expected and increase monotonically with $\Delta_{PQ}$.

By contrast, the stau signal with leptons but no jet has relatively small cross section. This is also expected because such signals are produced entirely by electro-weak processes, and so are not connected with the decay of a squark or gluino, which have much higher production cross sections when compared with charginos and neutralinos. In addition, note that this signal has very little dependence on $\Delta_{PQ}$, which is to be expected because the gaugino masses are independent of $\Delta_{PQ}$.

\subsection{Inclusive Lepton Signals}

In the above analysis we have focussed on signatures which aim to distinguish staus from muons. Even when such staus cannot be distinguished from muons, they can still generate signatures with low Standard Model background. Since Standard Model processes copiously produce leptons, such signatures require either the presence of multiple leptons, or specific charge combinations of leptons combined with a large $p_T$ cut.

In this subsection we study inclusive lepton signals where we do not explicitly attempt to differentiate staus from muons. In other words, we now consider any stau as a ``muon" if it satisfies:
\begin{itemize}
\item $\beta > 0.67$
\item $p_T > 20$ GeV, $|\eta| < 2.5$
\end{itemize}
With respect to these cuts which allow fast-moving staus, the inclusive lepton signatures we consider are:
\begin{itemize}
\item SS: A pair of same-sign isolated leptons.
\item $3l$: Three isolated lepton candidates.
\item $4l+$: Four or more isolated lepton candidates.
\end{itemize}
In order to reduce background, we also require these signal events to have:
\begin{itemize}
\item At least two hard leptons with $p_T > 100$ GeV
\item At least two hard jets with $p_T > 150$ GeV.
\end{itemize}
Since staus passing the ``muon" cut should always have $p_T\gtrsim 100$~GeV, the $p_T$ cut on the two hardest leptons only mildly affect the signals, but can significantly reduce Standard Model background. For the SS signal, the Standard Model background is negligible. For the $3l$ signal, the primary source of Standard Model background is from $WZ$ and $t\bar t$ production. We reduce the $WZ$ background by requiring that no pair of charged leptons reconstructs to an invariant mass $m_Z \pm 10$~GeV. The $t\bar t$ background is potentially large if the muon from a b quark decay is isolated. To suppress it, we require two hard leptons with $p_T>100$~GeV to reduce the acceptance of leptons from $W$ decay, and additionally require two hard jets with $p_T> 150$~GeV. Alternatively, if one is interested in the signal from electro-weakino or slepton productions, one can also use a jet veto to suppress the $t\bar t$ background, which we will not discuss here. For the $4l+$ signal, the Standard Model background is pre-dominantly from $ZZ$ production. We reduce this background using the same type of cut as in the $3l$ signal.

\begin{figure}[htbp] 
   \centering
   \includegraphics[width=5.in]{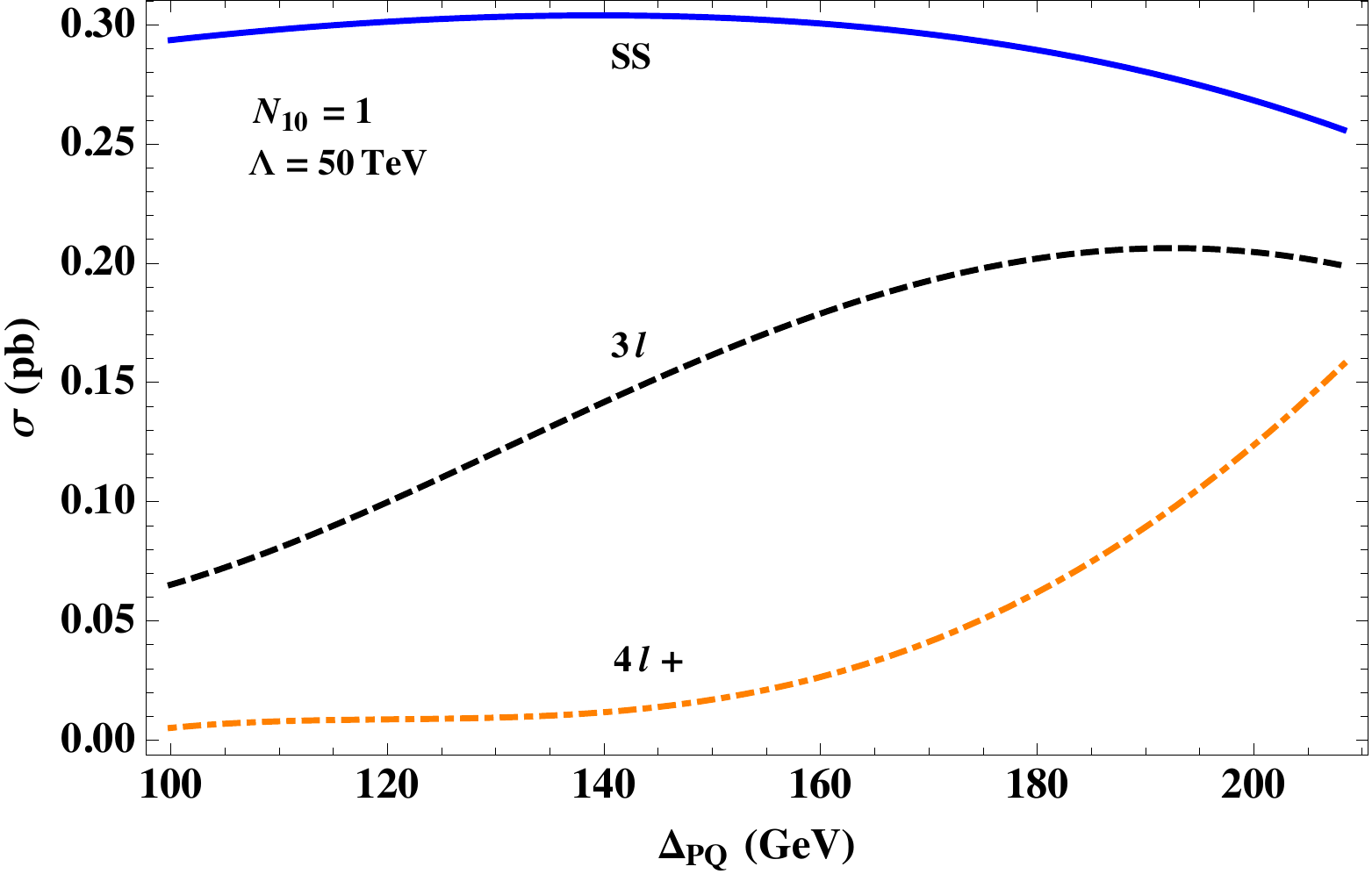}
   \includegraphics[width=5.in]{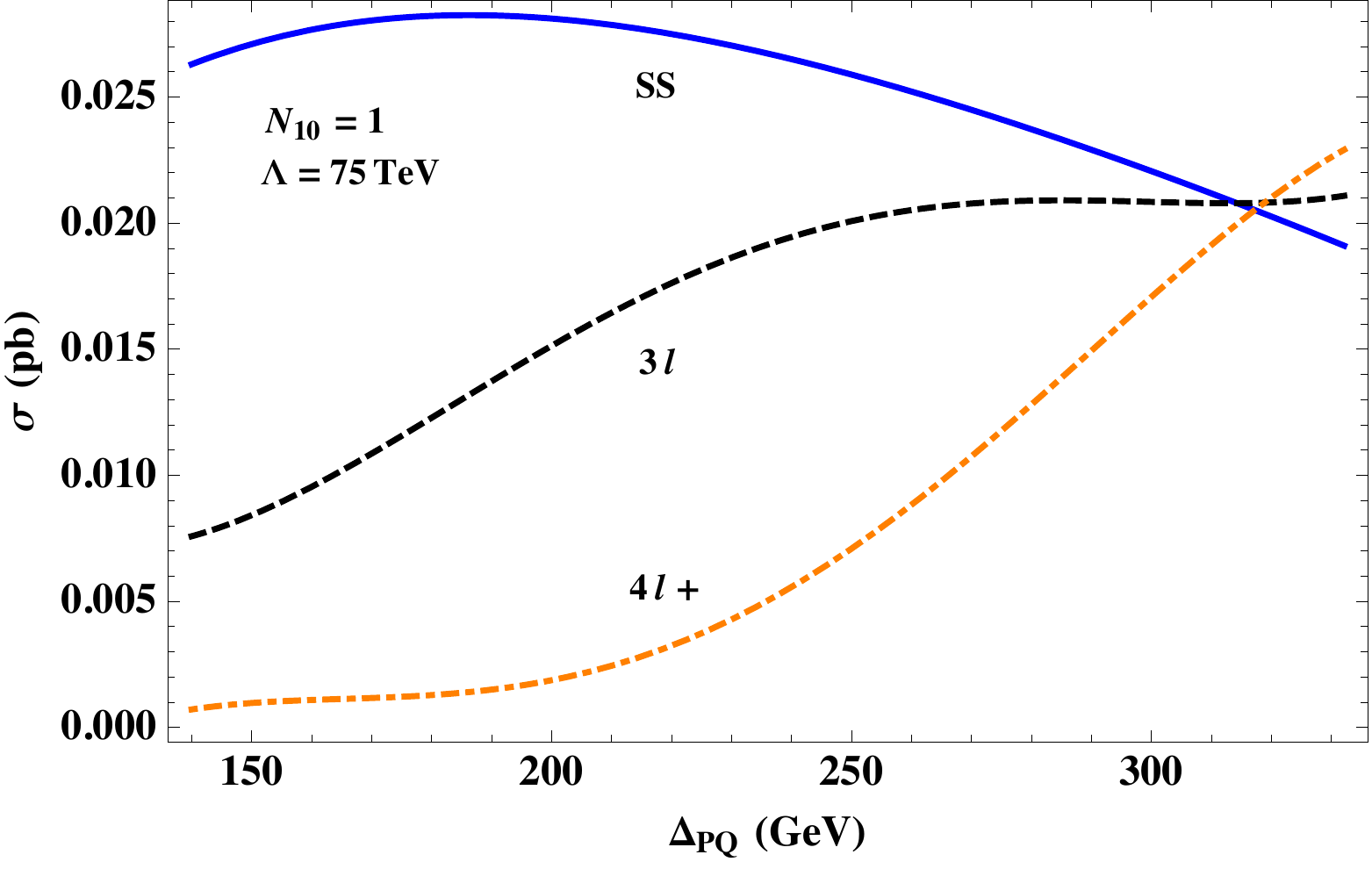}
   \caption{Plots of the cross sections of inclusive lepton events in which fast-moving staus are identified as leptons for two F-theory GUT scenarios with $N_{10} = 1$ at $\Lambda = 50$ TeV (upper panel) and $\Lambda = 75$ TeV (lower panel). The same-sign lepton (SS) tri-lepton ($3l$) and at least four lepton ($4l+$) signals respectively correspond to roughly increasing and decreasing functions of $\Delta_{PQ}$.}
   \label{fig:inclepton1}
\end{figure}

\begin{figure}[htbp] 
   \centering
   \includegraphics[width=5.in]{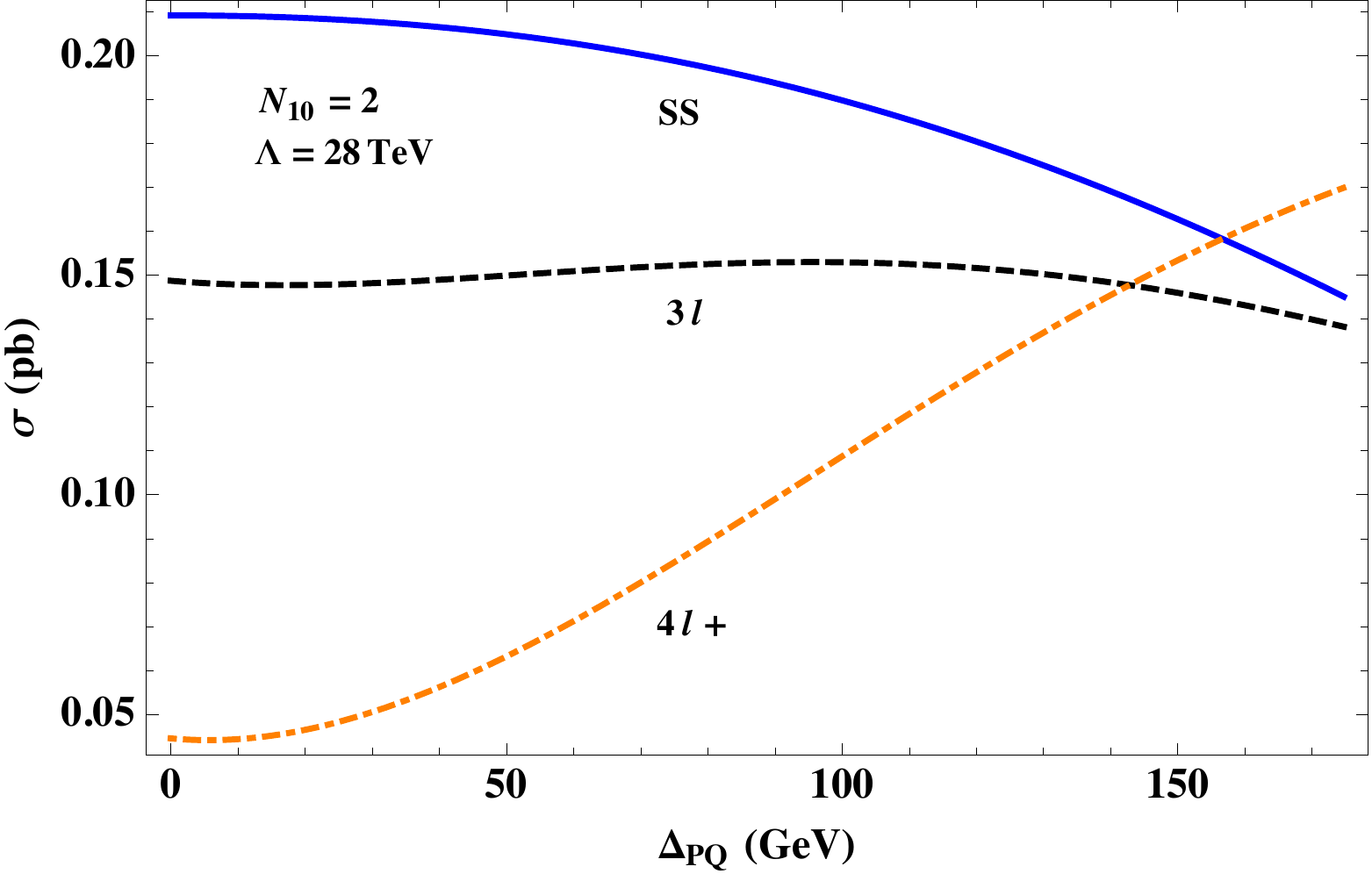}
   \includegraphics[width=5.in]{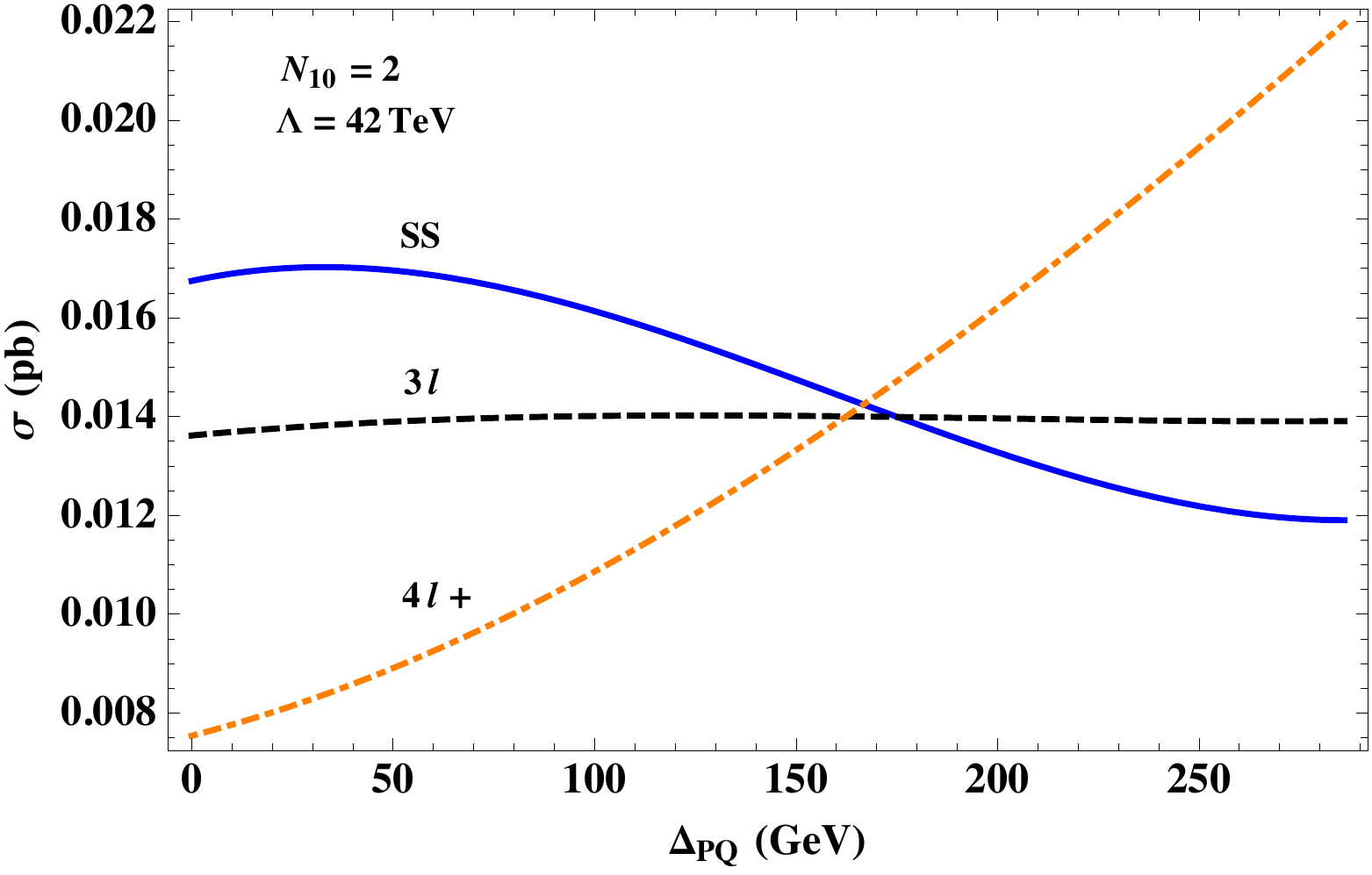}
   \caption{Plots of the cross sections of inclusive lepton events in which fast-moving staus are identified as leptons for two F-theory GUT scenarios with $N_{10} = 2$ at $\Lambda = 28$ TeV (upper panel) and $\Lambda = 42$ TeV (lower panel). The same-sign lepton (SS) tri-lepton ($3l$) and at least four lepton ($4l+$) signals respectively correspond to roughly increasing, constant, and decreasing functions of $\Delta_{PQ}$.}
   \label{fig:inclepton2}
\end{figure}

Let us now discuss the primary origin of these signatures and our theoretical expectations for the corresponding dependence on the PQ deformation. For the same-sign (SS) isolated lepton signal, gluinos and squarks can decay to pairs of same sign lightest charginos $\widetilde{\chi}^{\pm}_{1}\widetilde{\chi}^{\pm}_{1}$ and pairs of second lightest neutralinos $\widetilde{\chi}^{0}_{2}\widetilde{\chi}^{0}_{2}$. The subsequent decay of the lightest chargino for example via $\widetilde{\chi}^{\pm}_{1} \rightarrow \widetilde{\tau}^{\pm}_{1} + \nu_{\tau}$ and the second neutralino for example via $\widetilde{\chi}^{0}_{2} \rightarrow \widetilde{\tau}^{\pm}_{1} + \tau^{\mp}$ with the $\tau$ decaying hadronically, will generate SS lepton signals.

As can be seen from figures \ref{fig:inclepton1} and \ref{fig:inclepton2}, the SS signal is roughly a decreasing function of $\Delta_{PQ}$. To see how this comes about, note that in the case of same sign staus generated by lightest chargino and second neutralino decays, as $\Delta_{PQ}$ increases, the branching fraction to states without additional charged leptons decreases. Note that increasing the value of $\Delta_{PQ}$ simply diverts these decays through channels which generate additional charged leptons in the final state.

Tri-lepton ($3l$) and four or more lepton ($4l+$) signatures correspond to processes involving two fast-moving staus, as well as additional $e$ or $\mu$ leptons, and can be generated in a variety of different cascade decays. For example, focussing on the contribution from a single branch of a cascade decay, the $3l$ signal can be generated by decays involving $\widetilde{\chi}_1^{+} \rightarrow {\widetilde l} + \nu$, $\widetilde{\chi}_1^{+} \rightarrow \widetilde \nu + l$, $\widetilde{\chi}_2^0 \rightarrow \widetilde{\tau}^{\pm}_{1} + \tau^{\mp}$ or other channels where $\widetilde{\chi}^{0}_{2}$ decays to a stau and some taus. Note that we can get contributions to the $3l$ and $4l+$ signatures depending on whether the taus in the final state decay hadronically or leptonically. Figure \ref{fig:inclepton1} shows that for $N_{10} = 1$, both of these signals are roughly increasing functions of $\Delta_{PQ}$. As $\Delta_{PQ}$ increases, the masses of $\widetilde{e}_L$, $\widetilde{\mu}_L$, $\widetilde{\tau}_2$, $\widetilde{\nu}_{e}$, $\widetilde{\nu}_{\mu}$ and $\widetilde{\nu}_{\tau}$ decrease and eventually drop below the mass of $\widetilde{\chi}_2^0$ and $\widetilde{\chi}_1^{\pm}$, while $\widetilde{e}_R$ and $\widetilde{\mu}_R$ both become lighter than $\widetilde{\chi}^{0}_{1}$.\footnote{Note that in the case of $N_{10} = 2$ F-theory GUTs, this statement already holds for vanishing PQ deformation.} Consequently, the branching ratios of $\widetilde{\chi}_2^0$ and $\widetilde{\chi}_1^{\pm}$ into these states increases. Since these states typically decay to $\widetilde{\tau}_1$ with additional leptons, the increase of $\Delta_{PQ}$ also increases the $3l$ and $4l+$ signals.

In the case of $N_{10} = 2$ models, figure \ref{fig:inclepton2} shows that the $3l$ signal is roughly constant, while the increase in $4l+$ signatures becomes more prominent than for $N_{10} = 1$. The difference in the $N_{10} = 1$ and $N_{10} = 2$ cases is due to a subtle tradeoff in the increase and decrease of the branching fractions of heavier sparticle states. For example, the branching fraction of $\widetilde{\chi}_1^{+}$'s via $\widetilde{\chi}_1^{+} \rightarrow {\widetilde l} + \nu$ increases as $\Delta_{PQ}$ increases, while the branching fraction of $\widetilde{\chi}_2^0 \rightarrow \widetilde{\tau}^{\pm}_{1} + \tau^{\mp}$ decreases. For $N_{10} = 1$, the net contribution from these effects leads to an increase in the overall number of $3l$ signals, while for $N_{10} = 2$, the positive and negative contributions to this branching fraction more exactly balance out, leading to a constant $\Delta_{PQ}$ dependence.

\section{Stopped Staus}\label{sec:STOPPED}

Up to this point, we have treated the stau NLSP as stable on timescales probed by colliders. In this section we study the prospects for F-theory GUT stau NLSPs to become stopped in or near the detector, and to decay at some later time to a tau and a gravitino. Searching for stopped staus through their decay products would provide additional evidence that the quasi-stable massive charged particle is in fact a stau. When the origin of the stopped stau can be reconstructed, observing its decay would provide a measurement of its lifetime, and consequently the scale of supersymmetry breaking \cite{Buchmuller:2004rq, Hamaguchi:2004df, Feng:2004yi, DeRoeck:2005bw, Hamaguchi:2006vu}.

\begin{figure}[ptb] 
   \centering
   \includegraphics[width=5.5in]{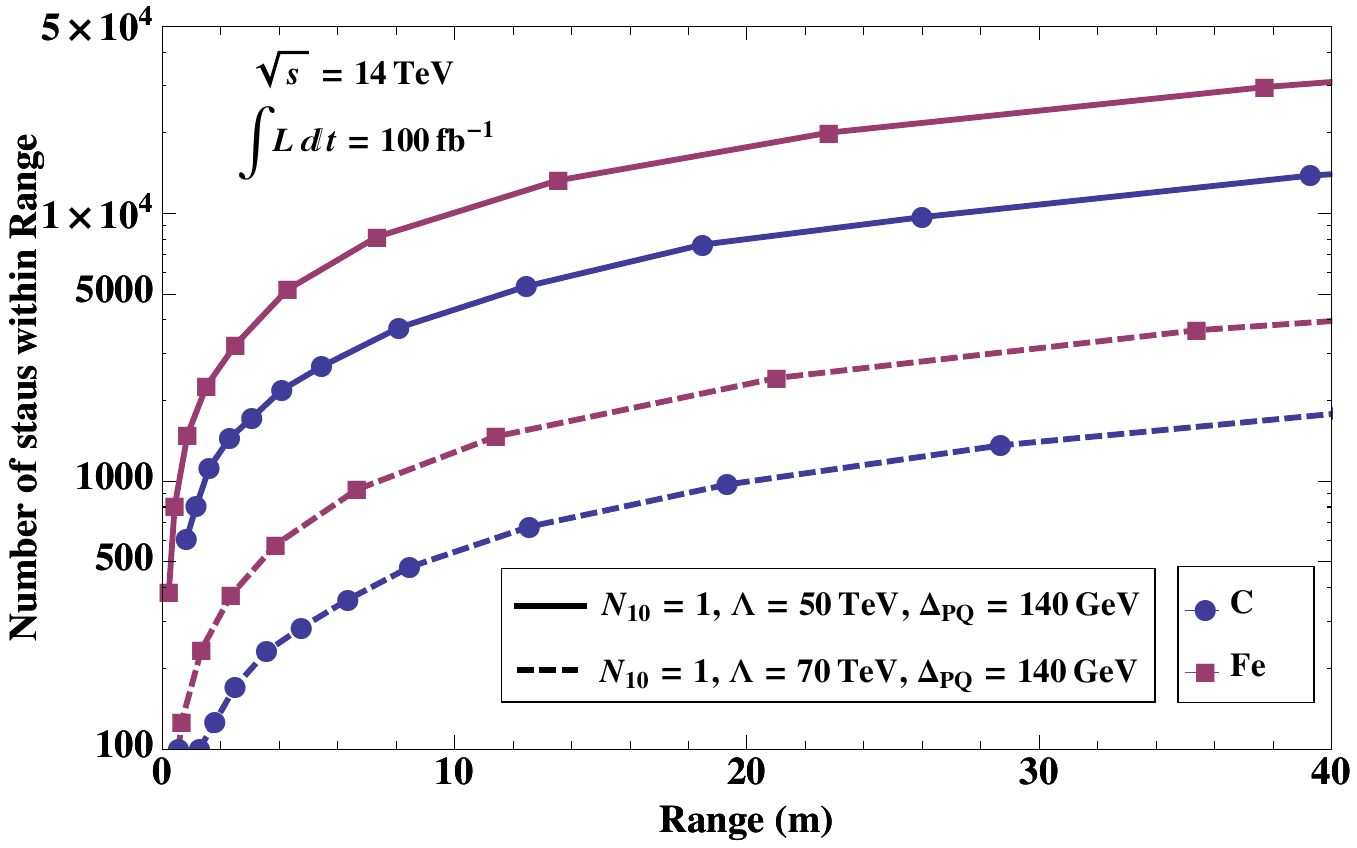}
   \caption{Plot of the number of stopped staus expected from $100$ fb$^{-1}$ of simulated LHC data in two F-theory GUT scenarios as a function of distance from the collision point in an idealization where surrounding the collision point there is a large ball of either carbon (C) or iron (Fe). As we integrate over larger distances, the number of stopped staus necessarily increases.}
   \label{fig:staurange}
\end{figure}

An analysis including all details of the ATLAS and CMS detectors, and in particular the ways that they are situated in their respective caverns is beyond the scope of the present paper. To extract the main physical features of stopped stau scenarios, we consider an idealization where after the initial production of staus, these particles enter a large spherical ball of either iron or carbon. Using the overall production cross section of staus as well as the associated velocity distribution obtained in section \ref{sec:LongLive}, we simulate $100$ fb$^{-1}$ of LHC data and compute the total number of stopped staus $N(R)$ contained within a distance $R$ from the collision point. Here we focus on candidate staus with $p_T > 20$ GeV and $|\eta| < 2.5$. As expected, we see that integrating over an ever larger ball leads to more stopped staus. Though small, the number of such stopped staus is not completely negligible. In figure \ref{fig:staurange} we show a plot of $N(R)$ for stopping in iron and carbon for the $\mathrm{Maj}_{\mathrm{MID}}^{(1)}$ scenario. In this plot, we see that the expected number of stopped stau events is on the order of a few thousand after passing through $\sim 5$ meters of either iron or carbon. More generally, increasing $\Lambda$ tends to decrease the total production cross section of staus, and thus the total number of staus which could become stopped.

The effects of the PQ deformation on the number of stopped staus is more subtle. Fixing the mass of the other sparticles while lowering the mass of the stau will cause the staus which are produced through a decay chain to be more energetic, which in turn skews the velocity distribution to faster moving staus. Though this does reduce the total number of slow-moving staus, the stopping length of the stau depends linearly on its mass. In other words, though there are fewer slow-moving staus, they are more easily stopped. We find that when both effects are taken into account, the effective number of stopped staus at a given distance from the collision point increases as the stau mass decreases. For example, comparing the benchmark Maj$_{\mathrm{MID}}^{(1)}$ scenario with $\Delta_{PQ} = 140$ GeV to the benchmark Maj$_{\mathrm{HI}}^{(1)}$ scenario with $\Delta_{PQ} = 209$ GeV, we find the number of stopped staus at a given distance is larger in the Maj$_{\mathrm{HI}}^{(1)}$ model by a factor of $\sim 1.3$.

We now discuss how such stopped staus could in practice be observed. First consider the case of staus stuck in the rock surrounding the detector. In this case, a full reconstruction of the timing of the stau decay with that of the initial production can likely not be achieved, but the observation of the decay product, given by a leptonically decaying tau could still be observed. The decay of a tau to a muon can be measured by the muon chamber, and in some cases also the inner tracking chamber.

The primary source of Standard Model background for such muon events is from cosmic rays. Such cosmic rays enter from the sky above, and so will correspond to muons passing \textit{downwards} through the detector. By contrast, there are no cosmic rays passing through the other side of the Earth, so that contamination from \textit{upwards} moving cosmic rays will be negligible. Hence, we focus on the prospects for detecting upwards moving muons from stau decays. The number of such upward moving muons is estimated to be $\sim 0.7\%$ of the total number of stopped staus \cite{DeRoeck:2005bw}. Assuming the number of stopped staus can be estimated by the number of staus with $\beta < 0.5$, we can expect ${\cal O}(50)$ events in the benchmark ${\rm Maj}_{\rm MID}^{(1)}$ scenario. We note that this is an order of magnitude estimate, and should be improved with a more realistic simulation.

Even for upwards moving muons, however, there is still a non-negligible background from atmospheric neutrinos which pass through the Earth. When these neutrinos (and anti-neutrinos) strike nuclei underneath the detector, they can generate a conversion $\nu \rightarrow \mu$, with roughly equal likelihood for generating $\mu^{+}$ and $\mu^{-}$. Using the results of the MACRO experiment \cite{Ambrosio:2004ig}, the number of such upward-going muons  passing through a $20 \mathrm{m} \times 20 \mathrm{m}$ detector is expected to be $\sim 80$ events/year. Using the same selection cuts on upward passing muons as in \cite{Ambrosio:2004ig}, it was found there that the muon energy distribution from this background has a broad peak around $50$ GeV. The kinematic upper bound on the energy of a muon generated by the decay $\widetilde{\tau}_{1} \rightarrow \widetilde{G}\nu_{\tau} \nu_{\mu} \mu$ will have energy $E_{\mu} < m_{\widetilde{\tau}_{1}}/2 \sim 50 - 100$ GeV, which is itself on the tail of the distribution of this three-body process. This means that of the roughly $80$ background muons, less than half actually constitute background for stau decays. Further, though this background is not completely negligible, the signal can still be statistically significant if the velocity and direction of the background is reliably calculated. In addition, the tracking information of a stau event can be helpful in identifying these events \cite{DeRoeck:2005bw}. This is because the momentum and the direction of the slow-moving staus can be recorded during their escape from the detector. This information can be used to infer the trajectory along which a slow-moving stau could be stopped. Looking for muons coming from a particular direction, this would provide a means to significantly reduce the background of such events.

Staus can also become stuck in the detector. For tau leptons which decay hadronically, the optimal location for a stau to be stopped would be in the hadronic calorimeter of either the ATLAS or CMS detector. This type of search is similar in spirit to looking for stopped gluinos in split supersymmetry models \cite{StoppedGluinos}. From figure \ref{fig:staurange} we see there are $\sim 3000$ (resp. $5000$) staus stopped in $2$ meters (resp. $4$ meters) of iron out from the collision point for the low $\Lambda$ model. This implies that a similar number of staus should be stopped in the calorimeter. Let us first consider the case that those staus stopped inside the hadronic calorimeter (HCAL) decay hadronically via the tau decay product of the stau. This possibility is interesting since the tau jet is typically narrow, and therefore can deposit all its energy in one or very few calorimeter cells. For a $200$~GeV mass stau, the tau jet will have energy $\sim 100$~GeV. In this case, the hadronic calorimeter would light up, but no track would be seen in the tracking chamber, and no
energy would be deposited in the electromagnetic calorimeter (ECAL). This is in contrast to QCD jets which would leave tracks as well as have
energy deposited in both the ECAL and HCAL. This would seem to indicate that QCD background can be reduced significantly.
Even so, given the large number of QCD events, any leakage to the signal could be dangerous, which must be studied in detail
separately. Cosmic rays are another potential source of background. However, most of the cosmic rays are muons and can be detected
in the muon chamber. If they are neutral particles, they are likely to be stopped by the outer material of the detector.
Stopped staus stuck in the HCAL can also decay into muons, although the chances are smaller at around ~10\%.
If the muon goes outward from the hadronic calorimeter, the muon can only be detected in the muon chamber when there is no activity in other
parts of the detector. Standard Model objects cannot mimic this type of process, and so hardware fluctuations or
malfunctions would appear to provide the main source of background.

Stopped staus can in principle also be used to deduce the scale of supersymmetry breaking \cite{Hamaguchi:2004df, Feng:2004yi, DeRoeck:2005bw, Hamaguchi:2006vu}. Recall that the decay rate of the stau NLSP to a gravitino and tau $\Gamma_{\widetilde{\tau}_{1}}$ is given in terms of the scale of supersymmetry breaking $\sqrt{F_0}$ felt by the gravitino and the mass of the stau through the relation (see for example \cite{GiudiceSUSYReview}):
\begin{equation}\label{eq:DECAY}
 \Gamma_{\widetilde{\tau}_{1}} = \left(\frac{m_{\widetilde{\tau}_{1}}}{100\;{\rm GeV}}\right)^5 \left(\frac{100\;{\rm TeV}}{\sqrt{F_0}}\right)^4 \cdot
 (2 \times 10^{-3} \; {\rm eV}).
\end{equation}
Measuring the scale $\sqrt{F_0}$ would then allow us to also determine the mass of the gravitino. Let us note that although here we have distinguished between $\sqrt{F_0}$ and the scale of supersymmetry breaking $\sqrt{F}$ felt by the messengers, in the context of F-theory GUTs it is most natural to set $F = F_0$. In F-theory GUTs, $F \sim 10^{17}$ GeV$^{2}$, and the lifetime of the NLSP is on the order of one second to an hour \cite{HVLHC}.

Methods to measure the lifetime of the stopped stau typically rely on being able to transport the material in which the stau is stopped to a remote location, where its decay can then be carefully studied. For example, in \cite{Hamaguchi:2004df, Hamaguchi:2006vu} it was proposed that a
``stopper detector'' be added which could surround the ordinary detector, and in \cite{Feng:2004yi} it was proposed to use a large water tank as a collector for stopped staus. Staus which become lodged in the neighboring cavern rock could in principle also be removed \cite{DeRoeck:2005bw}. Both methods work best for staus which are stable on the order of at least a day to a week. In the context of F-theory GUTs, however, where the stau lifetime is far less than a day, such detection methods do not appear as practical. Other methods have been proposed for detecting quasi-stable staus which decay on shorter timescales, as for example in \cite{Asai:2009ka}.\footnote{Though not the case studied here, for staus which decay with an average time of $10^{-8} - 10^{-5}$ seconds, it is possible to exploit the fact that there is a distribution of timescales for the decay of the stau, so that some staus will decay inside the detector \cite{Ambrosanio:2000ik}. For F-theory GUTs, this method is not available because the stau has a lifetime of one second to an hour.} For example, in the related context of stopped gluino scenarios \cite{StoppedGluinos}, it was proposed that if a pair of gluinos become stopped in the detector or neighboring cavern, correlating the time separation between the decays of these pairs would then provide another means to extract the lifetime, and could in principle be adapted to the case of stopped stau scenarios.

Though a full study is beyond the scope of the present paper, we now speculate on potential ways to measure the lifetime
of staus in the context of F-theory GUTs. The primary challenge in extracting the lifetime of the stau stems from
correlating the initial production of the stau with its final decay. Although staus typically decay out of time with the bunch
crossing of the original production event, the information on the stau (for example information of tracking and energy
deposit in both the HCAL and ECAL) may already be recorded in the data if the production event is hard enough (e.g. with hard jets
or leptons) to be triggered on. In fortuitous circumstances, the direction and velocity of the stau could then be reconstructed
from an off-line analysis from the energy deposit $dE/dx$ in the calorimeters and monitored drift tube data for staus stopped in the muon chamber. In
such cases, it would then be possible to deduce where such a stau would possibly be stopped. For example, if the stau stopped in
the middle of the CMS muon chamber\footnote{Compared to ATLAS, the CMS muon tracking system has a far higher density of iron, making it much more likely that a stau can become trapped there. We thank members of the CMS group for discussions on this point.}, the tracking apparatus would detect
a muon track traversing and then ending in some section of the muon chamber, and it would become possible to estimate the position of
the candidate stopped stau. If at some later time a muon was emitted in the vicinity of the stopped stau, and provided there is no other activity in that part of the detector, this would likely correspond to a signal for a stopped stau. The time difference between the decay event
and production event would then provide a measure of the lifetime of the long-lived stau. For staus stuck in the
HCAL a similar type of correlation analysis may also be available. Given the tantalizing prospect of obtaining
direct access to the scale of supersymmetry breaking, it would seem important to investigate this issue further.

\section{Mass Spectra and Distinguishing Look-Alikes}\label{sec:distinguish}

Our focus in much of this paper has been on signatures of stau NLSP F-theory GUTs. Assuming evidence has been found for such a scenario, it is important to study what classes of models reproduce the same observables. This broader question based on the LHC inverse problem has been studied from different vantagepoints for example in \cite{LHCInverse, Hubisz:2008gg}. One approach to this problem is based on the general footprint method \cite{Kane:2006yi,KaneFootprint}, which directly compares the collider observables of different classes of models to determine possible differences. This method has been applied to F-theory GUT models with a bino NLSP in \cite{HVLHC} to compare particular classes of well-motivated supersymmetry breaking scenarios.

In this section we consider a simplified version of this problem by studying possible degeneracies in the mass spectrum of F-theory GUTs with other models with an MSSM-like spectrum. More precisely, we study how mass measurements extracted from stau NLSP scenarios can be used to distinguish between F-theory GUTs and minimal gauge mediated supersymmetry breaking (mGMSB) models. To this end, we first review how to extract details of the sparticle mass spectrum in stau NLSP scenarios which are similar to F-theory GUTs. Next, we use the high precision of these mass measurements as a means to distinguish between F-theory GUTs and mGMSB models.

\subsection{Mass Reconstruction with Staus}

Following \cite{Ellis:2006vu} and \cite{Ito:2009xy}, we now discuss how quasi-stable stau scenarios allow one to deduce much of the sparticle mass spectrum. We can extract the mass of the stau through the relation:
\begin{equation}
m_{\widetilde{\tau}_{1}} = \frac{p}{\beta\gamma}
\end{equation}
where $\beta$ is measured with the time-of-flight in the precision muon chambers, and $p$ is reconstructed through a measurement of $p_T$ and $\eta$ \cite{Aad:2009wy}. Due to the uncertainty in the measurement of $p$ and $\beta$, the reconstructed mass distribution will show a broadening around the correct mass. From the resolution of $p$ given in equation (\ref{eq:resolution}) of section \ref{sec:LongLive}, we see that the uncertainty in $p$ is proportional to $p^2$ for large stau momentum. This illustrates that the mass reconstruction is more accurate at low momenta and velocities. Since the relative number of low velocity staus is lower compared with the number of high velocity staus, increasing the total number of such events requires a relatively large integrated luminosity. It was shown in \cite{Ellis:2006vu} that the stau mass can be determined up to an uncertainty of $\sim 0.02$ GeV using $30\; {\rm fb}^{-1}$ of integrated luminosity.

\begin{figure}[t] 
   \centering
   \includegraphics[width=5in]{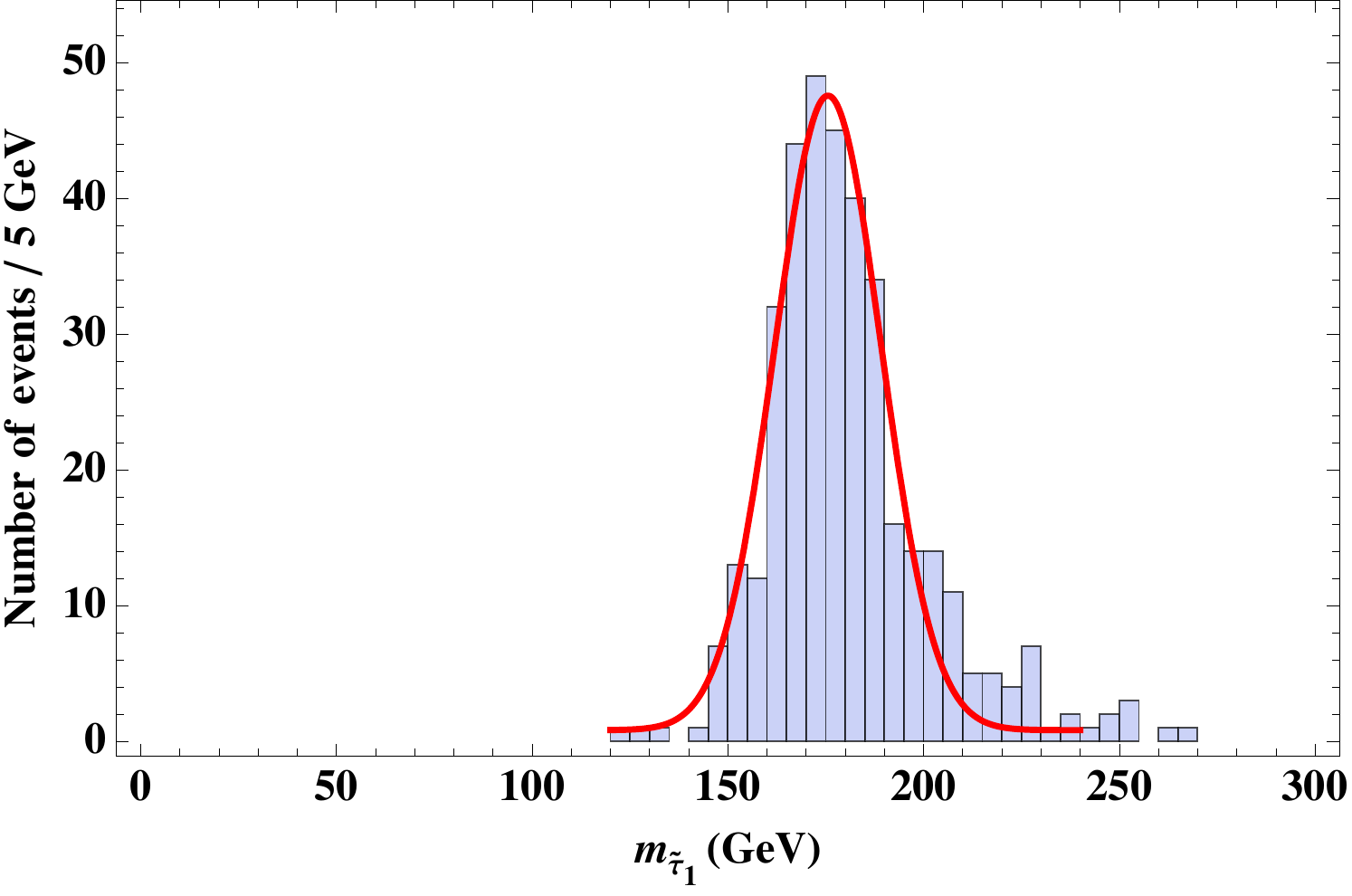}
   \caption{Distribution of the reconstructed stau mass for the benchmark model ${\rm Maj}_{\rm MID}^{(1)}$ defined in subsection \ref{ssec:BENCHMARK}. $500$ events are generated and staus are selected with velocity $0.6< \beta < 0.91$. The Gaussian fit gives $m_{\tilde \tau}^{fit}=175.59\pm 0.47$~GeV, while the actual stau mass is $175$~GeV. }
   \label{fig:mstau}
\end{figure}

Reasonable estimates of the stau mass can also be obtained for lower integrated luminosity. For example, at $\sqrt{s} = 14$ TeV and $\sim 500$ pb$^{-1}$ of simulated LHC data, figure \ref{fig:xsec-cme} of section \ref{sec:LongLive} illustrates that for the $\rm{{Maj}_{MID}^{(1)}}$ benchmark model, the total cross section for supersymmetric production of events is $\sim 1$ pb, indicating there will be about $500$ supersymmetric events with staus. In this case, we find that the best fit of the stau mass is $m_{\tilde \tau}^{fit}=175.59\pm 0.47$~GeV 
using a Gaussian fit (see figure \ref{fig:mstau}). In the analysis, we selected staus with velocity $0.6< \beta < 0.91$. With $100\;{\rm pb}^{-1}$ data, we find only $69$ event stau candidates remain, but the fit of the stau mass is almost as good as the previous case, with the error bar increased slightly to $0.63$~GeV.

The presence of quasi-stable staus in the final state provides a convenient means to reconstruct the masses of other sparticles further up a decay chain. For example, the mass of $\widetilde{\chi}^{0}_{1}$ can be reconstructed by exploiting the fact that over the entire parameter space of stau NLSP F-theory GUT scenarios, there is a significant branching fraction to $\widetilde{\chi}^{0}_{1} \rightarrow \widetilde{\tau}^{\pm}_{1} \tau$. By selecting final states which contain opposite sign staus and two opposite sign $\tau$'s, we can reconstruct the resonance associated to the invariant mass of a pair of $\widetilde{\tau}_{1}^{\pm} \tau^{\mp}$ to a $\widetilde{\chi}^{0}_{1}$. Note that a similar resonance will also occur for $\widetilde{\chi}^{0}_{2}$, providing a means to reconstruct the mass of this sparticle as well. In \cite{Ellis:2006vu} this type of strategy was employed to work backwards up cascade decays to deduce the masses of many sparticles. In particular, the ``$\varepsilon$ model'' considered in \cite{Ellis:2006vu} has a similar mass spectrum to that of the Maj$_{\text{MID}}^{(1)}$ benchmark model (though different enough to be distinguishable!), and so rather than perform a complete analysis of our case, we shall simply use the same precision estimates based on $30$ fb$^{-1}$ of integrated luminosity at $\sqrt{s} = 14$ TeV:
\begin{align}
\Delta m_{\widetilde{\tau}_{1}}  & =0.021\text{ GeV, }\Delta m_{\widetilde
{\nu}_{\tau}}=1.2\text{ GeV, }\Delta m_{\widetilde{l}_{L}}=2.0\text{ GeV} \nonumber\\
\Delta m_{\widetilde{\chi}_{1}^{0}}  & =0.9\text{ GeV, }\Delta m_{\widetilde
{\chi}_{2}^{0}}=2.0\text{ GeV,}\nonumber \\
\Delta m_{\widetilde{q}_{R}}  & =2.8\text{ GeV, }\Delta m_{\widetilde{q}_{L}%
}=3.7\text{ GeV, }\Delta m_{\widetilde{b}_{1}}=57.7\text{ GeV.} \label{eq:mass-resolution}
\end{align}
Note that for all of the non-colored sparticles, the accuracy is within the range of $2$ GeV or less. In addition, the mass difference for first and second generation squarks are both less then $4$ GeV. The reason for the larger uncertainty in the sbottom mass is that events from $\tilde b_1$ and $\tilde b_2$ are mixed together since both of them can decay to $b\tilde \chi_1^0$ \cite{Ellis:2006vu}. The result only serves as an estimate of the combination of $\tilde b_1$ and $\tilde b_2$ masses.

To distinguish between models which can mimic the spectrum of F-theory GUTs, we will also find it convenient to reconstruct the mass of the first and second generation right-handed sleptons $\widetilde{l}_{R}$. In this case, the decay chain of interest depends on the relative masses of $\widetilde{\chi}^{0}_{1}$ and $\widetilde{l}_{R}$. Indeed, we have seen that for $N_{10} = 1$ scenarios at moderate PQ deformation $m_{\widetilde{l}_{R}} > m_{\widetilde{\chi}^{0}_{1}}$, whereas at larger PQ deformation for $N_{10} = 1$ and for all values of $\Delta_{PQ}$ in $N_{10} = 2$ scenarios, we instead have $m_{\widetilde{l}_{R}} < m_{\widetilde{\chi}^{0}_{1}}$. The latter case has been studied in detail in \cite{Ito:2009xy} where it was found that with $100$ fb$^{-1}$ of integrated luminosity at $\sqrt{s} = 14$ TeV, $\Delta m_{\widetilde{l}_{R}} \sim 1$ GeV.

\begin{figure}[t] 
   \centering
   \includegraphics[width=5.in]{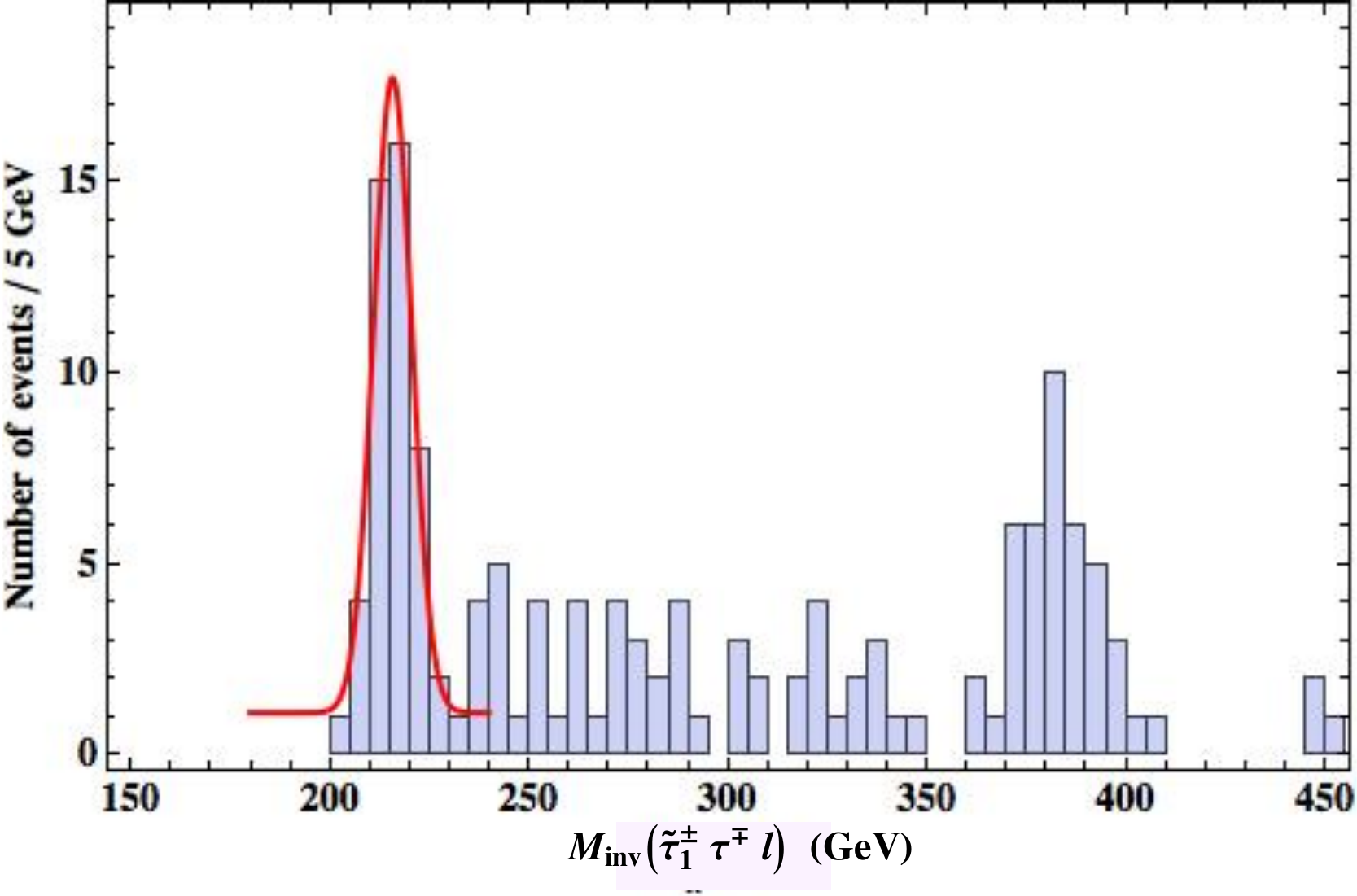}
   \caption{Plot of the lepton-tau-stau invariant mass in the benchmark Maj$_{\text{MID}}^{(1)}$ model using the event selections discussed below equation \eqref{eq:GAUSSIAN}. The peak at $\sim 215$ GeV corresponds to the resonance from decays of the right-handed selectron $\widetilde{e}_{R}$ and smuon $\widetilde{\mu}_{R}$. The somewhat lower and broader peak at $\sim 380$ GeV corresponds to the resonance from decays of the left-handed selectron $\widetilde{e}_{L}$ and smuon $\widetilde{\mu}_{L}$.}
   \label{fig:sleptonmass}
\end{figure}

Let us now return to scenarios where $m_{\widetilde{l}_{R}} > m_{\widetilde{\chi}^{0}_{1}}$ so that right-handed sleptons dominantly decay via
\begin{equation}
\widetilde{l}_{R} \rightarrow \widetilde{\chi}^{0}_{1} l \rightarrow \widetilde{\tau}^{\pm}_{1} \tau^{\mp} l.
\end{equation}
Since the actual production of right-handed sleptons is somewhat low in models such as the Maj$_{\text{MID}}^{(1)}$ benchmark model, generating a sufficient number of events requires a large integrated luminosity, in part because heavier sparticles rarely decay to $\widetilde{l}_{R}$, and because the direct production cross section $\sigma_{\widetilde{l}_{R}^{+} \widetilde{l}_{R}^{-}} = 1.3 \times 10^{-2}$ pb is relatively small. To demonstrate the reconstruction of the right-handed slepton mass
can be done with small enough uncertainty, we perform an analysis with $100\;{\rm fb}^{-1}$ integrated luminosity of simulated data for the benchmark model ${\rm Maj}_{\rm MID}^{(1)}$. First, we selected events with at least one stau candidate, one tau jet and no jets. Then the invariant mass $M_{inv}(\widetilde{\tau}^{\pm}_{1} \tau^{\mp})$ of any opposite-sign tau-stau pair is calculated.\footnote{Here we assume that the tau four momentum
can be fully constructed. In practice, a calibration of the tau momentum can be done as in \cite{Ellis:2006vu}.} Only if there is one invariant mass within $5$~GeV around the true mass of $\tilde \chi_1^0$, the event is selected. Finally, we take all possible combinations of leptons in the event with the tau-stau pair, and the invariant mass of the lepton-tau-stau system is calculated, giving rise to the distribution in figure \ref{fig:sleptonmass}.
Fitting the peaks with a Gaussian plus a constant background
\begin{equation}\label{eq:GAUSSIAN}
A \exp \left [ - \frac{1}{\sigma^2} (M_{inv}(\widetilde{\tau}^{\pm}_{1} \tau^{\mp} l) - M^{peak})^2 \right] + B
\end{equation}
with $A$, $B$, $\sigma$ and $M^{peak}$ fitting parameters, we find that the peak position is $215.8\pm 0.6$~GeV.
This is in good agreement with the actual $\tilde l_R$ mass $214.3$~GeV. Of course, we should also include uncertainties which propagate from the
mass measurement of $\tilde \tau_{1}$ and $\tilde \chi_1^0$. Since these uncertainties are typically $\lesssim 1$~GeV, including them does not significantly affect the overall uncertainty in the mass measurement of $\tilde l_{R}$.

Let us note that in the above we have quoted results from the literature which typically utilize integrated luminosities on the order of $30 - 100$ fb$^{-1}$. We have also seen that cruder mass reconstructions can be performed at lower integrated luminosity. It would be interesting to study the overall expected mass resolution of such stau NLSP scenarios as a function of integrated luminosity. Further detailed information on mass measurements would also be readily available with the ILC.

\subsection{Comparison with Minimal Gauge Mediation}

Measuring the masses of the sparticles provides a very precise way to potentially distinguish models with similar spectra.
Indeed, fitting to an F-theory GUT model leads to a quite specific range of possible values for $N_{10}$ and $\Lambda$. Moreover, because the PQ charge assignments are different in the Majorana and Dirac neutrino scenarios (see section \ref{sec:ParamSpace}), measuring the masses of the sleptons and fitting to the rest of the spectrum would allow us to distinguish between these two possibilities. This would then provide even more precise information on both the size of $\Delta_{PQ}$, as well as the PQ charge assignments of the sparticles.

In this subsection we consider minimal gauge mediated supersymmetry breaking (mGMSB) models such that the lightest bosonic and fermionic superpartners respectively given by a quasi-stable $\widetilde{\tau}_1$ and $\widetilde{\chi}^{0}_{1}$ have the same mass as their F-theory counterparts. We analyze to what extent such look-alikes can mimic the rest of an F-theory GUT spectrum. In minimal gauge mediation, the available parameters which can be varied are the supersymmetry breaking scale $\sqrt{F}$ in the messenger sector, the analogous scale $\sqrt{F_0}$ which generates the mass of the gravitino,\footnote{As noted previously, in the context of F-theory GUTs, it is most natural to set $F = F_0$.} the messenger scale $M_{mess}$, the effective number of messengers in the $5 \oplus \overline{5}$ $N_{5}$, and $\tan \beta$. Here we assume that the mass of the messengers are determined through the supersymmetry breaking vev of a single GUT singlet. The parameters $F$ and $M_{mess}$ fix the characteristic scale $\Lambda = F/M_{mess}$. Parameters such as $\mu$ and $B \mu$ are determined by the condition that proper electroweak symmetry breaking is achieved for a given value of $\tan \beta$. Let us note that a much broader range of sparticle masses can be achieved in more general gauge mediation setups. See for example \cite{GGMI,GGMII} for recent discussion.

As a first step in identifying potential mGMSB look-alikes, we must require that the stau is the NLSP, and that it is quasi-stable on timescales probed by the LHC. Assuming that $\sqrt{F} \sim \sqrt{F_0}$, equation \eqref{eq:DECAY} now implies that $\sqrt{F} \gtrsim 10^{6}$ GeV. In particular, this implies that the ratio $F / M^{2}_{mess} = \Lambda^{2} / F \lesssim 0.01$. Hence, we can reliably use an approximation for the mGMSB candidate look-alike in which all soft masses at the messenger scale are proportional to $\Lambda$. Later in this subsection we shall consider a potential look-alike where we drop this assumption and allow lower messenger scale models, but we will again see that this candidate does not effectively mimic F-theory GUTs.

In quasi-stable mGMSB models with $\sqrt{F}\sim\sqrt{F_{0}}$, the primary
deviation between F-theory GUTs and minimal gauge mediation models comes from
the specific scale of supersymmetry breaking, and the effects of the PQ
deformation on the soft scalar masses. As a first step towards finding
potential look-alikes, we focus on the masses of sparticles where the effects
of the PQ deformation are largely unimportant, in particular the gauginos and
the first and second generation squarks. Assuming that we have measured enough of the gaugino masses, we therefore require:\footnote{Note that in general, there will be some mixing between the neutral $SU(2)_L \times U(1)_Y$ gauginos and the neutral Higgsinos, and the charged gauginos and the charged Higgisinos. This can in principle complicate the identification of the gaugino mass measurement, though in the present class of models, this is a small effect which we shall ignore.}
\begin{equation}
3N_{10}^{(F)}\Lambda^{(F)} \simeq N_{5}^{(m)}\Lambda^{(m)},\label{eq:gauginorel}%
\end{equation}
where the superscript ``$F$'' denotes the F-theory model and ``$m$'' is for the
\textquotedblleft mimic\textquotedblright. Next consider the masses of the
squarks. There is little left-right mixing for the first two generations of
left and right-handed squarks, and so we focus on this case here. As for all
scalar partners in mGMSB models, the masses of the squarks at the messenger
scale are:
\begin{equation}
m_{\widetilde{q}}=\frac{\alpha}{4\pi}C_{\widetilde{q}}\sqrt{N_{5}^{(m)}%
}\Lambda^{(m)},
\end{equation}
where we have again schematically indicated the dependence on the various fine
structure constants and quadratic Casimirs of representations through $\alpha$
and $C_{\widetilde{q}}$. The possible ways that an F-theory GUT could
reproduce the same squark masses but with a different number of messengers and
$\Lambda$ is by turning on $\Delta_{PQ}$, and adjusting the messenger scale.
As we have already seen in section \ref{sec:NNLSP}, the effects of the PQ
deformation are far milder for heavy squarks. In particular, the overall shift
in the mass of the squarks is on the order of $\Delta_{PQ}^{2}/m_{\widetilde
{q}}\sim O(100$ GeV$)^{2}/O(1$ TeV$)$ $\sim O(10)$ GeV, which is a very slight
shift. Varying the messenger scale $M_{mess}$ produces logarithmic
contribution to the mass-squared, and in minimal gauge mediation models, only
shifts the squark masses by an effect which is at most on the order of
$O(100)$ GeV. This yields the additional constraint:
\begin{equation}
\frac{\alpha}{4\pi}C_{\widetilde{q}}\sqrt{3N_{10}^{(F)}}\Lambda^{(F)}%
=\frac{\alpha}{4\pi}C_{\widetilde{q}}\sqrt{N_{5}^{(m)}}\Lambda^{(m)}%
+O(100)\;\text{GeV.}%
\end{equation}
Since the squark masses are on the order of a TeV, we conclude that the
messenger number and $\Lambda$ must agree:
\begin{equation}
3N_{10}^{(F)}=N_{5}^{(m)},\;\;\;\;\Lambda^{(F)}\sim\Lambda^{(m)}.
\end{equation}

Fixing the parameters $N_{5}$ and $\Lambda$ to agree with the F-theory GUT
model allows it to mimic the masses of the gauginos and the first and second
generation squarks. We now discuss whether it is possible to adjust the values
of the remaining parameters $M_{mess}$ and $\tan\beta$ to exactly match the
rest of the spectrum. In the class of models we are considering, the effects
of $\tan\beta$ affect the third generation of squarks and sleptons by inducing
more mixing between left and right-handed sparticles. To isolate the effects
of $M_{mess}$, we therefore focus on the masses of the left and right-handed
sleptons $\widetilde{e}_{L}$ and $\widetilde{e}_{R}$. Our strategy will be to
compare the needed Messenger scale for the two soft mass relations in the
presence of the PQ deformation. We find that at most one of these masses can
be matched by adjusting $M_{mess}^{(m)}$.

\begin{table}[t]
\caption{Mass spectra of F-theory benchmark model ${\rm Maj}_{\rm MID}^{(1)}$ and
some candidate mGMSB look-alike models with $N_{5} = 3$ messengers. All masses are in GeV, and we have taken
the top mass to be $172.4$ GeV in our calculation.}
\begin{center}
\begin{tabular}{ c c c c }
\hline parameter\rule{0pt}{3.0ex}\rule[-1.5ex]{0pt}{0pt} &
${\rm Maj}_{\rm MID}^{(1)}$ & mGMSB1 & mGMSB2 \\ \hline
$M_{mess}$ \rule{0pt}{3.0ex} & $10^{12}$ & $10^{12}$ & $2\times 10^{9}$ \\
$\sqrt{F}$ \rule{0pt}{3.0ex} & $2.2\times 10^{8}$ & $2.2\times 10^{8}$ & $10^{7}$ \\
$\tan\beta$ & 24.05 & 34.7 & 24.5 \\\hline
$m_{\tilde g}$                   & 1112   &  1113   & 1116  \\
$m_{\widetilde \chi_1^0}$ & 198.6  & 199.0 & 199.3  \\
$m_{\widetilde \chi_2^0}$ & 377.1  & 379.4 & 378.0 \\
$m_{\widetilde \chi_1^{\pm}}$ & 380.3 & 382.3 & 381.2 \\
$m_{\tilde u_L}$ & 1106 & 1112 & 1102 \\
$m_{\tilde u_R}$ & 1059 & 1066 & 1063 \\
$m_{\tilde t_1}$ & 857.6 & 866.7  &  898.1 \\
$m_{\tilde t_2}$ & 1050 &  1047 &  1059 \\
$m_{\tilde b_1}$ & 997.2 & 982.2 & 1014 \\
$m_{\tilde b_2}$ & 1032 &  1032 &  1046 \\
$m_{\tilde e_L,\tilde \mu_L}$ & 383.0 & 421.7 &  382.2 \\
$m_{\tilde \nu_{e},\tilde \nu_{\mu}}$ & 372.5 & 412.1 & 371.6 \\
$m_{\tilde e_R,\tilde \mu_R}$ & 214.3 & 246.9 & 204.9 \\
$m_{\tilde \tau_1}$ & 175.0 & 174.8 & 174.7 \\
$m_{\tilde \nu_{\tau}}$ & 366.1 & 400.4 & 367.7 \\
$m_{\tilde \tau_2}$ & 384.0 & 422.3 & 385.1 \\
$m_h$ & $114.3$ & 114.3 & 113.8 \\
$m_A$ & $693.1$ & $614.2$ & $623.4$\\
\hline
\end{tabular}%
\end{center}
\label{tab:comp-Majmid}
\end{table}%

To arrive at this result, we now study in more precise terms the masses of the
left and right-handed selectrons as a function of the messenger scale. Letting
$\mathcal{M}^{2}(t)$ denote the soft masses of a minimal gauge mediation
model with $N_{5}$ effective messengers and characteristic scale $\Lambda$ evaluated
at a renormalization group (RG) time $t=\log(M_{mess}^{2}/Q^{2})$, the condition
for the selectrons of the F-theory GUT and look-alike to have the same masses is:
\begin{equation}
\mathcal{M}_{\widetilde{l}}^{2}(t^{(m)}) = \mathcal{M}_{\widetilde{l}}^{2}(t^{(F)}) + q_{\widetilde{l}} \cdot \Delta_{PQ}^{2},
\end{equation}
where $t^{(F)}$ and $t^{(m)}$ respectively correspond to the RG time evolution from the messenger
scale to the weak scale for the F-theory GUT and the mimic. Here we are implicitly using the
fact that the Yukawas of the first two generations are small, so that
the PQ deformation can be treated as an additive shift for $\widetilde{l} = \widetilde{e}_{R}$
and $\widetilde{e}_{L}$. Eliminating $\Delta_{PQ}$ from this system of equations,
we therefore find the condition:
\begin{equation}
\frac{q_{\widetilde{e}_{L}}}{q_{\widetilde{e}_{R}}} = \frac{\mathcal{M}_{\widetilde{e}_{L}}^{2}(t^{(F)}) - \mathcal{M}_{\widetilde{e}_{L}}^{2}(t^{(m)})}
{\mathcal{M}_{\widetilde{e}_{R}}^{2}(t^{(F)}) - \mathcal{M}_{\widetilde{e}_{R}}^{2}(t^{(m)})}.
\end{equation}
In Appendix B we compute the leading order behavior of the righthand side of this equation, the end result of which is:
\begin{equation}
\frac{q_{\widetilde{e}_{L}}}{q_{\widetilde{e}_{R}}} \simeq\frac{1+N_{5}}{1+5N_{5}/33}\times\frac{9}{220}\times\left(
\frac{\alpha_{2}}{\alpha_{1}}\right)  ^{3}.
\end{equation}
In our conventions the weak scale values of the couplings $\alpha_i$ are $\alpha_1 \sim 1/100$ and $\alpha_{2} \sim 1/30$ which implies:
\begin{equation}
\frac{q_{\widetilde{e}_{L}}}{q_{\widetilde{e}_{R}}}\sim 1.5\times\frac{1+N_{5}}{1+5N_{5}/33},
\end{equation}
which for $N_{5}=3$ and $6$ yields the condition:%
\begin{align}
N_{5}  & =3\Longrightarrow\frac{q_{\widetilde{e}_{L}}}{q_{\widetilde{e}_{R}}%
}\sim 4,\\
N_{5}  & =6\Longrightarrow\frac{q_{\widetilde{e}_{L}}}{q_{\widetilde{e}_{R}}%
}\sim 6. \label{eq:mismatch}
\end{align}
In other words, since the actual ratio of $q$'s is $2$ for Majorana scenarios, and
$1$ for Dirac scenarios, a candidate look-alike cannot be found where the
masses exactly agree. For example, in an F-theory GUT Majorana scenario with
$N_{10} = 1$, this would lead to a mismatch by a factor of $2$ in the $q$'s.

In principle, uncertainties in the mass measurement could still allow a
candidate look-alike to remain compatible with an F-theory GUT, up to
experimental error. In the present context, we can view this as reflecting an
uncertainty in determining the ratio of the $q$'s. Since the mimics of $N_{10} = 1$
Majorana scenarios lead to a smaller deviation from the actual ratio of the $q$'s,
we focus on this case. Rescaling $q$ by the factor of $2$ discussed below equation \eqref{eq:mismatch}
translates to a shift in the mass squared of order $2 \times
\Delta_{PQ}^{2}$. We now ask whether this type of shift could be explained
by an uncertainty in the mass measurement. Using the relation:%
\begin{equation}
m^{actual} \simeq m^{observed}\pm\Delta m
\end{equation}
and squaring, we obtain:%
\begin{equation}
\left(  m^{actual}\right)  ^{2}\simeq\left(  m^{observed}\right)  ^{2}%
\pm2\times m\times\Delta m.
\end{equation}
In other words F-theory GUTs which can be mimicked have $\Delta_{PQ}$ which satisfies the upper bound:%
\begin{equation}
2 \times\Delta_{PQ}^{2}\lesssim2\times m\times\Delta m.
\end{equation}
In typical F-theory GUTs, the masses of the right-handed sleptons satisfy $m\sim300$ GeV. Combined with
the uncertainty in the mass measurement $\Delta m\lesssim2$ GeV reviewed in
the previous subsection, we find:
\begin{equation}
\Delta_{PQ}\lesssim30\text{ GeV},
\end{equation}
which should be viewed as an order of magnitude estimate. In other words, the
high precision of the mass measurement translates into a tight upper bound on
the size of the PQ deformation which can be mimicked.

\begin{table}[t]
\caption{Mass spectra of F-theory benchmark model ${\rm Maj}_{\rm HI}^{(1)}$ and some candidate mGMSB look-alike models with $N_{5} = 3$ messengers.
All masses are in GeV, and we have taken the top mass to be $172.4$ GeV in our calculation.}
\begin{center}
\begin{tabular}{cccc}
\hline parameter\rule{0pt}{3.0ex}\rule[-1.5ex]{0pt}{0pt} &
${\rm Maj}_{\rm HI}^{(1)}$ & mGMSB3 & mGMSB4 \\ \hline
$M_{mess}$ \rule{0pt}{3.0ex} & $10^{12}$ & $10^{8}$ & $3\times 10^{5}$ \\
$\sqrt{F}$ \rule{0pt}{3.0ex} & $2.2\times 10^{8}$ & $2.2\times 10^{6}$ & $1.22\times 10^{5}$ \\
$\tan\beta$ & 24.9 & 40.2 & 40.7 \\\hline
$m_{\tilde g}$                   & 1110   &  1118   & 1128  \\
$m_{\widetilde \chi_1^0}$ & 198.3  & 199.7 & 200.0  \\
$m_{\widetilde \chi_2^0}$ & 373.8  & 376.1 & 350.9 \\
$m_{\widetilde \chi_1^{\pm}}$ & 377.5 & 378.7 & 350.3 \\
$m_{\tilde u_L}$ & 1098 & 1098 & 1091 \\
$m_{\tilde u_R}$ & 1050 & 1062 & 1062 \\
$m_{\tilde t_1}$ & 846.1 & 916.3  &  968.4 \\
$m_{\tilde t_2}$ & 1041 &  1049 &  1062\\
$m_{\tilde b_1}$ & 979.8 & 986.3 & 1008 \\
$m_{\tilde b_2}$ & 1019 &  1037 & 1052 \\
$m_{\tilde e_L,\tilde \mu_L}$ & 329.7 & 363.7 &  330.0 \\
$m_{\tilde \nu_{e},\tilde \nu_{\mu}}$ & 317.4 & 352.9 & 318.4 \\
$m_{\tilde e_R,\tilde \mu_R}$ & 166.6 & 188.9 & 163.3 \\
$m_{\tilde \tau_1}$ & 104.6 & 104.5 & 104.7 \\
$m_{\tilde \nu_{\tau}}$ & 309.5 & 345.2 & 314.5 \\
$m_{\tilde \tau_2}$ & 331.5 & 373.2 & 340.5 \\
$m_h$ & $114.3$ & 113.3 & 112.8 \\
$m_A$ & $701.8$ & $497.5$ & $395.9$\\
\hline
\end{tabular}%
\end{center}
\label{tab:comp-Majhi}
\end{table}%

To further support this general analysis, we now consider in greater detail
potential look-alikes for the benchmark model $\mathrm{Maj}_{\mathrm{MID}}^{(1)}$. As a first
example, consider mGMSB models with $M_{mess} = 10^{12}$ GeV, $\Lambda= 50$
TeV, and $N_{5} = 3$. The only remaining parameter of an mGMSB model which we
can vary is $\tan\beta$, which changes the amount of left-right mixing for staus, allowing the lightest
stau to become degenerate in mass with the F-theory GUT value. By increasing $\tan\beta$ to $34.7$, one can achieve
a stau NLSP with mass very close to that of the F-theory GUT. The mass
spectrum of this model (mGMSB1) is shown in Table \ref{tab:comp-Majmid}.
Though the gaugino and squark masses are within a few GeV, the slepton masses
in mGMSB1 deviate from the $\mathrm{Maj}_{\mathrm{MID}}^{(1)}$ values by as
much as $\sim30$~GeV. Returning to equation \eqref{eq:mass-resolution}, we
conclude that this difference is large enough to distinguish these two models
with sufficient LHC data.

Next consider variations of the messenger scale $M_{mess}$. Table
\ref{tab:comp-Majmid} displays the mass spectrum of an mGMSB model (mGMSB2)
where the model parameters are the same as the benchmark $\mathrm{Maj}%
_{\mathrm{MID}}^{(1)}$ except $M_{mess}=2\times10^{9}$~GeV and $\tan
\beta=24.5$. One can see that in addition to matching the gauginos and first
and second generation squark masses, the left-handed slepton masses are also
very close to the $\mathrm{Maj}_{\mathrm{MID}}^{(1)}$ model. The notable
differences are the right-handed slepton mass and the third-generation
squarks, as well as the Higgs masses. The difference in the third-generation
squark masses is mainly due to the lower messenger scale which shortens the contribution
from renormalization group flow effects.

For F-theory models with larger $\Delta_{PQ}$, the deformation effects are
larger. As we have seen, this makes it more difficult to find an mGMSB
look-alike. As an example we consider the benchmark model $\mathrm{Maj}%
_{\mathrm{HI}}^{(1)}$ where $N_{10}=1$, $\Lambda=50$ TeV and
$\Delta_{PQ}=209$~GeV. Even when the messenger scale is as low as $\sim10^{8}$
GeV in the model mGMSB3, table \ref{tab:comp-Majhi} shows that the slepton
masses typically differ by as much as $\sim35$~GeV, which can be resolved with
the mass measurement procedure detailed above. For even lower messenger scale,
the stau NLSP in mGMSB model would decay inside the detector, so long as we
assume $F_{0}=F$.

Even when we allow the possibility that $F_{0} \neq F$ so that the stau decays
outside the detector, we find that representative low scale messenger models
do not correctly mimic the spectrum of the Maj$_{HI}^{(1)}$ scenario. An example of this type
is shown as the model mGMSB4 in table \ref{tab:comp-Majhi}. Let us note that when $F/M_{mess}^{2} \sim O(1)$,
the theoretical analysis presented earlier in this subsection and in Appendix C is no longer accurate.
Table \ref{tab:comp-Majhi} shows that even though the slepton masses can be quite
degenerate with the above mentioned F-theory model, the masses of
$\widetilde{\chi}_{2}^{0}$ and $\widetilde{\chi}_{1}^{\pm}$ would be $\sim
20$~GeV lighter than the corresponding masses in the F-theory model. The
reason for this is mainly due to the lowered messenger scale where the running
of $m_{H_{u}}^{2}$ is reduced and therefore become less negative. From the
relation
\begin{equation}
|\mu|^{2} \approx- m_{H_{u}}^{2} - \frac{1}{2} m_{Z}^{2},
\end{equation}
we can see the $\mu$ term decreases in this situation, and therefore the
higgsino masses become closer to the gaugino masses. The increased mixing
between gauginos and higgsinos decreases the physical masses of $\widetilde
{\chi}_{2}^{0}$ and $\widetilde{\chi}_{1}^{\pm}$. This mismatch in the
neutralino and chargino masses is by itself enough to resolve the potential degeneracy.

Our numerical analysis has mainly been based on two benchmark models with $N_{10}=1$.
The basic idea is similar for other models, as well as for $N_{10}=2$ scenarios.
Though beyond the scope of the present paper, it would be interesting to perform a more
complete scan over other well-motivated look-alikes.

\newpage

\section{Conclusions}\label{sec:CONCLUDE}

In this paper we have seen that over a broad range of parameter space in minimal F-theory GUTs, a quasi-stable
stau is the NLSP. The consequences for collider physics are quite striking, since in stau NLSP scenarios, this particle will appear as a massive quasi-stable charged particle at the LHC. We have seen that with the center-of-mass energies and integrated luminosity expected at the LHC, signatures of this scenario can be discovered. We have also seen how the corresponding number of stau events correlates
with the number of leptonic final states as a function of F-theory GUT parameters. Focussing on inclusive one and two stau events, it is
possible to constrain the characteristic scale of gauge mediation $\Lambda$, and using more exclusive signatures such as Drell-Yan production of
staus, it is possible to constrain the size of the stringy PQ deformation parameter present in F-theory GUTs. Slow-moving staus can also become stuck in or near the detector, and we have analyzed in an idealized situation the number of such events to expect. Finally, we have
used the high precision expected in mass measurements in stau scenarios as a way to compare the spectra of
F-theory GUTs with minimal gauge mediation models.

In the remainder of this section, we briefly speculate on some possible avenues of further investigation. In this work we have focussed on the potential for discovering quasi-stable stau NLSP F-theory GUT scenarios. Quasi-stable stau NLSPs are also a feature of other models with an MSSM-like spectrum, and we have seen that there are some measurable differences between minimal gauge mediation models and F-theory GUTs. Broadening the range of possible models both to more general gauge mediation models, and more general scenarios with a stau NLSP, it would be interesting to further study the level at which F-theory GUTs can be distinguished from other well-motivated possibilities.

We have seen that in the context of F-theory GUTs, staus can be become stuck in or near the detector. This holds out the exciting prospect of measuring the scale of supersymmetry breaking. It would be interesting to further investigate with more realistic simulations how to use the ATLAS and CMS detectors to detect stopped staus, and measure their lifetimes.

\section*{Acknowledgements}

We thank B.S. Acharya, J. Hubisz, G.L. Kane, M. Papucci, M. Perelstein, A. Pierce, G. Polesello,
A. Rizzi, A. De Roeck and M. Spiropulu for helpful discussions. We also thank
members of the ATLAS and CMS groups for comments. JJH and CV thank the CERN theory group for hospitality
during part of this work. JJH thanks the Harvard
high energy theory group for hospitality during part of this work. The work of JJH
is supported by NSF grant PHY-0503584. The work of JS is supported
by Syracuse University College of Arts and Sciences. The work of CV is
supported by NSF grant PHY-0244281.

\appendix

\section*{Appendix A: Simulation Details}

We simulate SUSY events using \texttt{PYTHIA} \cite{PYTHIA}. The analysis is done based on the
parton-level objects
\begin{itemize}
  \item  lepton:  $p_T > 5$~GeV and $|\eta|< 2.5$ and the transverse
energy deposited in a cone $\Delta R$= 0.4 around the lepton is less than $5$~GeV.
  Note: staus which have large velocity $\beta> 0.91$ are also treated as muons.
  \item tau jet: $p_T>20$~GeV and $|\eta|< 3.0$, with efficiency $\epsilon_\tau=0.5$.  \\
  leptonic decaying taus are identified as leptons based on the lepton cuts.
  \item  jet: $p_T>20$~GeV and $|\eta|< 3.0$.

\end{itemize}
We shall often augment these basic cuts by additional analysis cuts.

We include the detector resolution of the momentum and energy for
muon, lepton and jet, which are parameterized as follows in the ATLAS detector \cite{Aad:2009wy}:
\begin{itemize}
\item For muon the momentum resolution is:
\begin{equation}
\frac{\sigma_{p_T}}{p_T}= \frac{a p_T} {\rm TeV} \oplus \frac{b}{\sqrt{\sin\theta}}
\end{equation}
Here we take $a=36\%$ and $b=1.3\%$,

\item For electron the energy resolution is:
\begin{equation}
\frac{\sigma_{E}}{E}= \frac{a} {\sqrt{E/{\rm GeV}}} \oplus b
\end{equation}
Here we take $a=10\%$ and $b=0.4\%$,

\item For jet, the energy resolution is similar to electron, but with  $a=80\%$ and $b=15\%$.

\end{itemize}

\section*{Appendix B: Selectron Mass Comparison}

In this Appendix we compare in more precise terms the masses of the
left and right-handed selectrons in F-theory GUTs and minimal gauge mediated supersymmetry breaking.
In minimal gauge mediation, the first and second generation
soft masses squared at a renormalization group (RG) time $t=\log(M_{mess}^{2}/Q^{2})$ are
given by (see for example \cite{GiudiceSUSYReview}):%
\begin{equation}
\mathcal{M}^{2}(t)=2\underset{r=1}{\overset{3}{%
{\displaystyle\sum}
}}C_{r}k_{r}\left(  \frac{\alpha_{r}(t)}{4\pi}\right)  ^{2}N_{5}\Lambda^{2}\left[
\left[  1-\frac{\alpha_{r}(t)}{4\pi}b_{r}t\right]  ^{-2}+\frac{N_{5}k_{r}%
}{b_{r}}\left[  \left[  1-\frac{\alpha_{r}(t)}{4\pi}b_{r}t\right]
^{-2}-1\right]  \right]  ,
\end{equation}
where we have dropped running effects due to the Yukawa couplings, which are small for
the first two generations. In the above, $r$ is an index labelling the three gauge groups, $C_{r}$ is the
quadratic Casimir associated to the representation of the scalar, $k_{1}=5/3$,
$k_{2}= k_3 = 1$ so that $\alpha_{i}(t)k_{i}$ unify at the GUT scale, and in this
normalization $b_{r}=11,1,-3$ denote the beta functions for $r=1,2,3$,
respectively. Further, the renormalized fine structure constant satisfies:%
\begin{equation}
\alpha_{r}(0)=\alpha_{r}(t)\left[  1-\frac{\alpha_{r}(t)}{4\pi}b_{r}t\right]
^{-1}.
\end{equation}
Here, $t=0$ corresponds to the Messenger scale, and we shall be interested in
evaluating both the F-theory GUT and candidate look-alike at the weak scale.
Because of this, we can replace each $\alpha_{r}(t)$ by its weak scale value,
which we denote by $\alpha_{r}$. In this case, all of the difference from
RG\ flow effects is captured by the terms in brackets
containing $\alpha_{r}b_{r}t/4 \pi$. Letting $t^{(F)}$ and $t^{(m)}$ denote the
respective values of the RG\ time for the F-theory GUT and the candidate
look-alike, it follows that the right-handed selectrons and sleptons of a
candidate mimic must satisfy the relations:%
\begin{equation}
q_{\widetilde{l}}\Delta_{PQ}^{2}=2\underset{r=1}{\overset{3}{%
{\displaystyle\sum}
}}C_{r}^{(\widetilde{l})}k_{r}\left(  \frac{\alpha_{r}}{4\pi}\right)
^{2}N_{5}\left(  1+\frac{N_{5}k_{r}}{b_{r}}\right)  \Lambda^{2}\left[  \left[
1-\frac{\alpha_{r}}{4\pi}b_{r}t^{(m)}\right]  ^{-2}-\left[  1-\frac{\alpha
_{r}}{4\pi}b_{r}t^{(F)}\right]  ^{-2}\right]  .
\end{equation}
For the right-handed selectrons, the sum on $r$ truncates to the term $r=1$,
and for the left-handed selectrons, the $r=2$ term dominates. Taking the ratio
of these two contributions, we find:%
\begin{align}
\frac{q_{\widetilde{e}_{L}}}{q_{\widetilde{e}_{R}}}  & =\frac{1+N_{5}%
}{1+5N_{5}/33}\times\frac{k_{2}C_{2}^{(\widetilde{e}_{L})}}{k_{1}C_{1}^{(\widetilde
{e}_{R})}}\times\left(  \frac{\alpha_{2}}{\alpha_{1}}\right)  ^{2}%
\frac{\left[  \left[  1-\frac{\alpha_{2}}{4\pi}b_{2}t^{(m)}\right]
^{-2}-\left[  1-\frac{\alpha_{2}}{4\pi}b_{2}t^{(F)}\right]  ^{-2}\right]
}{\left[  \left[  1-\frac{\alpha_{1}}{4\pi}b_{1}t^{(m)}\right]  ^{-2}-\left[
1-\frac{\alpha_{1}}{4\pi}b_{1}t^{(F)}\right]  ^{-2}\right]  }\\
& \simeq\frac{1+N_{5}}{1+5N_{5}/33}\times\frac{C_{2}^{(\widetilde{e}_{L})}%
}{C_{1}^{(\widetilde{e}_{R})}}\times\frac{k_{2}b_{2}}{k_{1}b_{1}}\times\left(
\frac{\alpha_{2}}{\alpha_{1}}\right)  ^{3}\\
& = \frac{1+N_{5}}{1+5N_{5}/33}\times\frac{9}{220}\times\left(
\frac{\alpha_{2}}{\alpha_{1}}\right)  ^{3},
\end{align}
where in the second to last equality we have kept only the leading logarithms. Note that
this final expression is independent of the logarithmic dependence on the
messenger scale, as well as $\Lambda$.

\newpage

\section*{Appendix C: Miscellaneous Plots \label{MISCPLOT}}

In this Appendix we collect additional plots of potential interest.

\begin{figure}[h]
\begin{center}
\includegraphics[
height=3.87in,
width=6.9246in
]%
{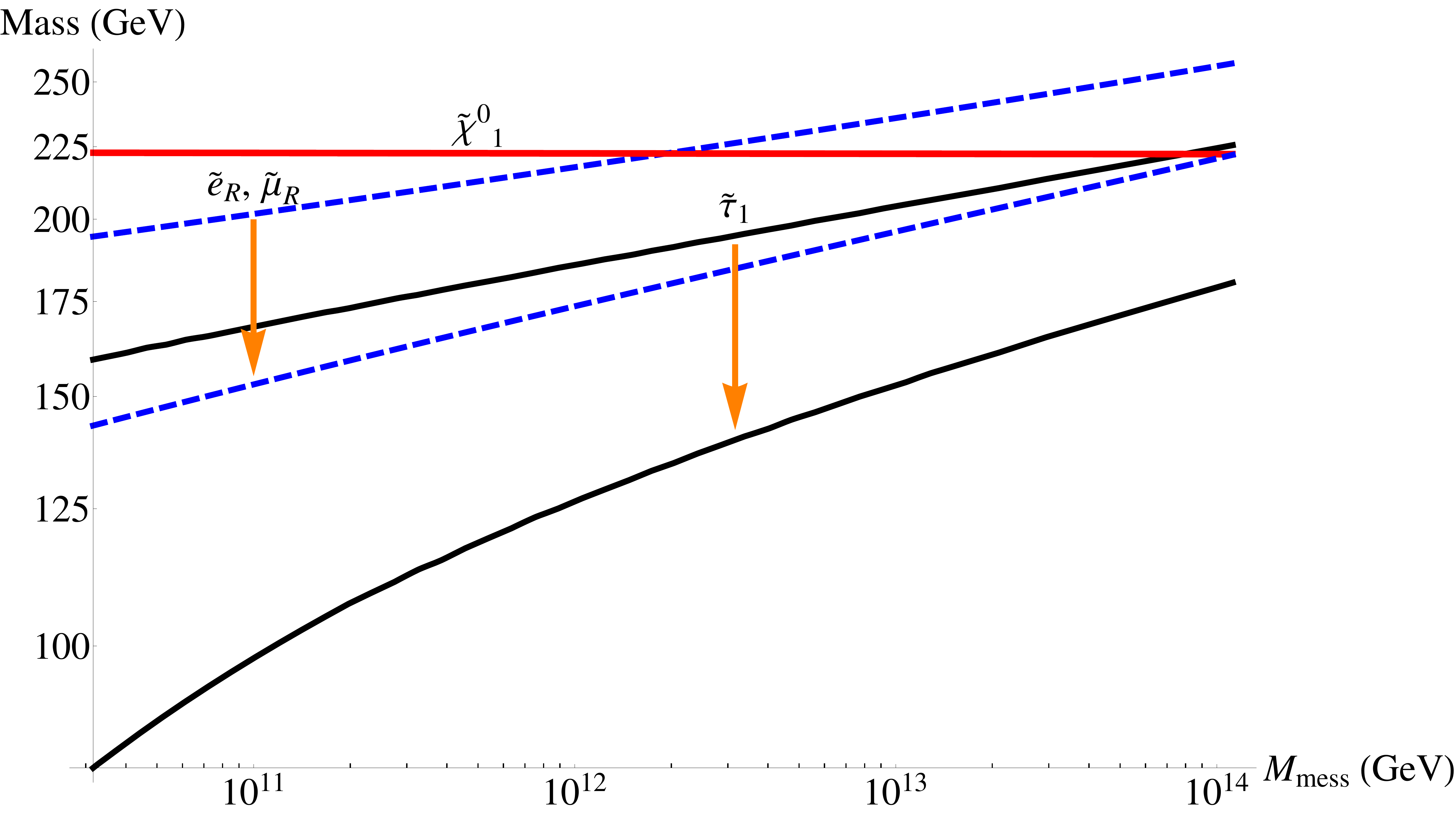}%
\caption{Plot of the four lightest sparticle masses as a function of messenger
scale $M_{mess}$. Here $N_{10}=2$, $\Lambda=28$ TeV, and we have
displayed the values for zero PQ deformation\ (upper lines) and $\Delta
_{PQ}=150$ GeV (lower lines).\ Note that the bino mass remains constant, both
as a function of messenger scale and of PQ deformation.}%
\label{messtwotenscan}%
\end{center}
\end{figure}

\begin{figure}
[ptb]
\begin{center}
\includegraphics[
height=3.8692in,
width=6.0035in
]%
{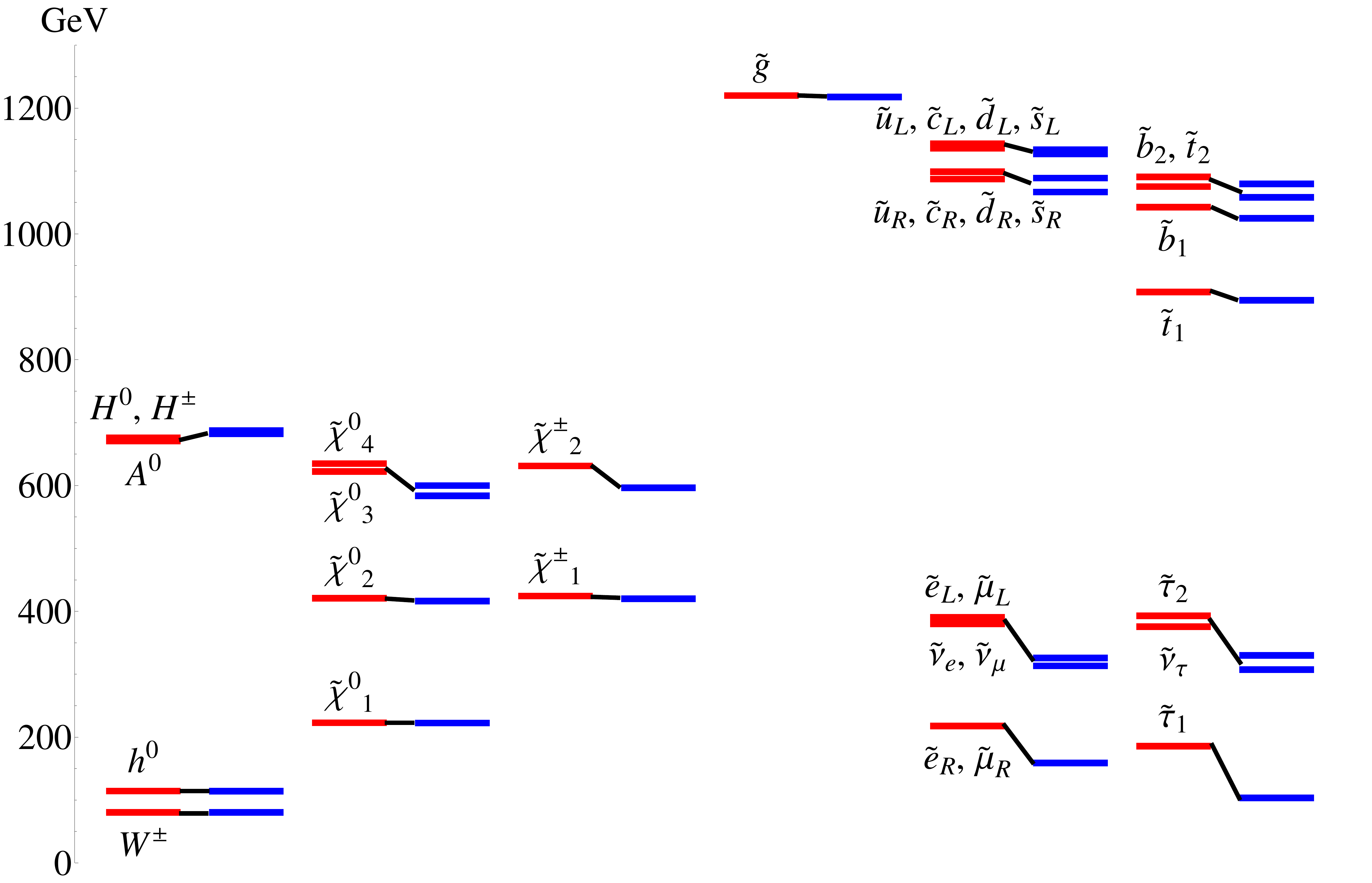}%
\caption{Spectrum of an F-theory GUT Majorana neutrino scenario with
$N_{10}=2$, $\Lambda=28$ TeV for minimal (red left columns) and
maximal (blue right columns) PQ deformation. As opposed to the case of
$N_{10}=1$ models, for larger numbers of messengers the stau is always the
NLSP\ for F-theory GUTs.}%
\label{n10eq2together}%
\end{center}
\end{figure}

\begin{figure}
[ptb]
\begin{center}
\includegraphics[
height=3.87in,
width=6.6106in
]%
{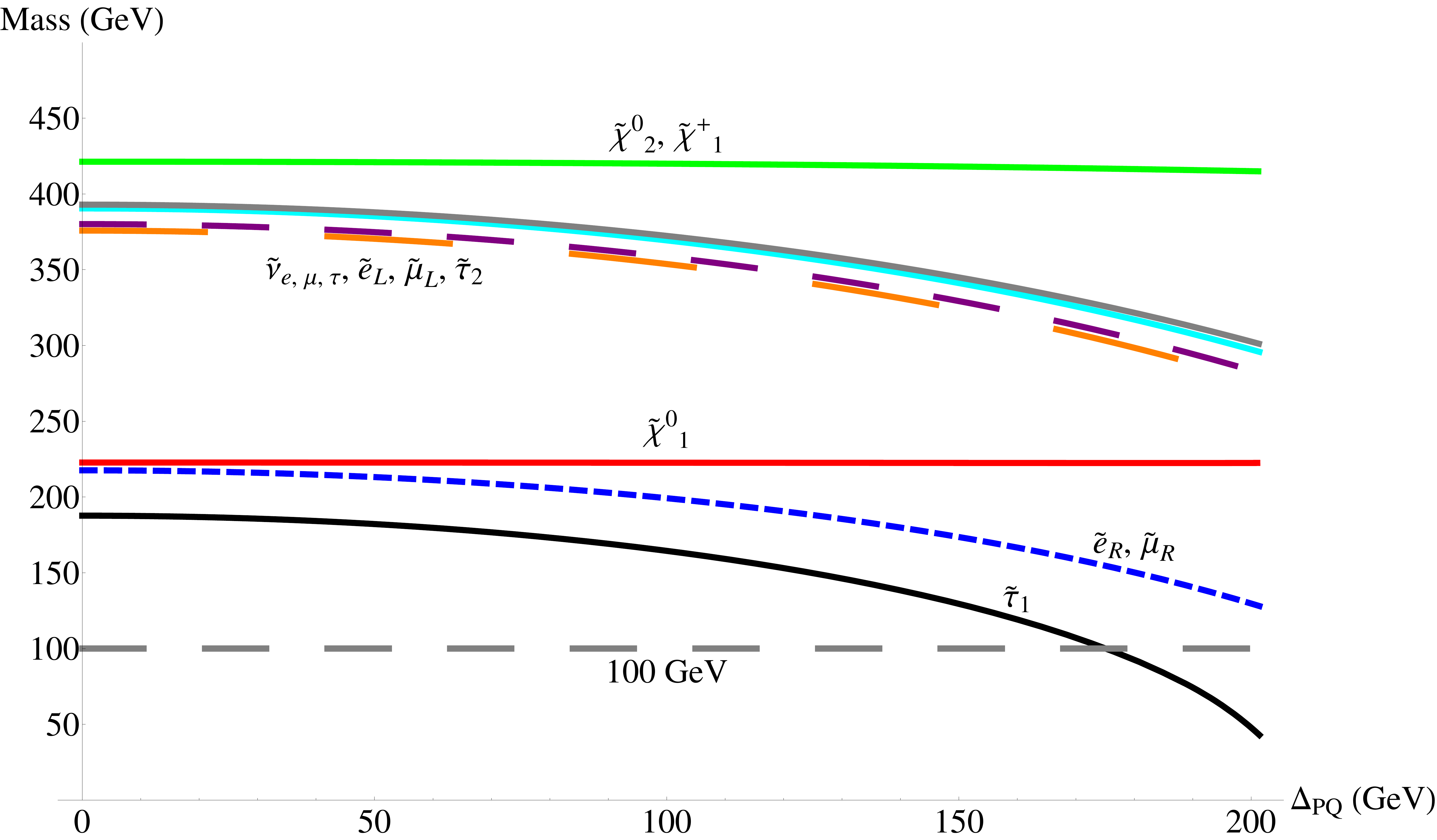}%
\caption{Mass spectrum of the sleptons, lightest chargino and two lightest neutralinos with
$N_{10}=2$, $\Lambda=28$ TeV as a function of $\Delta_{PQ}$. The
dashed grey line at $100$ GeV denotes the present experimental bound on the mass of a quasi-stable
stau.}%
\label{imprvmasslowtwotenscan}%
\end{center}
\end{figure}

\begin{figure}
[ptb]
\begin{center}
\includegraphics[
height=3.87in,
width=6.7187in
]%
{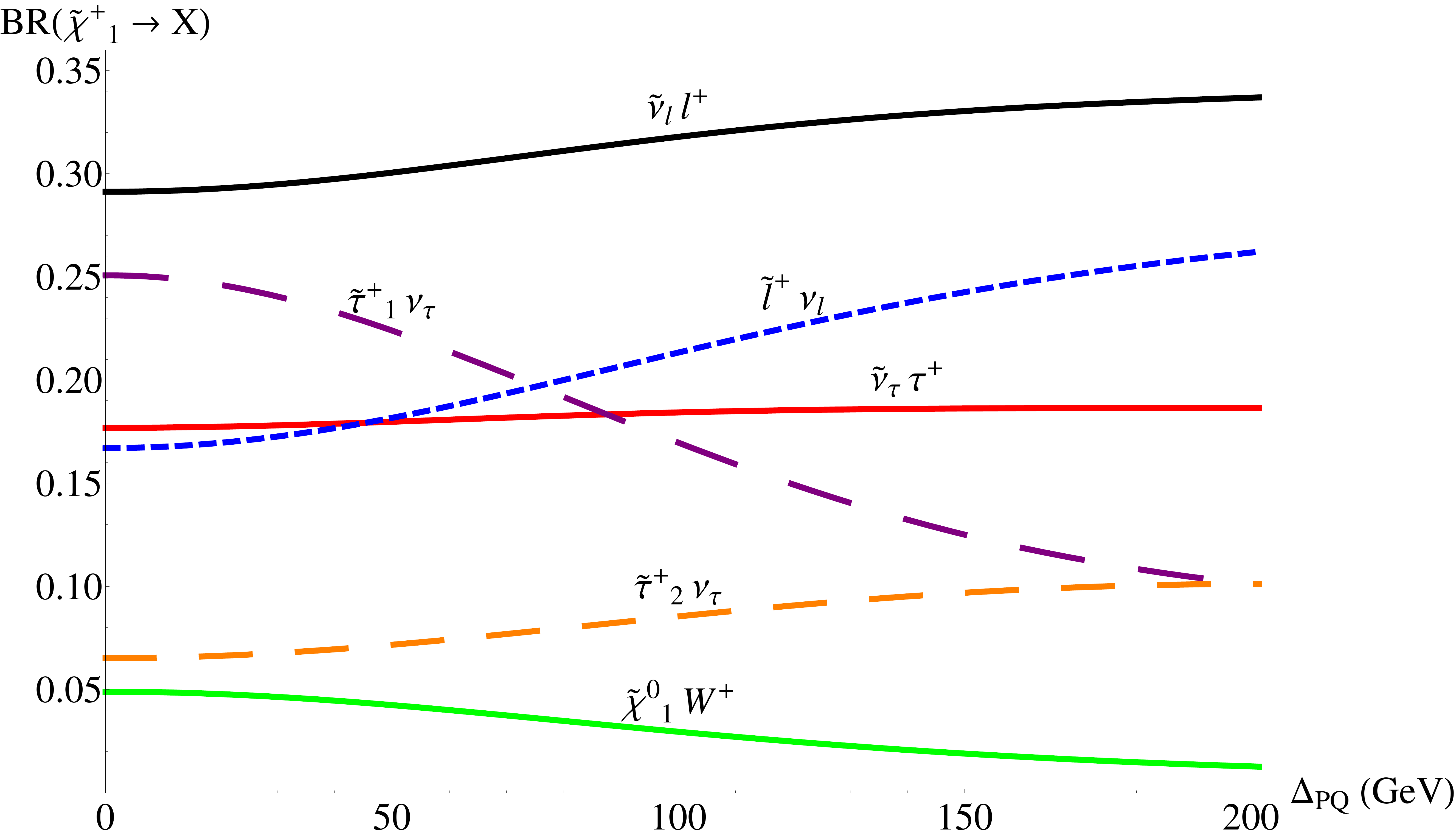}%
\caption{Dominant branching fractions of $\widetilde{\chi}_{1}^{\pm}$ as a
function of $\Delta_{PQ}$ with $N_{10}=2$ and $\Lambda=28$ TeV.
As opposed to the case $N_{10}=1$, the overall $\Delta_{PQ}$ dependence is
smaller. In addition, the decay channel to sneutrinos is available even for
$\Delta_{PQ}=0$.}%
\label{imprvbranchlowtwotenchp1}%
\end{center}
\end{figure}

\begin{figure}
[ptb]
\begin{center}
\includegraphics[
height=3.8692in,
width=6.6911in
]%
{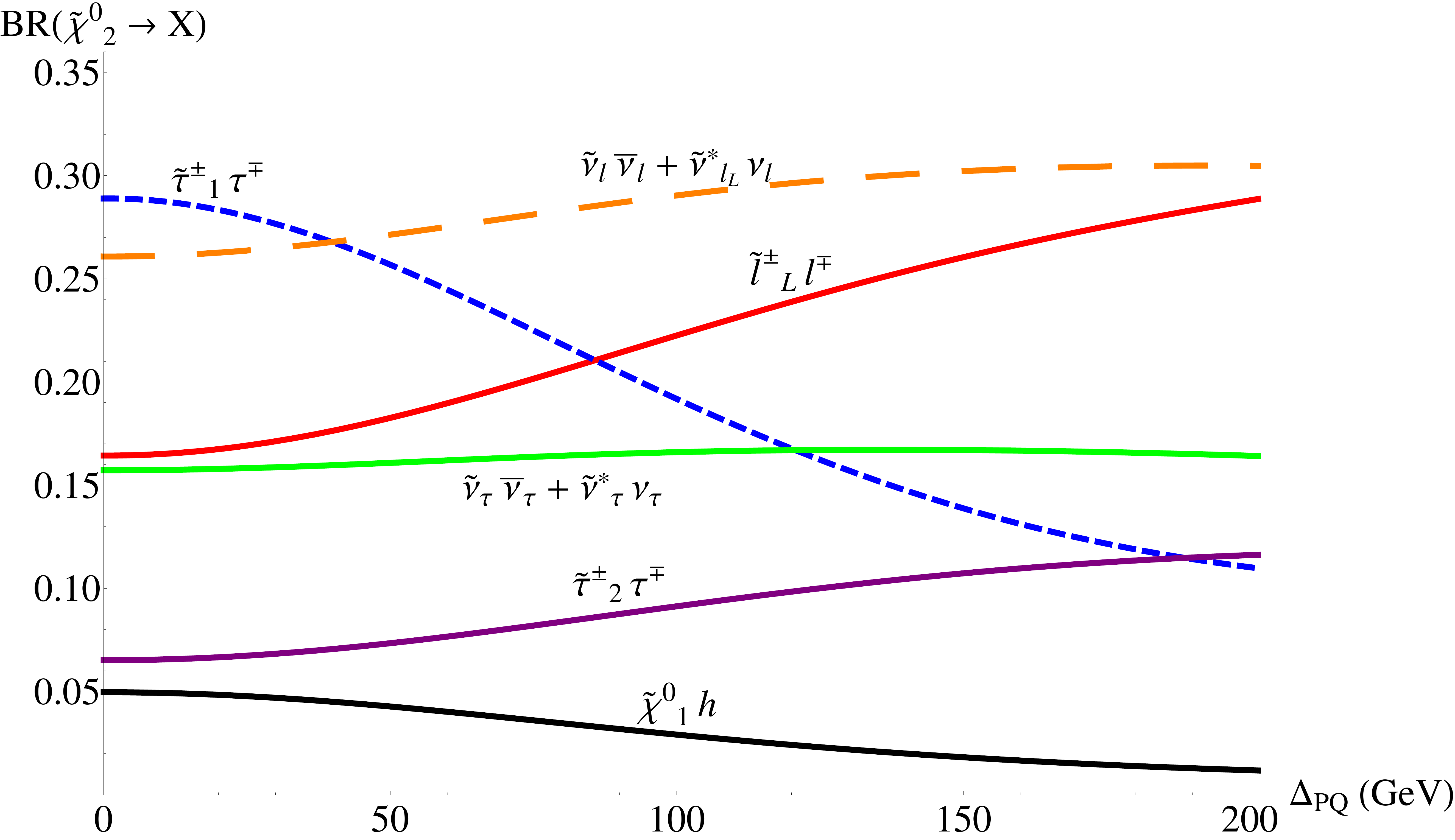}%
\caption{Dominant branching fractions of $\widetilde{\chi}_{2}^{0}$ as a
function of $\Delta_{PQ}$ with $N_{10}=2$ and $\Lambda=28$ TeV.
As opposed to the case $N_{10}=1$, the overall $\Delta_{PQ}$ dependence is
smaller. In addition, the decay channel to sneutrinos is available even for
$\Delta_{PQ}=0$.}%
\label{imprvbranchlowtwotenzchi2}%
\end{center}
\end{figure}

\begin{figure}[htbp] 
   \centering
   \includegraphics[width=5.5in]{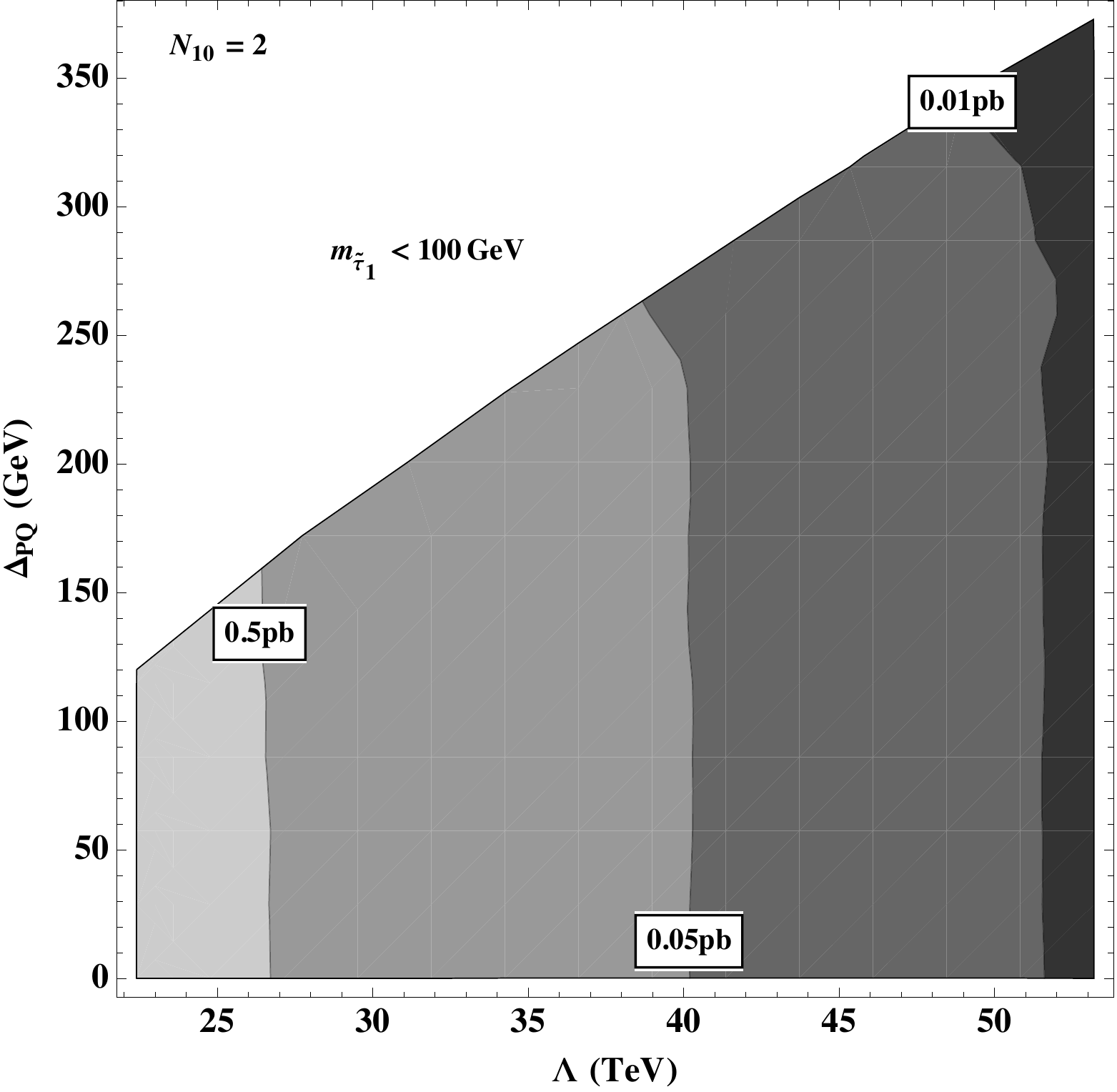}
   \caption{Contour plot of the inclusive $\geq 1 {\tilde \tau}_{1}$ signal in parameter space $\Lambda$ and $\Delta_{PQ}$ for $\sqrt{s}=14$~TeV and $N_{10} = 2$ messengers. The three contours correspond to the value of the cross section at $0.5$ pb, $0.05$ pb, and $0.01$ pb. To detect five inclusive one or two stau events, this amounts to an integrated luminosity of respectively $10$ pb$^{-1}$, $100$ pb$^{-1}$ and $0.5$ fb$^{-1}$.}
   \label{fig:dispotlTWO}
\end{figure}

\begin{figure}[htbp] 
   \centering
   \includegraphics[width=5in]{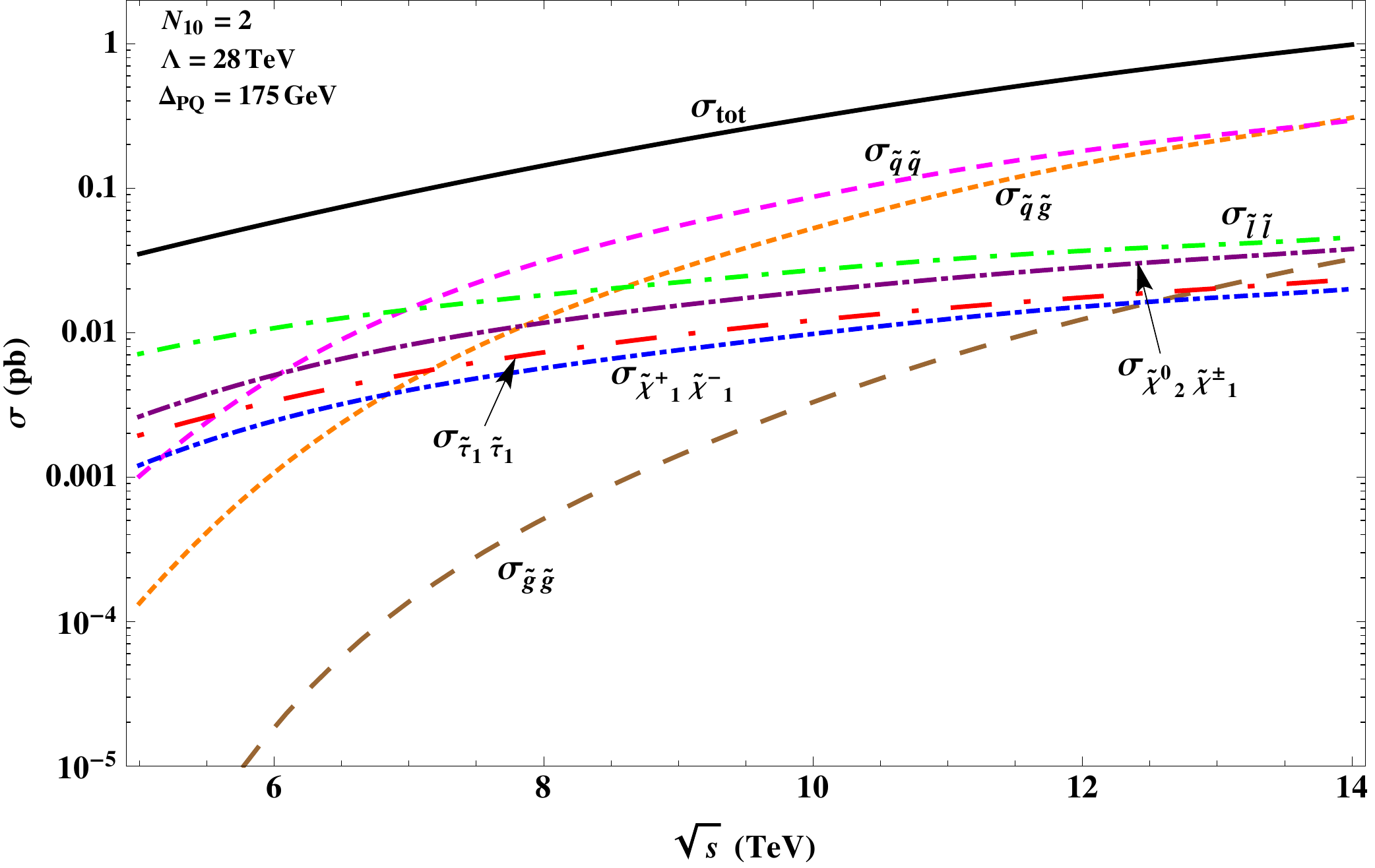}
   \caption{Plot of the leading order cross sections of various production channels as functions of center-of-mass energy $\sqrt{s}$ for the benchmark $\rm{{Maj}_{HI}^{(2)}}$ scenario.}
   \label{fig:xsec-cme2}
\end{figure}

\begin{figure}[htbp] 
   \centering
   \includegraphics[width=5in]{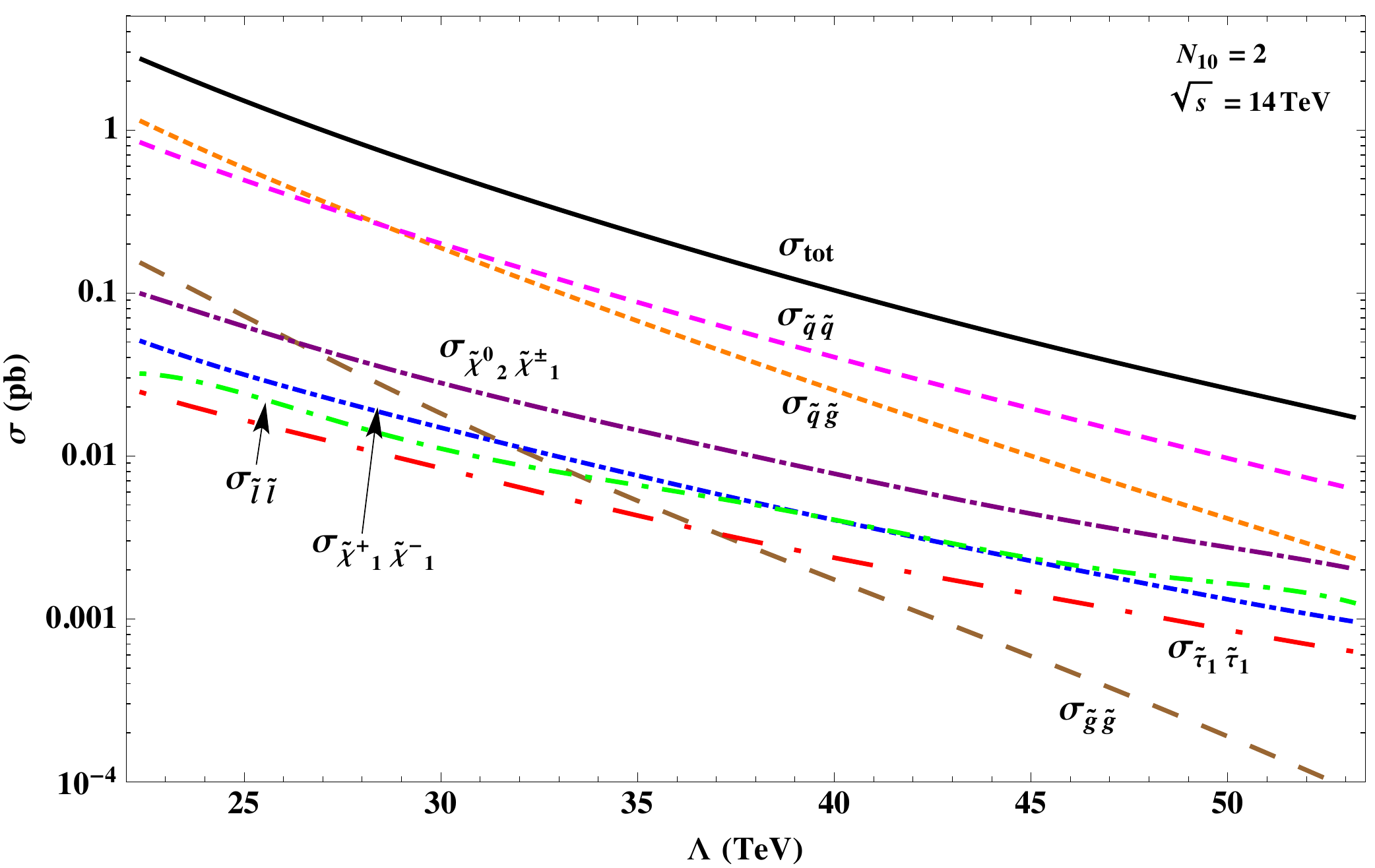}
   \caption{Plot of the leading order cross sections of various production channels for F-theory GUTs with $N_{10} = 2$ as a function of $\Lambda$ at $\Delta_{PQ} = 0$.}
   \label{fig:xsec-lbd-N10eq2}
\end{figure}

\newpage
\bibliographystyle{ssg}
\bibliography{e8coll}

\end{document}